\documentclass[twocolumn,twocolappendix]{aastex63}
\usepackage{hyperref}
\usepackage{verbatim}
\usepackage{tabularx}
\usepackage{breqn}

\newcommand{\otwo}{{[O\,II]\,}}

\def\D{{\Delta}}
\def\L{{\Lambda}}

\def\2pr{^{\prime \prime}}

\def\geqsim{\lower.73ex\hbox{$\sim$}\llap{\raise.4ex\hbox{$>$}}$\,$}
\def\leqsim{\lower.73ex\hbox{$\sim$}\llap{\raise.4ex\hbox{$<$}}$\,$}

\newcommand{\oii}{[\ion{O}{2}]~}

\newcommand{\oiii}{[\ion{O}{3}]~} % added by anand
\newcommand{\revisionforreviewer}[1]{{{#1}}} % revision based on Reviewer comments

\newcommand{\eg}{\textit{e.g.}}
\newcommand{\ie}{\textit{i.e.}}

\begin{document}

\title{The Early Data Release of the Dark Energy Spectroscopic Instrument}
% Author list file generated with: mkauthlist 1.3.0 
% mkauthlist -j apj --orcid -c DESI Collaboration DESI-2021-0215_author_list_sortedb.csv DESI-2021-0215_author_list_sorted_apj.tex 

\author{DESI Collaboration: A.~G.~Adame}
\affiliation{Instituto de F\'{\i}sica Te\'{o}rica (IFT) UAM/CSIC, Universidad Aut\'{o}noma de Madrid, Cantoblanco, E-28049, Madrid, Spain}
\author{J.~Aguilar}
\affiliation{Lawrence Berkeley National Laboratory, 1 Cyclotron Road, Berkeley, CA 94720, USA}
\author[0000-0001-6098-7247]{S.~Ahlen}
\affiliation{Physics Dept., Boston University, 590 Commonwealth Avenue, Boston, MA 02215, USA}
\author[0000-0002-3757-6359]{S.~Alam}
\affiliation{Tata Institute of Fundamental Research, Homi Bhabha Road, Mumbai 400005, India}
\author{G.~Aldering}
\affiliation{Lawrence Berkeley National Laboratory, 1 Cyclotron Road, Berkeley, CA 94720, USA}
\author[0000-0002-5896-6313]{D.~M.~Alexander}
\affiliation{Centre for Extragalactic Astronomy, Department of Physics, Durham University, South Road, Durham, DH1 3LE, UK}
\affiliation{Institute for Computational Cosmology, Department of Physics, Durham University, South Road, Durham DH1 3LE, UK}
\author{R.~Alfarsy}
\affiliation{Institute of Cosmology \& Gravitation, University of Portsmouth, Dennis Sciama Building, Portsmouth, PO1 3FX, UK}
\author{C.~Allende~Prieto}
\affiliation{Departamento de Astrof\'{\i}sica, Universidad de La Laguna (ULL), E-38206, La Laguna, Tenerife, Spain}
\affiliation{Instituto de Astrof\'{i}sica de Canarias, C/ Vía L\'{a}ctea, s/n, E-38205 La Laguna, Tenerife, Spain}
\author{M.~Alvarez}
\affiliation{Lawrence Berkeley National Laboratory, 1 Cyclotron Road, Berkeley, CA 94720, USA}
\author{O.~Alves}
\affiliation{University of Michigan, Ann Arbor, MI 48109, USA}
% ...

\submitjournal{\aj}

% A hack to get abstract to start on new page; a \clearpage here doesn't seem to work
\vspace{0.6in}

\begin{abstract}
The Dark Energy Spectroscopic Instrument (DESI) completed its five-month Survey Validation in May 2021.
Spectra of stellar and extragalactic targets from Survey Validation constitute the first major data sample from the DESI survey.
This paper describes the public release of those spectra, the catalogs of derived properties, and the intermediate data products.
In total, the public release includes good-quality spectral information from 466,447 objects targeted as part of the Milky Way Survey, 428,758 as part of the Bright Galaxy Survey, 227,318 as part of the Luminous Red Galaxy sample, 437,664 as part of the Emission Line Galaxy sample, and 76,079 as part of the Quasar sample.
In addition, the release includes spectral information from 137,148 objects that expand the scope beyond the primary samples as part of a series of secondary programs.
Here, we describe the spectral data, data quality, data products, Large-Scale Structure science catalogs, access to the data, and references that provide relevant background to using these spectra.
\\
\end{abstract}

%%%%%%%%%%%%%%%%%%%%%%%%%%%%%%%%%%%%%%%%%%%%%%%%%%

% Table of contents (not required by ApJ)
% \begingroup
% \let\clearpage\relax
% \tableofcontents
% \endgroup

%%%%%%%%%%%%%%%%%%%%%%%%%%%%%%%%%%%%%%%%%%%%%%%%%%

\setcounter{footnote}{0} 
\section{Introduction}

Wide-field imaging and spectroscopy enable a host of astrophysical studies that range from the largest cosmological scales to the local environment of the Milky Way galaxy.
Starting in 2000, the Sloan Digital Sky Survey \citep[SDSS;][]{sdss2000} represents the largest such program.
The \revisionforreviewer{largest public release of SDSS spectroscopic data}, DR17, includes 5,580,057 optical and near-infrared spectra passing quality cuts \citep{sdssDR17}. 
Data from SDSS has been used in more than 11,200 peer-reviewed publications.\footnote{http://tinyurl.com/SDSSPapers1}
Following the precedent of SDSS, recent releases from the Dark Energy Survey \citep{DESY3Gold,DESDR2}, {\it Gaia} collaboration \citep{gaia}, Hyper Suprime-Cam Subaru Strategic Program \citep{HSC}, and the Galaxy And Mass Assembly team \citep{gama} represent broader efforts of wide-field survey teams to provide well-calibrated data with comprehensive documentation to the public.

The Dark Energy Spectroscopic Instrument \citep[DESI;][]{desi16a,desi16b} began science observations in December 2020, making it the first Stage-IV \citep{detf} dark energy program to begin operations.
DESI will obtain spectra of stars, galaxies, and quasars over approximately 14,000 deg$^2$.
Data from the Milky Way Survey program \citep[MWS;][]{cooper22a}, comprised of more than 7 million spectroscopically-confirmed stars, will be used to characterize the assembly history and mass profile of our Galaxy.
The extragalactic spectroscopic sample, consisting of nearly 14 million bright galaxies \citep[BGS;][]{hahn22a}, 7.5 million luminous red galaxies \citep[LRG;][]{zhou22a}, 15.5 million emission line galaxies \citep[ELG;][]{raichoor22a}, and 3 million quasars \citep[QSO;][]{chaussidon22a}, will be used to explore the fundamental physics that governs the evolution of the Universe.
The DESI sample size will be ten times larger than the totality of the SDSS spectroscopic programs for extragalactic targets.

In this paper, we describe the public release of the first sample of DESI spectroscopic data, the ``Early Data Release'' (EDR).
The data in this release originate from the ``Survey Validation'' (SV) of DESI \citep{DESI23a} that took place between December 2020 and May 2021, prior to the start of the DESI Main Survey.
The first phase of SV, ``Target Selection Validation'' (abbreviated SV1), was comprised of observations made to refine and validate the selection of targets for the MWS, BGS, LRG, ELG, and QSO samples \citep{myers23a}.
Compared to the DESI Main Survey, this phase used looser target selection cuts to span a larger range of observed properties and observed these targets to higher signal-to-noise (S/N).
This enabled building truth samples, optimizing target selection cuts, and tuning the necessary signal-to-noise to meet the survey requirements \citep{DESI23a}.
After a brief ``Operations Development'' phase (SV2), DESI finished SV with the
``One-Percent Survey'' (SV3), which further optimized the efficiency of observing procedures and
produced samples with very high fiber assignment completeness for clustering studies over an area
that is approximately 1\% of the final DESI Main Survey.

This paper is organized as follows.
In \S\ref{sec:data}, we present the DESI instrument, the design of SV observations, and a brief summary of the target classes contained in this data release.
A full description of target classes can be found in Appendix~\ref{app:primarytargets} and Appendix~\ref{app:secondarytargets}.
In \S\ref{sec:products}, we discuss the spectral processing, data quality, and data included in this release. Next, in \S\ref{sec:lsscat}, we detail the creation and uses of the large-scale structure (LSS) value-added catalogs of the One-Percent Survey that accompany this release.
In \S\ref{sec:access}, we describe online access to the data and tutorials with examples of working with the data.
Finally, in \S\ref{sec:conclusion}, we provide a brief summary of results produced with these SV data and the plans for future releases.

\section{Data Acquisition}\label{sec:data}

The DESI spectrographs were built to obtain spectra of roughly 40 million galaxies and quasars over a five-year period to study dark energy through measurements of large-scale structure.
The maps produced with these spectroscopic samples are expected to allow volume-averaged measurements of the baryon acoustic oscillation (BAO) feature at a precision better than 0.5\% over each of the intervals $0.0 < z < 1.1$, $1.1 < z < 1.9$, and $1.9 < z < 3.7$.
These maps will also allow percent-level precision measurements of redshift space distortions (RSD) over each interval $0.0 < z < 1.1$ and $1.1 < z < 1.9$.
Here, we present an overview of the instrument design, observing strategy, phases of SV observing, and the SV samples that were used to inform the strategy to make these cosmological measurements.

%----------------------------------------------------------------------------------------------------------
\subsection{Instrument Design}
\label{sec:instdesign}
DESI requires a wide field of view that was made possible with a new prime focus corrector at the
NOIRLab's\footnote{Formerly named the National Optical Astronomy Observatory (NOAO).}
4-m Mayall telescope at Kitt Peak National Observatory in Arizona.
These optics allow spectroscopy over a 0.8-meter diameter focal plane \citep{Miller23a} located at the prime focus, corresponding to roughly 3.2 degrees on the sky and a field of view just over 8~deg${}^2$.
Installed in this focal plane are 5,020 robotically-controlled fiber positioners \citep{silber22a}, each holding a unique fiber with a core diameter of 107 $\mu$m $\sim 1.5$ arcsec \citep{poppett23a}. Twenty fibers direct light to a camera to monitor the sky brightness, while the remaining 5,000 fibers direct the light of a targeted object from the primary focus to one of ten spectrographs.
The focal plane is constructed of 10 ``petals'', with each petal of 500 science fibers mapping to a single spectrograph that measures all 500 targets simultaneously.
These spectrographs have three cameras, denoted as B (3600--5800~\AA), R (5760--7620~\AA), and
Z (7520--9824~\AA),\footnote{Future DESI data releases may adjust the exact wavelength grid extracted from the data.}
that provide a resolving power of roughly 2000 at 3600~\AA, increasing to roughly 5500 at 9800~\AA~\citep{jelinsky23a}.

The full system has the sensitivity to measure and resolve the [O\textsc{ii}] doublet down to fluxes
of $8 \times 10^{-17}~\mathrm{erg/s/cm}^2$ in effective exposure times of 1000 seconds for galaxies $0.6<z<1.6$. 
Here effective exposure time corresponds to an exposure time in reference conditions --- zenith, dark sky, FWHM seeing of 1.1~arcsecond, and no Galactic extinction (see \S\ref{sec:efftimespec} for a summary and \S4.14 of \citealt{guy22a} for details).  
At this effective exposure time, and accounting for increased overhead due to Galactic extinction, airmass, weather, operational overheads, and engineering downtime, the DESI instrument can be used to complete a 14,000 deg$^2$ survey in five years.
A comprehensive description of the completed instrument can be found in \citet{desi-collaboration22a}.

%----------------------------------------------------------------------------------------------------------
\subsection{Observing Strategy}
\label{sec:obsstrategy}

DESI has five ``\textit{primary}'' target classes (MWS, BGS, LRG, ELG, and QSO in increasing order of mean redshift),
as well as many ``\textit{secondary}'' target classes which are generally used as filler samples for 
fibers that cannot reach a primary target.
Observations are based upon \textit{tiles}, which are a given pointing of
the telescope combined with assignments of each fiber to a specific target
for that telescope pointing. 
Tiles are associated with a single \textit{survey}, or phase of the DESI
operations.  Tiles are further grouped within a survey by their
\textit{program} indicating the observing conditions under which they
should be observed.  For example, BGS and MWS targets are assigned to
``bright'' program tiles, while fainter ELG, LRG, and QSO targets are
assigned to ``dark'' program tiles. 
The bright vs.~dark distinction is based upon survey speed, or how quickly the instrument can
accumulate signal-to-noise given the current observing conditions, as estimated by the
exposure time calculator \citep{kirkby23a}.  Bright tiles are observed when the survey speed
is $2.5\times$ worse than the reference conditions described in \S\ref{sec:instdesign}.
``Backup'' program tiles are used when conditions are too poor for bright tiles,
$12.5\times$ worse than reference conditions, where bright stars are targeted.
Multiple programs were interleaved during each survey, selected dynamically
based upon current observing conditions.
During SV, some bright tiles were purposefully observed under both bright and dark conditions to
provide comparison datasets for quality assurance tracking.

Each program is subdivided into one or more subprograms, called \textit{fiberassign programs} (data column \texttt{FAPRGRM}). For the majority of programs, there is only one fiberassign program of the same name, however, some programs like the SV1 survey's dark program contained multiple fiberassign programs to differentiate the various purposes of each set of tiles. More information on fiberassign programs will be described in \citet{raichoor23a}. Table~\ref{tab:surveyprograms} lists the surveys, programs, and fiberassign programs available in the EDR, with further details in Table~\ref{tab:secondary_tiles} and the Appendices.
With the exception of commissioning and specifically designed tiles in a ``special'' survey, each survey included dark, bright, and backup programs,
while Target Selection Validation (\texttt{SURVEY=sv1}) also includes an ``other'' program for tiles dedicated to secondary targets.

DESI files typically follow the convention of uppercase column names and lowercase string values, \eg~\texttt{PROGRAM=dark}.\footnote{A notable exception is the spectral classification \texttt{SPECTYPE} column, which has uppercase values \texttt{GALAXY}, \texttt{QSO}, and \texttt{STAR}.}
In the data files, tiles are tracked
with a unique integer \texttt{TILEID}, and each tile is associated with specific strings for \texttt{SURVEY}, \texttt{PROGRAM}, and \texttt{FAPRGRM}.

\begin{deluxetable}{ccc}[htb]
\caption{Surveys, programs, and fiberassign programs available in the EDR, in increasing order of specificity. Each fiberassign program represents a choice for how targets were selected for a given observation and under what conditions a tile should be nominally observed. See Table~\ref{tab:secondary_tiles} for more details on the non-standard fiberassign programs.}
\label{tab:surveyprograms}
\tablehead{\colhead{\texttt{SURVEY}} & \colhead{\texttt{PROGRAM}} & \colhead{\texttt{FAPRGRM}}}
\startdata
cmx & other & m33\\[4pt]\hline\noalign{\vskip 2pt}
special & dark & dark\\[4pt]\hline\noalign{\vskip 2pt}
sv1 & backup & backup1\\[4pt]\hline\noalign{\vskip 2pt}
sv1 & bright & bgsmws\\[4pt]\hline\noalign{\vskip 2pt}
sv1 & dark & \begin{tabular}{@{}c@{}c@{}c@{}}elg \\elgqso \\lrgqso \\lrgqso2  \end{tabular}\\[4pt]\noalign{\vskip 4pt}\hline\noalign{\vskip 2pt}
sv1 & other & \begin{tabular}{@{}c@{}c@{}c@{}c@{}c@{}c@{}c@{}c@{}c@{}c@{}c@{}}dc3r2 \\m31 \\mwclusgaldeep \\praesepe \\rosette \\scndcosmos \\scndhetdex \\ssv \\umaii \\unwisebluebright \\unwisebluefaint \\unwisegreen  \end{tabular}\\\noalign{\vskip 4pt}\hline\noalign{\vskip 2pt}
sv2 & backup & backup\\[4pt]\hline\noalign{\vskip 2pt}
sv2 & bright & bright\\[4pt]\hline\noalign{\vskip 2pt} 
sv2 & dark & dark\\[4pt]\hline\noalign{\vskip 2pt} 
sv3 & backup & backup\\[4pt]\hline\noalign{\vskip 2pt}
sv3 & bright & bright\\[4pt]\hline\noalign{\vskip 2pt}
sv3 & dark & dark\\
\enddata
\end{deluxetable}

% \begin{deluxetable*}{ccc}[htb]
% \caption{Surveys, programs, and fiberassign programs available in the EDR, in increasing order of specificity. Each fiberassign program represents a choice for how targets were selected for a given observation and under what 
% conditions a tile should be nominally observed under.}
% \label{tab:surveyprograms}
% \tablehead{\colhead{SURVEY} & \colhead{PROGRAM's} & \colhead{FAPRGRM's}}
% \startdata
% cmx & other & m33\\\\\hline \\
% special & dark & dark\\\\\hline \\
% sv1 & backup & backup1\\\\\hline \\
% sv1 & bright & bgsmws\\\\\hline \\
% sv1 & dark & \begin{tabular}{@{}c@{}c@{}c@{}}elg \\elgqso \\lrgqso \\lrgqso2  \end{tabular}\\\\\hline \\
% sv1 & other & \begin{tabular}{@{}c@{}c@{}c@{}c@{}c@{}c@{}c@{}c@{}c@{}c@{}c@{}}dc3r2 \\m31 \\mwclusgaldeep \\praesepe \\rosette \\scndcosmos \\scndhetdex \\ssv \\umaii \\unwisebluebright \\unwisebluefaint \\unwisegreen  \end{tabular}\\\\\hline \\
% sv2 & backup & backup\\\\\hline \\
% sv2 & bright & bright\\\\\hline \\
% sv2 & dark & dark\\\\\hline \\
% sv3 & backup & backup\\\\\hline \\
% sv3 & bright & bright\\\\\hline \\
% sv3 & dark & dark\\\\
% \enddata
% \end{deluxetable*}

Tiles overlap on the sky, enabling both greater fiber assignment completeness
for dense targets such as ELGs, as well as the opportunity to observe
fainter targets to higher signal-to-noise by combining observations across
multiple tiles, \eg~for high redshift quasars.  Each target has a unique
integer \texttt{TARGETID} to track their observations across multiple tiles,
and targeting bitmasks\footnote{A bitmask is an integer generated from setting multiple predefined bits to either 0 or 1 to indicate False or True respectively. The integer is then computed as $\sum_{i} 2^{i}b_{i}$, with $b_{i}$ being 0 or 1 and~$i$~being the bit number. For more information, see \url{https://github.com/desihub/desitarget/blob/2.5.0/doc/nb/target-selection-bits-and-bitmasks.ipynb}.} to track the reason(s) that they were selected for
observation, as detailed in \cite{myers23a}, and Appendix \ref{app:primarytargets} and Appendix \ref{app:secondarytargets}.

A given \texttt{TARGETID} could be assigned to a single tile;
multiple tiles of the same \texttt{SURVEY} and \texttt{PROGRAM} (such as QSO targets); or multiple tiles of different \texttt{SURVEY}s or \texttt{PROGRAM}s
(such as brighter LRGs on \texttt{PROGRAM=dark} tiles also being selected as
BGS targets on \texttt{PROGRAM=bright} tiles).
To preserve the data uniformity within a (survey,~program) combination, target selection and fiber assignments within a (survey, program) are independent of whether the target is also selected and assigned in a different (survey,~program), even when this results in additional observations of the same target.  
Similarly, these data are also processed independently such that spectra are not coadded or fit across different surveys and programs.

\subsection{SV Observations}\label{sec:svobservations}
\begin{figure}
    \centering 
    \includegraphics[width=1.01\columnwidth]{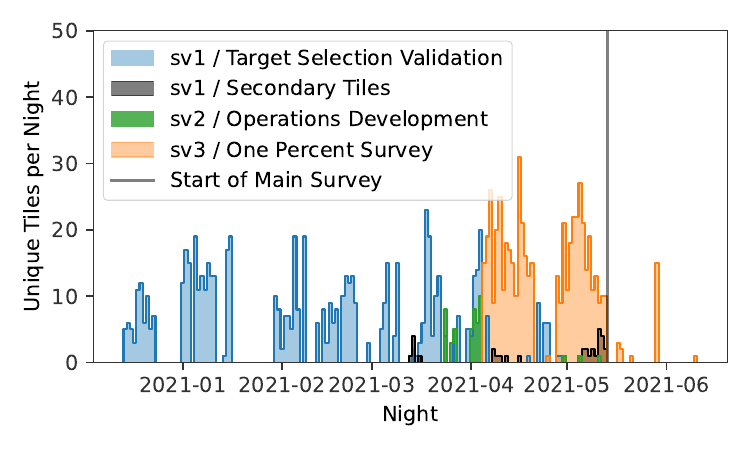}
    \caption{The number of unique tiles per night observed during Survey Validation.  The same tile can be observed on multiple nights.}
    \label{fig:tiles_per_night}
\end{figure}

DESI Survey Validation observations began on December 14, 2020 and nominally concluded on May 13, 2021 (see \citealt{DESI23a} and \S2.3 of \citealt{myers23a} for further details).
% bin/extra_obs for counting these SV tiles observed during the Main Survey
An additional 22 SV-designed tiles were observed on 5 nights after the
start of the Main Survey on May 14, 2021; with the final observations taking place on June 10, 2021. 
The additional tiles were observed to improve the completeness of the
One-Percent Survey areas, and are included in the EDR.
Table~\ref{tab:survey_nights_tiles_exp} lists the number of nights, tiles, exposures,
effective exposure times, and area covered by tiles for each survey included in the EDR.
Figure~\ref{fig:tiles_per_night} shows the number of unique tiles per night
for each of the three phases of Survey Validation.  Although there are distinct
boundaries for when each survey began, there is an overlap in dates as incomplete tiles from a previous survey were sometimes completed after the start of the next survey.
A given tile can be observed on multiple nights, and thus contributes to multiple
bins, but if it was observed multiple times on a single night it is only counted once for
that night in Figure~\ref{fig:tiles_per_night}.

% from edrpaper/bin/count_obs
\begin{deluxetable*}{ccccccc}[htb]
\tablecaption{Number of nights, tiles, exposures, effective exposure time, and approximate area
covered by tiles for surveys included in the Early Data Release.
\texttt{SURVEY=sv1} includes both Target Selection Validation tiles and
tiles dedicated to secondary targets.
The area covered by tiles is larger than the true effective area available to targets due to
bright star exclusions, focal plane geometry, hardware configuration, and higher priority targets blocking
lower priority targets.
}
\label{tab:survey_nights_tiles_exp}
\tablehead{Survey   & Description & Nights & Tiles & Exposures & Effective Hours & Covered Area \\ 
                    &             &        &       &           &                 & [deg$^2$] }
\startdata
cmx      & Commissioning                  &  1 &    1 &      4 &    0.9 & 8 \\
special  & Test Tiles                     &  2 &   16 &     21 &    0.3 & 75 \\
sv1      & Target Selection Validation    & 73 &  175 &   1568 &  137.5 & 1040 \\
sv1      & Secondary Tiles                & 18 &   13 &    105 &   37.9 & 67 \\
sv2      & Operations Development             & 10 &   39 &     72 &    6.4 & 102 \\
sv3      & One-Percent Survey             & 38 &  488 &    710 &  102.2 & 197 \\
\enddata
\end{deluxetable*}

% from edrpaper/bin/count_obs

The covered area listed in Table~\ref{tab:survey_nights_tiles_exp} is simply the unique area that
is overlapped by any tile in that survey, while not double-counting area covered by more than one tile.
This gives a sense of the scope of the Survey Validation observations,
but note that these areas are larger than the true effective area due to gaps in the focal plane coverage, disabled or broken hardware, and target assignment priorities, as will be discussed in \S\ref{sec:lss_inputs}.

%%% sv1
\subsubsection{Target Selection Validation}\label{sec:sv1}
The first phase of SV observations, Target Selection Validation, 
optimized the DESI survey strategy and target selection algorithms. 
Target Selection Validation used \texttt{SURVEY=sv1}.\footnote{We use capital ``SV1'' as the acronym in text descriptions and column names, but lowercase
\texttt{sv1} for values in data files, as well as directory and file names on disk.}

A key product of the observations for Target Selection Validation was the calibration of effective exposure times.
As stated previously, effective exposure times account for varying throughput and background and provide a standard metric of exposure depth that corresponds to an exposure time at airmass 1, zero Galactic extinction, 1.1 arcsecond FWHM seeing, and zenith dark sky. 
DESI uses two effective times, \texttt{EFFTIME_ETC} and \texttt{EFFTIME_SPEC}, which are used for determining when to stop an exposure of a tile at the telescope and determining when a tile has been observed enough to meet survey specifications, respectively. 
\texttt{EFFTIME_SPEC} is based on the offline spectroscopic data and will be described in \S\ref{sec:efftimespec}. 
\texttt{EFFTIME_ETC} is based on active monitoring of the sky conditions and location on the sky \citep{kirkby23a}.
The sky brightness is monitored with sky monitor fibers along the outer rim of the focal plane, which send light to a dedicated imaging system that is read out regularly during the spectroscopic exposure. 
\revisionforreviewer{These are independent of the sky targets used in the spectroscopic data processing. Sky targets used in the data processing are assigned to robotic positoners on the focal plane along with the other target types, and their light is acquired using one of the DESI spectrographs.}
The image quality and sky transparency are derived from the guide focus assembly (GFA) system. 
For more information about these components, see \cite{desi-collaboration22a}. 
For the Main Survey, a calibrated algorithm called the Exposure Time Calculator (ETC) determines when the exposure is estimated to be complete \citep{kirkby23a}.
For SV1, there was no calibration of sky conditions to effective time, so a power law: $t_{\rm exp} = t_0X^{1.25}$~was used. 
Here $X$ is the airmass, $t_0$~is the nominal time, and the relation is empirically derived from BOSS/eBOSS data. 
These data were used to calibrate the \texttt{EFFTIME_SPEC} of the offline pipeline, which in turn was used to calibrate \texttt{EFFTIME_ETC}.

During Target Selection Validation, requested effective exposure times were increased by approximately a factor of four relative to the survey design to provide high signal-to-noise spectra.
These data were typically collected over four different nights to allow tests of redshift classification using subsets of data acquired under different observing conditions. 

In addition to the four-epoch strategy, Target Selection Validation observations included even deeper exposures of 
some tiles. The deepest tiles for each of the five primary DESI target classes (MWS, BGS, LRG, ELG, and QSO) are listed
in Table~\ref{tab:deeptiles}.
These tiles contain spectra that are much higher signal-to-noise than most data expected from the program and can therefore be used for unique studies of stellar, galaxy, or quasar astrophysics.
A subset of these deep tiles (including tiles not listed in Table~\ref{tab:deeptiles}) also have visual inspections; see \S\ref{sec:vac-files}.

% from edrpaper/bin/sv1_tile_depth
\begin{table}
\caption{The 5 deepest tiles for the five primary target classes in DESI.\label{tab:deeptiles}}
\begin{tabular}{ccc}
% \begin{tabular}{lll}
\hline
\hline
\texttt{TILEID} & Targets & Effective Time [hours] \\
\hline
80613 & MWS,BGS & 0.85 \\
80736 & MWS & 0.95 \\
80607 & LRG,QSO & 2.6 \\
80608 & ELG & 4.2 \\
80711 & ELG,QSO & 6.7 \\

\hline
\end{tabular}
\end{table}

% source: bin/count_obs for effective hours per survey
 In total, 137.5 effective hours during SV were dedicated to 175 Target Selection Validation (SV1) tiles.

%%% sv2
\subsubsection{Operations Development}\label{sec:sv2}
After Target Selection Validation, SV continued with
an Operations Development phase with \texttt{SURVEY=sv2}, in preparation for the One-Percent Survey.  The purpose of these observations was to validate the end-to-end operational procedures needed to schedule observations of targets in a tile, process those observations, identify successfully acquired redshifts, and determine which targets were completed or should be scheduled for new observations in additional overlapping tiles. A major focus of this phase was to establish the ``Merged Target List'' ledgers \citep[MTLs; see \S5 of][]{schlafly23a}, which track the observational state and redshift of each target and determine whether a target requires further observations.
% source: bin/count_obs
In total, 6.4 effective hours during SV were dedicated to 39 Operations Development tiles.

%%% sv3
\subsubsection{One-Percent Survey}\label{sec:sv3}
The final phase of the SV program, the One-Percent Survey, (\texttt{SURVEY=sv3}) was used to validate final operational procedures and compile extensive samples of sources that could be used for clustering studies. The One-Percent Survey was conducted over the period April 5 to June 10, 2021, with the vast majority of observations occurring on or before May 13, 2021. In total, the One-Percent Survey covered 20 fields, each with a `rosette' pattern of $\sim$10-11 overlapping tiles in bright time and $\sim$12-13 tiles in dark time \revisionforreviewer{whose centers were offset in a circle of radius 0.12 deg from the field center. The overlapping tiles provided high fiber assignment completeness over an area of 6.48 deg$^2$; within which more than 95\% of MWS and ELG targets and more than 99\% of BGS, LRG, and QSO targets, received fibers.}
Targets covering an additional $\sim$2 deg$^2$ were observed with fewer visits and lower completeness in fiber assignment \revisionforreviewer{because of fewer overlapping tiles in the outer edge of the rosette of each field and the hole in the center of the DESI focal plane. Figure \ref{fig:BGSntile} shows the tile coverage of one bright time rosette.} Large-Scale Structure catalogs were created covering the entire area and are detailed in \S\ref{sec:lsscat}.

One-Percent Survey rosettes were selected to cover major datasets from other surveys,
including the Cosmic Evolution Survey \citep[COSMOS;][]{cosmos2007_overview},
Hyper Suprime-Cam \citep[HSC;][]{hsc2018_overview},
Dark Energy Survey \citep[DES;][]{des2016_overview} deep fields,
Galaxy And Mass Assembly \citep[GAMA;][]{gama2011_overview},
Great Observatories Origins Deep Survey \citep[GOODS;][]{goods2003_overview},
and anticipated deep fields from future
Legacy Survey of Space and Time \citep[LSST;][]{ivezic19}
and Euclid \citep{euclid2022_overview} observations.
These are summarized in Table~\ref{tab:SV_fields}.
% source: bin/count_obs
In total, 102.2 effective hours during SV were dedicated to 488 tiles in the One-Percent Survey (SV3).

%%%%%
% VI and SV3 Tiles
% See https://desi.lbl.gov/trac/wiki/SurveyOps/OnePercent and bin/sv3_rosette_tiles
\begin{deluxetable*}{ccccc}[htb]
\tablecaption{Selected DESI tiles with visual inspections (VI) or
overlapping with datasets from other surveys.
``SV3~R$n$'' denotes rosette number $n$ from the One-Percent Survey,
with RA and DEC referring to the center of the Rosette rather than the center
of an individual tile.
}
\label{tab:SV_fields}
\tablehead{
    \colhead{Set} & \colhead{RA} & \colhead{DEC} & \colhead{\texttt{TILEID}(s)} & \colhead{Other Surveys}
}
\startdata
VI LRG+QSO  & 36.448   & -4.601  & 80605                & XMM-LSS \\
VI ELG      & 36.448   & -4.501  & 80606                & XMM-LSS \\
VI LRG+QSO  & 106.74   & 56.100  & 80607                & Lynx    \\
VI ELG      & 106.74   & 56.200  & 80608                & Lynx    \\
VI BGS      & 106.74   & 56.100  & 80613                & Lynx    \\
VI LRG+QSO  & 150.12   & 2.206   & 80609                & COSMOS  \\
VI ELG      & 150.12   & 2.306   & 80610                & COSMOS  \\
\hline
SV3 R0      & 150.10   &  2.182  & 1-21,23-24,442       &
    \begin{tabular}{c} COSMOS, DES deep, LSST deep, CFHTLS-D2 \\ HSC ultradeep, VVDS-F10, DESI SV1 80871 \end{tabular} \\
SV3 R1      & 179.60   &  0.000  & 28-53,445            & GAMA G12, KiDS-N, DESI SV1 80662 \\
SV3 R2      & 183.10   &  0.000  & 55-79,448            & GAMA G12, KiDS-N \\
SV3 R3      & 189.90   & 61.800  & 82-107,451-452       & GOODS-North \\
SV3 R4      & 194.75   & 28.200  & 109-129,131-133,454  & Coma cluster, DESI SV1 80707 \\
SV3 R5      & 210.00   &  5.000  & 136-156,158,457      & VVDS-F14 \\
SV3 R6      & 215.50   & 52.500  & 163-187,460          & DEEP2, CFHTLS-D3/W3, DESI SV1 80711 \& 80712 \\
SV3 R7      & 217.80   & 34.400  & 190-215,463          & Bootes NDWFS/AGES \\
SV3 R8      & 216.30   & -0.600  & 217-238,466          & GAMA G15, HSC DR2, KiDS-N \\
SV3 R9      & 219.80   & -0.600  & 244-265,469          & GAMA G15, HSC DR2, KiDS-N \\
SV3 R10     & 218.05   &  2.430  & 271-291,472          & GAMA G15, HSC DR2, KiDS-N \\
SV3 R11     & 242.75   & 54.980  & 298-321,475          & ELAIS N1, HSC deep field, DESI SV1 80865--80867 \\
SV3 R12     & 241.05   & 43.450  & 325-347,478          & HSC DR2 \\
SV3 R13     & 245.88   & 43.450  & 352-375,433          & HSC DR2 \\
SV3 R14     & 252.50   & 34.500  & 379-402,436-437      & XDEEP2 \\
SV3 R15     & 269.73   & 66.020  & 406-429,439          & Ecliptic pole, Euclid deep field \\
SV3 R16     & 194.75   & 24.700  & 481-491,495-504,506  & Coma cluster outskirts \\
SV3 R17     & 212.80   & -0.600  & 511-521,525-534      & GAMMA G15, HSC DR2, KiDS-N \\
SV3 R18     & 269.73   & 62.520  & 541-551,555-565      & Near ecliptic pole \\
SV3 R19     & 236.10   & 43.450  & 571-581,585-596      & HSC DR2 \\
\enddata
\end{deluxetable*}

\subsubsection{Other SV Observations}
\label{sec:svother}

In addition to Target Selection Validation, Operations Development, and One-Percent Survey tiles, Survey Validation
observed additional tiles dedicated to secondary targets proposed by members of the DESI collaboration to extend the scientific reach of this Early Data Release.  These tiles are listed in
Table~\ref{tab:secondary_tiles}, with the
target selection bits described in Appendix~\ref{app:secondarytargets}.  Secondary targets were also used
as low-priority filler targets on other tiles, to be used in cases when an available fiber could not
reach a primary target and was not needed for calibration targets (standard stars or sky locations).
% source: bin/count_obs
DESI observed 37.9 effective hours of these 13 dedicated secondary target tiles.

% https://desi.lbl.gov/trac/wiki/SurveyOps/TileDesigns#a80862-80872-Dedicatedsecondaryprograms
% https://desi.lbl.gov/trac/wiki/SurveyOps/TileDesigns#a80971-80976-Dedicatedsecondaryprogramscontinued
\begin{deluxetable*}{lp{3.5cm}p{1.5cm}p{1cm}p{7cm}}[htb]
% \begin{deluxetable*}{lp{2.5cm}p{1.5cm}p{1cm}p{8cm}}[htb]
%\begin{rotatetable*}
%\begin{deluxetable*}{lcccc}[htb]
%\begin{deluxetable*}{lcccc}
\tablecaption{Fiberassign programs, descriptions, TILEID ranges, effective exposure time and targeting bits for all tiles dedicated to secondary targets.  See Appendix~\ref{app:secondarytargets} for a description of the targeting bits for each program. \revisionforreviewer{Most of these fiberassign programs targeted specific secondary program objects, while `rosette' and `praesepe' used all available secondary targets and `dc3r2' used dark time primary targets as filler targets after assigning the desired secondary targets. }}
\label{tab:secondary_tiles}
% \tablehead{Description & Targeting bits & Tiles & Effective Hours}
\tablehead{FAPRGRM & Description & TILEID(s) & Effective Hours & Targeting Bit-names}
\startdata
%Milky Way star clusters and dwarf galaxies     & ... & 80862--80864 & ... \\
%... & ... & ... & ... \\
m33 & M33 & 80615 & 0.94 & \texttt{M33\_}\{\texttt{H2PN},  \texttt{GC},  \texttt{QSO},  \texttt{M33cen},  \texttt{M33out}\}\\
 & & & & \texttt{SV0\_\{WD,QSO,LRG,ELG\}}\\
m31 & M31 & 80715 & 0.50 & \texttt{M31\_KNOWN},  \texttt{M31\_QSO},  \texttt{M31\_STAR}\\
rosette & Rosette Nebula & 80718 & 0.01 & All secondary target \revisionforreviewer{bits}\\
praesepe & Praesepe (Beehive cluster) & 80719 & 0.03 & All secondary target \revisionforreviewer{bits} except \texttt{LOW\_Z}\\
umaii & Ursa Major II dwarf galaxy & 80720 & 0.03 & \texttt{MWS\_}\{\texttt{ANY}, \texttt{CALIB}, \texttt{MAIN\_CLUSTER\_SV}, \texttt{RRLYR}\}\\
 & & & & \texttt{BHB}, \texttt{BACKUP\_CALIB}\\
ssv & Stellar Survey Validation & 80721-80738 & 0.37 & \texttt{MWS\_}\{\texttt{NEARBY}, \texttt{MAIN\_BROAD}, \texttt{MAIN\_FAINT}\}\\
 & & & & \texttt{MWS\_}\{\texttt{WD}, \texttt{BHB}, \texttt{CALIB}\}\\
 & & & & \texttt{MWS\_}\{\texttt{MAIN\_CLUSTER\_SV},\texttt{RRLYR}\}\\
 & & & & \texttt{WD\_}\{\texttt{BINARIES\_BRIGHT}, \texttt{BINARIES\_DARK}\}\\
 & & & & \texttt{BACKUP\_}\{\texttt{FAINT}, \texttt{VERY\_FAINT}, \texttt{CALIB}\}\\
mwclusgaldeep & Star Clusters and Dwarf Galaxies & 80862-80863 & 2.38 & \texttt{MWS\_}\{\texttt{MAIN\_BROAD}, \texttt{NEARBY}, \texttt{MAIN\_FAINT}\}\\
 & & & & \texttt{MWS\_}\{\texttt{WD}, \texttt{BHB}, \texttt{CLUS\_GAL\_DEEP}\}\\
 & & & & \texttt{WD\_BINARIES\_DARK}\\
unwisegreen & n(z) calibration for CMB lensing cross-correlations & 80865 & 7.85 & \texttt{UNWISE\_GREEN\_II\_}\{\texttt{3700}, \texttt{II\_3800}, \texttt{3900}, \texttt{4000}\}\\
 & & & & \texttt{UNWISE\_BLUE\_FAINT\_II}\\
 & & & & \texttt{LOW\_MASS\_AGN},\texttt{LOW\_Z}\\
unwisebluebright & n(z) calibration for CMB lensing cross-correlations & 80866 & 0.68 & \texttt{UNWISE\_BLUE\_BRIGHT\_II}\\
unwisebluefaint & n(z) calibration for CMB lensing cross-correlations & 80867 & 1.79 & \texttt{UNWISE\_BLUE\_FAINT\_II}\\
scndhetdex & HETDEX follow-up and Ly-$\alpha$ tomography & 80869-80870 & 5.33 & \texttt{LBG\_TOMOG\_W3}, \texttt{HETDEX\_}\{\texttt{HP}, \texttt{MAIN}\}\\
 & & & & \texttt{LOW\_MASS\_AGN},\texttt{LOW\_Z}\\
scndcosmos & Various samples & 80871-80872 & 5.22 & \texttt{DESILBG\_}\{\texttt{TMG\_FINAL}, \texttt{BXU\_FINAL}, \texttt{G\_FINAL}\}\\
 & & & & \texttt{QSO}, \texttt{ISM\_CGM\_QGP}, \texttt{HSC\_HIZ\_SNE}\\
dc3r2 & Photo-z calibration & 80971-80975 & 1.40 & \texttt{DC3R2\_GAMA} and all DARK primary targets as filler\\
\enddata
\end{deluxetable*}
%\end{rotatetable*}

EDR includes one commissioning tile (\texttt{TILEID}=80615, \texttt{SURVEY=cmx})
covering M33 that was observed during the SV time period.
It also includes sixteen \texttt{SURVEY=special} tiles used for fiber assignment testing (\texttt{TILEID}s 81100--81115).
Although these special tiles used dark time targets (LRG, ELG, QSO), most were observed for
less than one minute of effective exposure time \revisionforreviewer{and have very low spectroscopic signal-to-noise. Only tiles 81100 and 81112 will be retained in
future data releases, as they have effective exposure times greater than two minutes and may provide sufficient signal for brighter targets.}
In total, 0.9 effective hours of exposure time were dedicated to commissioning observations
and 0.3 effective hours to \texttt{SURVEY=special} observations.

%%% Targeting bitmasks

% \begin{table*}[htb]
% \caption{The number of spectra obtained in SV, the One-Percent Survey, and dedicated programs}
% \begin{center}
% \begin{tabular}{lccc}
% \hline
% \hline
% Spectral Class & \# spectra & \# unique redshifts & average effective exposure time \\
% \hline
% \hline
% \multicolumn{4}{c}{\bf Deep SV Fields}
% \hline
% MWS & \\
% \hline
% \multicolumn{4}{c}{\bf Standard SV Fields}
% \hline
% MWS & \\
% \hline
% \multicolumn{4}{c}{\bf One-Percent Survey}
% \hline
% MWS & \\
% \hline
% \multicolumn{4}{c}{\bf Secondary Programs}
% \hline
% XXX & \\
% \hline
% \end{tabular}
% \end{center}
% \tablecomments{Comment.
% }
% \label{tab:spectra_stats}
% \end{table*}

%----------------------------------------------------------------------------------------------------------
\subsection{SV Target Samples}
\label{sec:sv_target_samples}

DESI primary targets (MWS, BGS, LRG, ELG, and QSO) are selected from Data Release~9 of the Legacy Imaging Surveys
\citep[LS/DR9;][]{Zou2017_BASS, dey19a, schlegel23a}, while secondary targets
could come from LS/DR9 or other sources.  
Targets identified for spectroscopic observations are recorded from the imaging data and documented for downstream redshift and clustering catalogs (see \citealt{myers23a} and Appendix \ref{app:primarytargets}).
The target selections used for the SV data in EDR are described
in \cite{allende-prieto20_tsmws_prelim} for MWS, \cite{ruiz-macias20_tsbgs_prelim} for BGS, \cite{zhou20_tslrg_prelim} for LRG, \cite{raichoor20_tselg_prelim} for ELG, and \cite{yeche20_tsqso_prelim} for QSO. 
Secondary targeting programs are briefly summarized in Appendix~\ref{app:secondarytargets} and references therein.
These algorithms were updated based on analysis of the EDR data resulting in the
DESI Main Survey final target selections documented for the MWS program in \citet{cooper22a}, BGS in \citet{hahn22a}, LRG in \citet{zhou22a}, ELG in \citet{raichoor22a}, and QSO in \citet{chaussidon22a}. 

The Early Data Release consists of 2,847,435 spectra unique to a given survey and program, when including science targets, standard stars, and sky fiber spectra. Of those, 2,757,937 are unique locations on the sky. Selecting only spectra that do not have hardware or observing flags yields 2,183,282 unique spectra. Further sub-selecting to science targets, the EDR contains 1,852,883 unique science spectra free of hardware (\eg~fiber, CCD pixels, positioner) flags and observing (\eg~poor positioning, low effective exposure time) flags. Finally, restricting to those that are free of redshift fitting (\eg~a bad fit) flags results in 1,712,004 unique, `good' target spectra and redshifts; including 1,125,635 \texttt{GALAXY}, 90,241 \texttt{QSO}, and 496,128 \texttt{STAR} spectral classifications. For more details about these selection choices see \S\ref{sec:qualitycuts}, and for more information on the redshift classification see \S\ref{sec:redrock} and \cite{bailey23a}. For the selection criteria used for the DESI large-scale structure catalogs, which are generally more restrictive, see \S\ref{sec:lsscat}.

\begin{figure}
     \centering 
     \includegraphics[width=1\columnwidth]{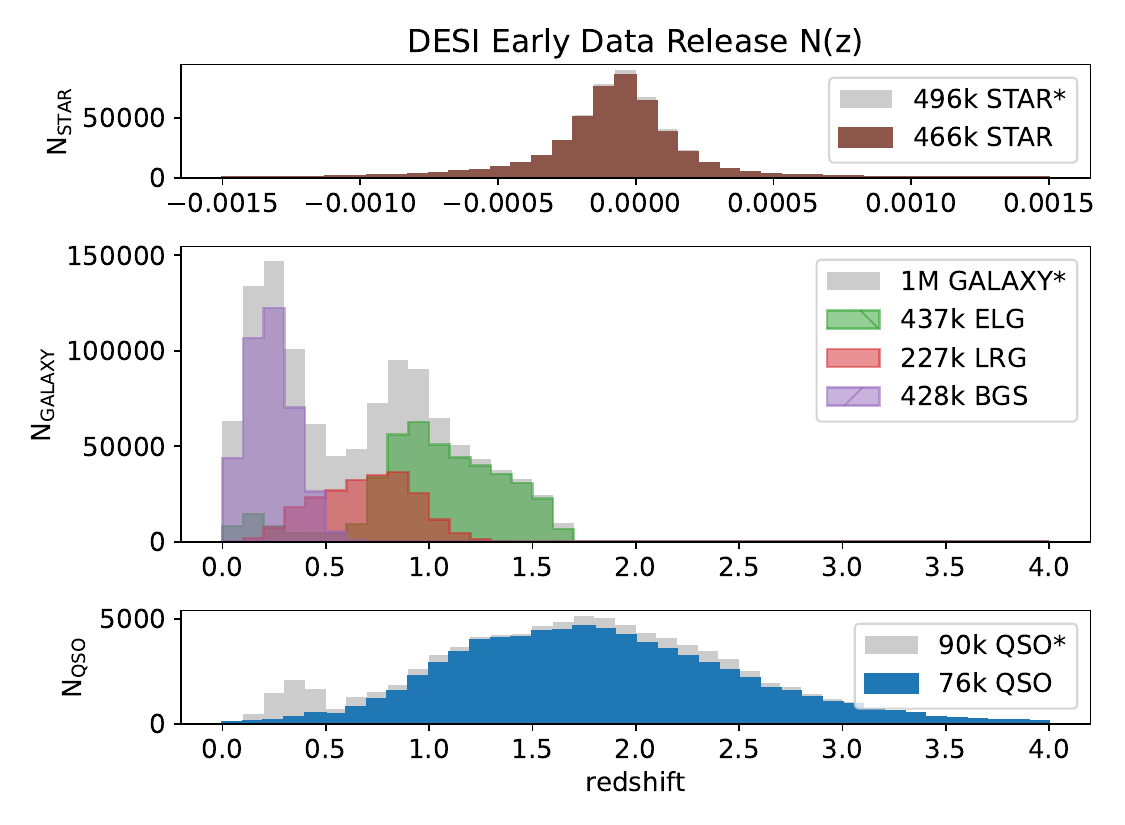}
     \caption{The number of good, unique target redshifts as a function of redshift for each tracer type as defined in \S\ref{sec:qualitycuts}. The reddish-brown distribution is for objects targeted to be a star and classified by Redrock to be \texttt{SPECTYPE==STAR}. The purple, green, and red histograms show objects targeted as BGS, ELG, and LRG respectively, and classified as a \texttt{GALAXY}. The blue distribution shows objects targeted as a QSO and classified as a \texttt{QSO}. The gray distributions depict all objects that were classified by Redrock as a \texttt{STAR}, \texttt{GALAXY}, or \texttt{QSO} for the top, middle, and bottom panels respectively. (*)Note that the gray differs from the colored histograms because of secondary targets and other target types that were classified to a different category (\eg~a QSO target that was classified as a \texttt{STAR}). Also note that an object can be targeted by two galaxy target classes, and such objects will appear in both distributions.}
     \label{fig:totalNz}
 \end{figure}

Table~\ref{tab:spectra_stats} summarizes the number of good objects per target class in each survey as well as in the total EDR sample, with the additional restriction that the target was classified as the intended target selection
type --- \texttt{GALAXY} for BGS, ELG, and LRG samples; \texttt{QSO} for QSO samples; and \texttt{STAR} for MWS samples.
Note that the sum of a column or row may not be equal to the total due to individual objects being observed in multiple surveys or individual objects being selected for multiple target classes. The numbers as a function of redshift for the full EDR sample are shown in Figure \ref{fig:totalNz}.
This includes both primary targets that were classified as their targeted type (colored histograms),
as well as confident classifications of any targets (primary or secondary) regardless of their expected
target type (gray histograms). For example, the bump at $z<0.5$~in the gray histogram for \texttt{QSO} classifications comes primarily from
BGS targets that were either quasars or Active Galactic Nucleus (AGN)-like galaxies that Redrock classified as \texttt{QSO} instead of galaxies.  Although they may be valid redshifts, these are not included
in Table~\ref{tab:spectra_stats}.

\begin{deluxetable*}{c|c|c|c|c|c|c}[htb]
\tablecaption{The number of ``good'' spectra obtained in each phase of SV, along with details for dedicated pointings (special) and commissioning (cmx). Here each target class is selected with 
the bitmasks for that tracer in that survey, and ``good'' refers
to science targets that have no Redrock \texttt{ZWARN} bits set and whose best fitting templates are consistent with the tracer (\texttt{GALAXY} for BGS, ELG, and LRG targets; \texttt{QSO} for QSO targets, and \texttt{STAR} for MWS).
Each row counts unique targets, but since some targets were observed under multiple surveys, the total number of unique
targets is less than the sum of the rows.
}
\label{tab:spectra_stats}
\tablehead{\colhead{SURVEY} & \colhead{N${}_{\mathrm{BGS}}$} & \colhead{N${}_{\mathrm{ELG}}$} & \colhead{N${}_{\mathrm{LRG}}$} & \colhead{N${}_{\mathrm{QSO}}$} & \colhead{N${}_{\mathrm{STAR}}$} & \colhead{N${}_{\mathrm{SCND}}$}}
\startdata
cmx & 247 & 761 & 1,037 & 275 & 468 & 0 \\
sv1 & 134,419 & 111,692 & 66,161 & 29,839 & 163,254 & 60,430 \\
sv2 & 46,628 & 12,308 & 22,151 & 11,032 & 10,506 & 0 \\
sv3 & 253,915 & 312,790 & 137,317 & 34,173 & 295,232 & 75,947 \\
special & 925 & 3,866 & 3,588 & 3,045 & 867 & 3,482 \\
\hline
Total & 428,758 & 437,664 & 227,318 & 76,079 & 466,447 & 137,148
\enddata
\end{deluxetable*}

\begin{figure}
     \centering 
     \includegraphics[width=1\columnwidth]{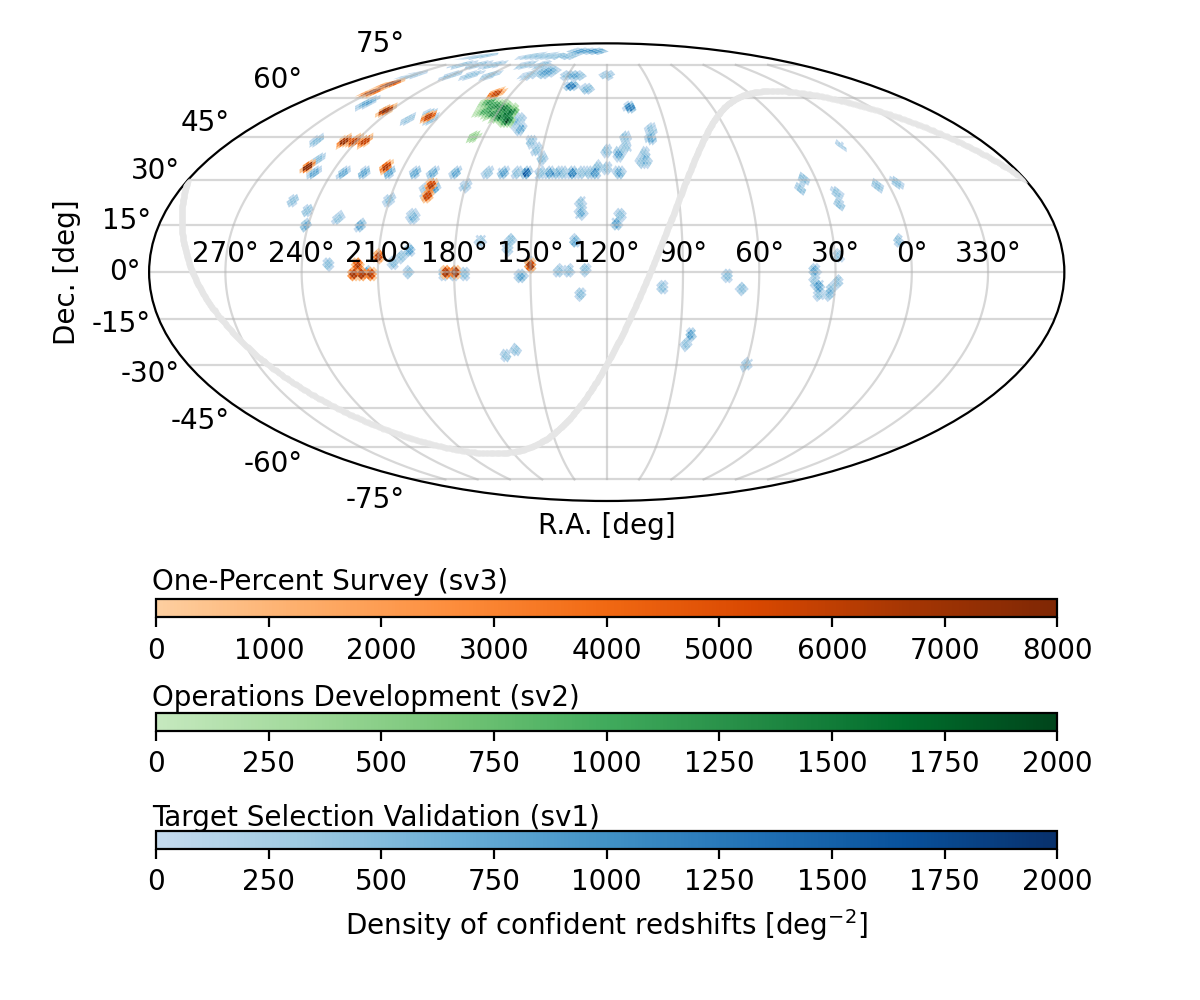}
     \caption{The density of good, unique target redshifts on the sky, split by the three primary phases of Survey Validation ---
     Target Selection Validation (sv1, blue), Operations Development (sv2, green), and the One-Percent Survey (sv3, orange).
     Note that the orange One-Percent Survey colormap goes to 4$\times$ higher density than the others, reflecting the much
     higher density of targets in the One-Percent Survey rosettes of overlapping tiles.}
     \label{fig:goodz_density}
\end{figure}

Figure \ref{fig:goodz_density} shows the density of good target redshifts on the sky for each of the three primary phases of Survey Validation ---
Target Selection Validation (sv1, blue), Operations Development (sv2, green), and the One-Percent Survey (sv3, orange).
Target Selection Validation has many tiles distributed over the sky, while the One-Percent Survey has a larger
number of tiles over a smaller area, leading to much higher good target densities (and thus much higher target completeness for those patches of sky) for the One-Percent Survey.

% from edrpaper/bin/plot_density_skymap

 Metadata about the targets included in each file are recorded in the \texttt{FIBERMAP} Header/Data Unit extension of the \texttt{FITS}\footnote{\url{https://fits.gsfc.nasa.gov/fits_standard.html}} files in the EDR.  
These include the input photometry used for target selection,\footnote{Targets that were secondary-only and not matched to LS/DR9 do not have photometry included.}
target selection bitmasks recording which target classes each target was selected for, and information about which
fiber each target was assigned to and how accurately that fiber was positioned.  This information
is also propagated into the final redshift catalogs.  
The target selection bitmasks for each of the surveys included in the EDR are described in Appendix~\ref{app:primarytargets}, with further details in \S2.4 of \cite{myers23a}.

\section{Data Reduction and Data Products}\label{sec:products}

\subsection{Data Reduction}
\label{sec:datareduction}

\subsubsection{Spectroscopic Calibration and Reduction}
\label{sec:specreduc}
The spectroscopic data reduction for DESI is performed with a newly developed, Python-based pipeline. 
\revisionforreviewer{The EDR is a re-processing of the raw data performed using a tagged version of the pipeline code.\footnote{\url{https://github.com/desihub/desispec/releases/tag/0.51.13}} 
The directory structure and filenames retain the internal name of the release, {\tt fuji}, named after the mountain and signifying the sixth internal release of data.} 
A detailed description of the DESI pipeline can be found in \citet{guy22a}; here we provide a short summary.

Spectroscopic data are transferred from the telescope at Kitt Peak in nearly real-time to the National Energy Research Scientific Computing Center (NERSC) for processing and archiving. 
Each night the data are processed from raw images to redshifts such that tiles identified as being observed to full depth can be approved as complete and the targets entered into the MTL ledgers as being observed (see \S5 of \citealt{schlafly23a}).
This allows for subsequent overlapping tiles to be designed and observed in the same region of sky with new targets (and repeated high-redshift QSO targets for those identified as Lyman-$\alpha$ forest sources). 

The DESI instrument consists of ten petals of 500 fibers each that send light to 10 corresponding spectrographs.
Each spectrograph has three arms: blue, red, and near-infrared (denoted B, R, and Z in the data files). 
In its nominal configuration, the instrument generates 10 4096x4096 pixel blue images and 20 4114x4128 pixel red and NIR images, with each containing the data for 500 fibers in one of the three wavelength ranges.
In the afternoon before each night of observation, calibration data are acquired and processed so that they can be applied to the scientific data taken throughout the night.
Twenty-five 0-second \emph{zero} exposures are taken with the shutter closed to generate a master bias for the night. This bias is removed from all other calibrations and science exposures for the night. 
Next, a 300-second \emph{dark} exposure, again with the shutters closed, is taken to identify new bad columns or pixels in the CCD and to test the nightly master bias against a template bias to determine which gives the smallest residuals in the processed image. 
If the template produces smaller residuals, then it is used instead of the master bias derived on that night.

Five \emph{arc} exposures are then taken with Ar, Cd, Hg, Kr, Ne, and Xe arc lamps. 
These are used for an initial wavelength calibration for each fiber throughout the night, in addition to generating a full two-dimensional point spread function (2D PSF) model for each image to be used for extracting the spectral signal from the 2D data using an implementation of the spectroperfectionism algorithm described in \citet{bolton10}.
These exposures are also used to determine the fiber traces representing the two-dimensional extent of each fiber on the image.
Finally, four sets of three \emph{flat} exposures are taken with LED lamps on a white dome screen and combined to generate a per fiber flat field correction that is applied to the subsequent science exposures on the night.

Science data taken throughout the night are first preprocessed to convert analog digital units into electron counts, identify and mask cosmic rays and bad pixels, remove detector bias and correct for dark current, apply a CCD flat field correction, and estimate the per-pixel variance in the image. 
Next, the wavelength solution derived from the \textit{arc} exposures is refined using sky lines, small trace shifts are identified, and a new 2D PSF model is generated for each exposure. 
Counts for each fiber are then extracted, along with estimates of the variance and the resolution matrix that encapsulates the mapping between the 2D model and the 1D, uncorrelated, linear wavelength binned output spectra. 
The output spectra are then flat fielded using the per fiber flat field vectors derived from the \textit{flat} exposures. 
The sky is then removed using fibers explicitly positioned to empty sky locations. 
These sky fibers are combined to create one high-resolution sky model for each spectrograph, before using the resolution matrix of each fiber to subtract the sky.

After sky subtraction, main sequence F stars are used for flux calibration.
The selected standard star fiber spectra are fit to all 3 cameras simultaneously using theoretical models for a wide range of stellar effective temperature ($\mathrm{T}_{\mathrm{eff}}$), surface gravity (log${}_{10}$\,g), and iron abundance ([Fe/H]). 
The resulting fits can be used to derive the throughput of the instrument by relating the measured counts to the expected flux from photometry, which enables the generation of calibration vectors for all fibers on the petal to convert counts into fluxes in units of $10^{-17}$~erg/s/cm$^2$/\AA.
To improve signal-to-noise, if multiple exposures of a single tile are observed on a night, then all of the exposures are used to jointly model standard stars, before each exposure is used independently to generate a calibration vector for that exposure.
The per-star calibration is used to derive the observed flux compared to the expected flux from imaging photometry,
and 3$\sigma$ outliers are rejected compared to the scatter observed for all standard stars across all spectrographs.
The final calibration vector per camera is an average over the remaining standard stars on that camera.
Finally, a crosstalk correction is applied to account for the fact that the 2D PSF of each fiber extends into the region of its neighbors before writing out the per-exposure per-fiber calibrated fluxes and variances.

During SV1, tiles were allocated with 80 sky fibers and a goal of 20 standard stars per petal.
Using these data, tests were performed to identify the minimum number of sky fibers and standard stars that could be used per petal before degradation in sky subtraction or flux calibration would be observed. 
It was shown that as few as 20 sky fibers per petal and 10 standard stars per petal were sufficient to maintain ELG redshift efficiency, which was used as a proxy for measuring the impact of sky subtraction residuals on the resulting spectra.
For SV3 and the Main Survey, DESI requires a minimum of 40 sky fibers and 10 standard star fibers per petal to be conservative. 

DESI tiles are observed to have roughly equal effective exposure time rather than raw exposure time. This results in varying raw exposure times and a varying number of exposures acquired for a given tile. For the EDR the median number of exposures for a tile is 1, with a mean of $\sim3.4$, and a maximum of 30 for deep SV1 BGS tile 80613.

\subsubsection{Spectroscopic Effective Exposure Time}
\label{sec:efftimespec}
The Exposure Time Calculator (ETC) provides a real-time estimated effective exposure time to determine when to end observations of a given tile, using information available during an exposure. 
However, a more ideal quantity would be an effective time that incorporates the instrumental effects on the spectra themselves and how a particular target class might be impacted by such effects. 
\texttt{EFFTIME_SPEC}, which is derived from the spectroscopic data themselves each night as the data are acquired, was designed to incorporate these features. This quantity is what is used by survey operations to determine if observations of a tile have achieved enough effective time to be marked as complete for the designated \texttt{PROGRAM}. 

The spectroscopic effective time estimate is detailed in \S4.14 of \cite{guy22a}. 
First, a template signal-to-noise ratio squared (TSNR${}^2$) is defined as the mean of the squared signal-to-noise for an ensemble of templates over a representative redshift range, which incorporates instrument and observational quantities such as fiber aperture losses, detector read noise, and sky residuals. 
Finally, the mean TSNR${}^2$~value for all fibers is multiplied by a constant to get the \texttt{EFFTIME_SPEC} for the exposure.
Each tracer class has a different morphology and redshift range, so this is done for each separately, with each target class having a different constant of proportionality.
The constants are empirically fit to equal the actual exposure time when observing in nominal conditions at zenith with dark sky, ideal transparency, no Galactic extinction, and median seeing of 1.1\arcsec. 
This is done for all target classes, but the reported \texttt{EFFTIME_SPEC} for dark time is based on the LRG value, while the bright time \texttt{EFFTIME_SPEC} is based on the BGS value.

\subsubsection{Redshift Fitting and Classifications}\label{sec:redrock}

Spectral classifications and redshifts are measured using the Redrock software package\footnote{\url{https://github.com/desihub/redrock/releases/tag/0.15.4}} \citep{bailey23a}.
Redrock performs a $\chi^2$ vs.~redshift scan, fitting a set of Principal Component Analysis (PCA) templates
to every target at every redshift.
The fit with the lowest $\chi^2$ determines the spectral classification
(\texttt{SPECTYPE=GALAXY}, \texttt{QSO}, or \texttt{STAR} for DESI) and redshift.
Each set of templates is fit on all spectra regardless of target selection type; including
standard stars, sky-subtracted sky fibers, and spectra from nonfunctioning
positioners that were not pointing at any known target.
This procedure is similar to the method used in SDSS/BOSS \citep{bolton12}, with improvements to the
underlying PCA templates, more exact error propagation, and more detailed modeling of the per-wavelength
per-fiber spectral resolution.  Although Redrock was originally developed for DESI, it was previously used by eBOSS for their final cosmology analyses \citep{ross20_eboss_lss}.

The primary outputs from Redrock are the
redshift (\texttt{Z}),
redshift uncertainty (\texttt{ZERR}),
spectral classification (\texttt{SPECTYPE}),
a warning bitmask (\texttt{ZWARN}),
the coefficients for the linear combination of the best-fitting templates (\texttt{COEFF}),
the $\chi^2$ of the fit (\texttt{CHI2}), \revisionforreviewer{and the value
$\Delta \chi^2$~(\texttt{DELTACHI2}) giving the difference between the best fit $\chi^2$ and that of the second best fit.}
Larger values of \texttt{DELTACHI2} represent greater statistical confidence that the best fit is correct.

In addition to classifications and redshifts, Redrock includes a per-object \texttt{ZWARN} bitmask 
indicating if there are any known problems with the data or the fit.  \texttt{ZWARN==0} means that there are
no known problems, and non-zero values encode the reasons for possible problems.
The meaning of the individual bits are documented in the Redrock code at
\url{https://github.com/desihub/redrock/blob/0.15.4/py/redrock/zwarning.py\#L14} and further described in
\citet{bailey23a}.  Some bits record problems with the input spectrum, \eg, that all flux values were masked, while
other bits record problems with the Redrock fit itself, \eg, a failed parabola fit to the $\chi^2$ vs.~$z$ minimum.
Most analyses should require \texttt{ZWARN==0} to obtain good quality results, which implicitly include the requirement
that \texttt{DELTACHI2}~$>9$. To obtain a purer sample of more confident redshifts, some analyses may place a
higher cut on \texttt{DELTACHI2}.

%%% SB note: actual distribution was 6.7x - 15x
% Target Selection Validation includes observations of tiles to $\sim$10$\times$ longer exposure time
% than planned for the DESI Main Survey.  Visual inspections were performed on deep coadds of these
% spectra for galaxies \citep{lan22a} and quasars \citep{alexander22a} to estimate Redrock performance
% and identify common failure modes to improve in subsequent internal data releases.  These data were
% also used to determine algorithms for selecting reliable redshifts for each individual target class
% as described in Section~\ref{sec:lsscat}.

\subsubsection{Post Redshift Value-Added Processing}
% \label{sec:postproc}
% In what follows, we describe the custom algorithm for determining the redshift of each target class.
% Using the SV sample, we report the statistics regarding total redshift efficiency, target redshift efficiency, catastrophic failure rate, and statistical redshift precision for each sample.
% We then use the custom algorithms for each target class to summarize redshift completeness as a function of exposure time, thus setting the conditions for the exposure sequence in the Main Survey.
% We conclude with a summary of the performance on all spectroscopic samples compared to the requirements that drove the instrument design as described in \citet{desi-collaboration22a}.

After redshift fitting, three additional steps were included in the pipeline to post-process the data and derive value-added quantities such as line identifications, line flux estimates, refined redshift fits, and further quasar classifications. 

In addition to the Redrock results, \texttt{emlinefit} provides simple fits of the major galaxy emission lines. The approach is purposefully simple; a more refined approach is performed with \eg~\texttt{Fastspecfit}.\footnote{\url{https://fastspecfit.readthedocs.io/en/latest/}}
The primary motivation is to fit the \oii doublet, and use it to identify reliable redshift measurements for ELG spectra (see \S\ref{sec:lss_fullcat} and Equation~\ref{eq:elg_zcrit}). 
For convenience, we also provide fits for the \oiii doublet ($\lambda \lambda$ 4960,5007~\AA),~ and for the H${}_\alpha$, H${}_\beta$, H${}_\gamma$, and H${}_\delta$ lines.
All lines are computed for all spectra. However, detailed studies should be performed prior to using quantities other than [\ion{O}{2}], or \oii on non-ELG targets, to assess their accuracy.
The fits are simple Gaussian fits at the expected position based on the Redrock best-fit results; the fitted flux is not forced to be positive, so negative values can be reported.
The continuum is estimated from the wavelengths 200 \AA~(in rest-frame) around the emission line (bluewards for the \oii doublet).
For the \oii doublet, the line ratio is left free during the fit; for the \oiii doublet, it is fixed.
For more details, see \S7 of \citet{raichoor22a}.

To improve the classification and the redshift determination for quasars, we also provide results from two additional codes: an Mg~II broadband fitter and a neural network classifier, QuasarNET \citep{Busca18, farr20}. The Mg~II fitter aims to classify spectra that exhibit a broad Mg~II emission line as QSO. The algorithm determines the width of the Mg~II emission by fitting a Gaussian in a 250\,\AA~window (observer-frame) centered at the position of the Mg~II line given by the redshift identified by Redrock. For the QSO classification, the Mg~II emission line is considered broad if the improvement of $\chi^2$ is better than 16, the width of the Gaussian is greater than 10\,\AA, and the significance of the amplitude of the Gaussian is greater than 3. 

Additionally, we run QuasarNET on all the targets. QuasarNET is a deep convolutional neural network (CNN) classifier designed explicitly to identify quasars and their redshifts. The input power spectrum is reduced by four layers of convolutions and is then passed to a fifth, fully connected layer before feeding into six line finder units: one for Ly$\alpha$, C IV, C II, Mg~II, H$\alpha$, and H$\beta$. Each line finder unit consists of a fully connected layer trained to identify a particular emission line. The output of each unit is a confidence level (between 0 and 1) to have found the desired line and the redshift at which it was found. The DESI large-scale structure analyses require at least one emission line to have a confidence level above 0.95 for a spectrum to be considered a QSO.
For each target identified by QuasarNET to be a QSO, Redrock is rerun using only QSO templates and a tophat redshift
prior of $\pm0.05$ to determine the final redshift.

\subsection{Suggested Quality Cuts}\label{sec:qualitycuts}

The choice of a `good' redshift is subjective and depends on the individual science case in question. In this paper, unless stated differently, we have elected to restrict to spectra with no hardware, observing, or redshift fitting flags (\texttt{ZWARN==0}). This is our generic recommendation, where some may choose to relax restrictions for specific bits if an analysis is robust to the implications of including such data, and some may choose to use additional selection criteria such as a cut on \texttt{DELTACHI2} or \eg~\texttt{TSNR2_LRG}. The large-scale structure analyses within DESI use cuts that are generally more stringent than this, as outlined in \S\ref{sec:lss_clusteringcat}. More details for each target class are available in the references cited in that section.

A less strict criterion would be to restrict based on the coadded spectrum's \textit{fiberstatus}, \texttt{COADD_FIBERSTATUS}, which encapsulates hardware and observing issues for all input data for that spectrum, but does not depend on redshift fitting. The bits are defined in the \texttt{desispec.maskbits} code.\footnote{\url{https://github.com/desihub/desispec/blob/0.51.13/py/desispec/maskbits.py\#L55}} Note that \texttt{ZWARN} includes a bit which is false if \texttt{COADD_FIBERSTATUS} equals 0 or 8, where 0 signifies no issues and~$2^3=8$~corresponds to a positioner that had a restricted range but could still reach the target location. Therefore selecting \texttt{ZWARN==0} implies selecting \texttt{COADD_FIBERSTATUS} $\tt \in [0,8]$~in addition to the redshift fitting flags.

\subsection{Data Products}

Full details of the directory organization and file formats in the EDR are given
in the DESI Data Model at \url{https://desidatamodel.readthedocs.io}. The directory structure is summarized in Table~\ref{tab:directory_structure}. 
The following subsections provide a conceptual overview of the structure of
the available data, starting from the root directory of the EDR;
see \S\ref{sec:access} for methods to access these data.

\begin{deluxetable*}{ll}[htb]
\tablecaption{Summary of the directory structure of data available in the EDR.
See \url{https://desidatamodel.readthedocs.io} for more details including subdirectory
structure underneath these directories, individual file formats, and additional
directories with pipeline inputs such as calibration files.
}
\label{tab:directory_structure}
%%% \tablehead{\colhead{Directory} & \colhead{Description} }
\tablehead{Directory & Description}
\startdata
\texttt{spectro/data/}          & Raw data \\
\texttt{spectro/redux/fuji/}    & Processed data \\
~~~~~~\texttt{healpix/}           & Spectra, classifications, and redshifts grouped by HEALPix \\
~~~~~~\texttt{tiles/}             & Spectra, classifications, and redshifts grouped by tile \\
~~~~~~\texttt{zcatalog/}          & Combined redshift catalogs \\
~~~~~~\texttt{exposures/}         & Intermediate processing files per exposure \\
\texttt{survey/}                & Survey operations bookkeeping \\
\texttt{target/catalogs/}       & Input target catalogs, same as ``Early Target Selection'' release \\
\texttt{target/fiberassign/}    & Assignments of targets to fibers per tile \\
\texttt{vac/edr/}               & Value Added Catalogs contributed by the DESI science collaboration \\
\enddata
\end{deluxetable*}

\subsubsection{Spectroscopic Data Processing Runs}

Production data processing runs are alphabetically named after mountains and
contained under \texttt{spectro/redux/}.  Each production run uses a defined
set of input data processed with a fixed set of tagged software.
The EDR contains a single production named ``Fuji'', available under
\texttt{spectro/redux/fuji/}.
Future data releases will contain one or more production runs, differing
by the input raw data, the software tags used, or both.

In the top-level production directory, \texttt{spectro/redux/fuji/tiles-fuji.fits} contains a catalog of all DESI tiles
included in Fuji.  This can be used for a quick assessment of the footprint
of the available DESI data and to filter available tiles by \texttt{SURVEY} and \texttt{PROGRAM}.
Since tiles may be observed on multiple exposures spanning multiple nights, more detailed per-exposure information is available
in \texttt{exposures-fuji.fits} if needed for time-domain studies or systematics comparisons of data on different nights.

\subsubsection{Spectra, Coadds, Classifications, and Redshifts}
\label{sec:spectra-files}

Spectra, coadditions (coadds) of those spectra, and classifications and redshifts fit to those coadds are available under two broad groups: per-tile and full-depth. 
Tile-based spectra under \texttt{spectro/redux/fuji/tiles/} combine information across multiple exposures
of the same tile, but not across different tiles even if the same target was observed
on multiple tiles (such as high redshift quasars).
Full-depth coadds combine exposures for targets on a given HEALPix \citep{Gorski05} pixel of sky, including combining data across tiles if the same target was observed on multiple tiles. 
These coadds are referred to as ``healpix'' coadds and redshifts since they are stored in files based on HEALPix pixel number (nested scheme, \texttt{NSIDE=64}) under \texttt{spectro/redux/fuji/healpix/}.

However, even in the healpix case, data are not combined across surveys
(\eg~sv1, sv2, sv3) and programs (\eg~dark, bright, backup) in order to prioritize
the uniformity of each (survey,~program) combination.  We anticipate that the
tile-based spectra will be of primary interest to analyses of
Target Selection Validation (\texttt{SURVEY=sv1}), while the HEALPix-based full-depth spectra will be
used more for the overlapping rosettes of the One-Percent Survey (\texttt{SURVEY=sv3})
and future releases of DESI Main Survey data.

Tile-based spectra, coadds, classifications, and redshifts come in additional
subgroupings under \texttt{spectro/redux/fuji/tiles/}.
The \texttt{cumulative/} directory tree contains all data for each tile,
coadded across exposures and nights.
The \texttt{pernight/} directory tree combines data within a night but not
across nights, enabling reproducibility studies of the same targets observed
under different conditions on different nights.
The \texttt{perexp/} directory contains classifications and redshift fits to
individual exposures to explore performance and reproducibility at even lower
signal-to-noise.  Additional custom coadds, classifications, and redshift fits
are available for selected tiles to match the expected depth of the Main Survey
(\texttt{1x\_depth}), four times the expected Main Survey depth
(\texttt{4x\_depth}), and coadds using only data from poor observing conditions
(\texttt{lowspeed}). 

If a tile was only observed on a single exposure on a single night, the
\texttt{cumulative/}, \texttt{pernight/}, and \texttt{perexp/} outputs
are identical, but they are still kept in all three directories so that each
can be used independently.

Future data releases will continue to support \texttt{tiles/cumulative/} and
\texttt{healpix/}, but other groupings of tiles-based spectra will not be
included for Main Survey data and may be dropped from reprocessing runs of
the SV data.

\subsubsection{Redshift Catalogs}
\label{sec:redshift-catalogs}

Redshift and classification catalogs for individual tiles and healpix are
available in the same directories as the spectra and coadds to which they were fit.  For convenience,
these catalogs are also combined across the thousands of individual files into stacked
redshift catalogs in \texttt{spectro/redux/fuji/zcatalog/}.  Like the spectra
and coadds, these come in multiple groups, \eg~combining all of the cumulative
tile-based redshifts for a given (survey,~program) into a single file, with
different files for different (survey,~program) combinations.

For analyses that simply want the recommended ``best'' redshift for a given
target regardless of the DESI-specific (survey,~program), \texttt{zall-pix-fuji.fits} combines all the HEALPix-based redshifts
across all programs into a single file, with a \texttt{ZCAT\_PRIMARY} boolean
column indicating which row is considered the best redshift for each target.
This uses the code \texttt{desispec.zcatalog.find_primary_spectra},\footnote{\url{https://github.com/desihub/desispec/blob/0.51.13/py/desispec/zcatalog.py\#L13}}
which could also be used by any analysis to subselect multiply-observed targets
from a custom selection of spectra to determine the recommended redshift.
It first prioritizes results with
Redrock \texttt{ZWARN==0}, then sorts by the LRG-optimized template signal-to-noise \texttt{TSNR2\_LRG},
although users can specify a different secondary sort column. 
A description of the template signal-to-noise ratio can be found in \S\ref{sec:efftimespec} with
further details in \S4.14 of \cite{guy22a}.

Similarly, \texttt{zall-tilecumulative-fuji.fits} provides all cumulative
tile-based redshifts across surveys and programs, with \texttt{ZCAT\_PRIMARY}
indicating the recommended best single tile-based redshift per target.

\subsubsection{Target Catalogs}
\label{sec:target-catalogs}

Target catalogs used as input for DESI observations were previously published
under the ``Early Target Selection'' release available at
\url{https://data.desi.lbl.gov/public/ets/} and described in \citet{myers23a}.
Although the EDR does not include any new target selection catalogs,
\texttt{edr/target/} links to the same directory tree as \texttt{ets/target/}
so that the EDR can be used as a self-contained release including target selection
information, without having to combine information across releases.

\subsubsection{Fiber Assignment Catalogs}
\label{sec:fiberassign-files}

Fiber assignment is the process of assigning individual targets to individual fibers.
In DESI, the assignment of targets to fibers for a given tile is designed by the fiberassign program, as described in \citet{raichoor23a}.
The output files, called \textit{fiberassign} files, that detail the fiber assignments used in the EDR can be found in \texttt{target/fiberassign/tiles/tags/0.5/}.
Within each fiberassign file, the \texttt{FIBERASSIGN} table contains the mapping
of assigned \texttt{TARGETID} to \texttt{FIBER} and \texttt{LOCATION};\footnote{\texttt{FIBER} [0-4999] tracks the position of the spectra on the spectrographs, with the spectrograph number = petal number = \texttt{int(FIBER/500)}.  \texttt{LOCATION} tracks the position of the fiber on the focal plane which is purposefully randomized with respect to \texttt{FIBER} to break degeneracies between systematics related to position on the focal plane vs.~position on the spectrographs.  Although there is a fixed 1:1 mapping between \texttt{FIBER} and \texttt{LOCATION}, both are recorded for convenience.}
the \texttt{TARGET\_RA} and \texttt{TARGET\_DEC} coordinates and proper motions \texttt{PMRA},
\texttt{PMDEC}, and \texttt{REF\_EPOCH}; \revisionforreviewer{which are in the International Celestial Reference System (ICRS) tied
to \textit{Gaia} DR2 \citep{GaiaDR2astrometry}}; the input photometry, object shapes, and quality flags used for primary target selection
\citep{myers23a};\footnote{
Secondary targets are allowed to come from any data source, and thus their exact input selection parameters are not tracked here.}
and the targeting bitmasks described
in sections~\ref{sec:sv1_target_bitmasks}, \ref{sec:sv2_target_bitmasks}, and \ref{sec:sv3_target_bitmasks}.

In addition to the target-to-fiber assignments, each file contains all possible assignments
in the \texttt{POTENTIAL\_ASSIGNMENTS} and \texttt{TARGETS} extensions.
The \texttt{TARGETS} table has one row per \texttt{TARGETID} and includes the targeting bitmasks for each potential target.
The \texttt{POTENTIAL\_ASSIGNMENTS} table only has \texttt{TARGETID}, \texttt{FIBER}, and \texttt{LOCATION} columns;
providing the mapping of which fibers could reach that target, with potentially multiple rows per \texttt{TARGETID}.
Although fiberassign files contain the input photometry for targets that were assigned,
they do not include full photometric information for unassigned targets covered by each tile, in order to keep the files to a manageable size. 
If photometric information is needed for unassigned targets, the \texttt{TARGETID}s must be cross-matched back to the input targeting catalogs.  
This is most easily done by using the ``LS/DR9 Photometry'' Value Added Catalog included in the EDR and discussed in section \ref{sec:vac-files}, which combines and standardizes the information from the multiple input targeting catalogs.

Each raw data exposure directory also contains a copy of the fiberassign file
that was used at the telescope at the time of observation. 
For some tiles, the photometry
for secondary targets was incorrect, and this was corrected post-facto in the
files under \texttt{target/fiberassign/tiles/tags/0.5/}, which should be considered
the most correct reference version.  
This was only done for catalog columns that were not actually used by observations.  
Spectroscopic data processing used these files to supersede the raw data versions when propagating
the information downstream via the \texttt{FIBERMAP} extensions.

\subsubsection{Value Added Catalogs}
\label{sec:vac-files}

%%% Following the tradition of SDSS,
DESI data releases will include
``Value Added Catalogs'' (VACs), which are additional data products contributed by the DESI science collaboration. VACs are built upon the core data products
(spectra, classifications, redshifts) from this spectroscopic data release.

%%% Unlike SDSS,
EDR will include a set of VACs with the initial release, but will also add additional VACs based upon EDR when they become available from the DESI science collaboration in the future.
The website \url{https://data.desi.lbl.gov/doc/vac/} will be kept up to date with details of the available VACs, including documentation and references to relevant journal articles. The files can be accessed in the same manner as the EDR data under the \texttt{vac/} subdirectory. In the rest of this section, we present two VACs in the EDR that are broadly applicable to many analyses using the EDR data --- the LS/DR9 Photometry VAC and the Survey Validation Visual Inspection VAC. The Large-scale Structure (LSS) Catalogs VAC used for DESI LSS analyses will also be described in \S\ref{sec:lsscat}. \revisionforreviewer{Additional VACs are available and described on the website.}

\underline{LS/DR9 Photometry VAC}:~This VAC delivers merged targeting catalogs (\textit{targetphot}) from DESI target selection \citep{myers23a} and Tractor\footnote{\url{https://github.com/dstndstn/tractor}} catalog photometry (\textit{tractorphot}) from the DESI Legacy Imaging Surveys Data Release 9 \citep[LS/DR9;][]{dey19a}\footnote{\url{https://www.legacysurvey.org/dr9/description}} for all observed and potential targets (excluding sky fibers) in the EDR. The observed targets in this VAC correspond to objects with at least one spectrum in the EDR, while the potential targets are the targets that DESI \textit{could have} observed in a given fiber assignment configuration (including the objects which were actually observed).\footnote{Note that these catalogs are distinct from the photometric target catalogs described in \S\ref{sec:target-catalogs}, which contain \textit{all} the possible photometric targets with no reference to DESI observations.} The construction and organization of the LS/DR9 VAC is fully documented at \url{https://github.com/moustakas/desi-photometry}; here, we briefly describe its contents.

The LS/DR9 VAC includes \textit{tractorphot} catalogs, which contain Tractor catalog photometry for every unique target in the \textit{targetphot} catalogs. These catalogs are ``value-added" compared to the information in catalogs described in \S\ref{sec:target-catalogs} in two key ways. First, the \textit{tractorphot} catalogs contain \textit{all} the photometric quantities measured by Tractor in the LS/DR9, not just the measurements included in the light-weight sweep catalogs used to select DESI targets (see also \citealt{myers23a}). Second, the \textit{tractorphot} catalogs include photometry for targets which were not necessarily targeted from the LS/DR9, such as secondary targets and targets of opportunity, by finding the LS/DR9 object within $1\arcsec$ of the observed (or potential) DESI target.

\underline{Survey Validation Visual Inspection VAC}:~During the Survey Validation period, in order to validate the performance of the DESI pipeline and assist the target selections of the Main Survey, DESI members visually inspected (VI'ed) the deep coadded spectra of 16,594 galaxy targets; including 2,718 BGS; 3,561 LRG; and 10,315 ELG targets; in addition to 5,496 QSO targets. 
Each spectrum had at least two inspectors. Each inspector reported the VI redshift, the redshift quality, the source type, issues observed in the spectrum, and any extra comments for each spectrum. The results from the inspectors were combined and reconciled by the VI chairperson if needed. This VI information was used to quantify the performance of the DESI operation and validate the design of the survey. The details of the compilation of the catalogs and the corresponding analyses are summarized in \cite{lan22a} for galaxies and \cite{alexander22a} for quasars.

These VI catalogs are provided as a VAC in the EDR. The catalogs are organized based on the target types. For galaxies, we provide the catalogs for BGS, LRG, and ELG separately. For QSO, we provide two catalogs: the quasar-survey deep-field VI catalog analogous to that provided for the galaxies, and the missed QSO catalog from a sparse VI campaign (see Tables 1 and 4 and \S2.3 of \citealt{alexander22a} for details). Each catalog contains target information, including \texttt{TARGETID}, \texttt{TILEID}, \texttt{FIBER}, \texttt{TARGET\_RA}, and \texttt{TARGET\_DEC}; and VI information, including the VI redshift (\texttt{VI\_Z}), the redshift quality (\texttt{VI\_QUALITY}), and the type of the source (\texttt{VI\_SPECTYPE}). Note that we do not include issues or comments reported by the inspectors since most of them reflect the status of the spectra processed by an early version of the pipeline, rather than the improved pipeline used for the EDR. With the target information, one can crossmatch the VI catalogs with other catalogs in the EDR and obtain information on the sources such as the photometric properties as well as the redshift information from the DESI pipeline. We recommend using \texttt{VI\_QUALITY}~$\geq2.5$ as a selection criterion for sources with confident VI redshifts; see section 2 of \cite{lan22a} and section 2.2 of \cite{alexander22a} for details of the \texttt{VI\_QUALITY} flags within the context of the galaxy and quasar catalogs, respectively. There are 2,640 BGS; 3,513 LRG; 7,856 ELG; and 4,890 QSO targets with any VI classification and \texttt{VI\_QUALITY}~$\geq2.5$. 
In total there are 14,939 objects identified as a \texttt{VI\_SPECTYPE==GALAXY}; 3,182 objects identified as a \texttt{QSO}; and 778 identified as a \texttt{STAR} with \texttt{VI\_QUALITY}~$\geq2.5$.  

\subsection{Other Files}
\label{sec:other-files}

In addition to the high-level user-facing data products, DESI data releases
also contain raw data and intermediate data products.

The original raw data are available in \texttt{spectro/data/NIGHT/EXPID/}
subdirectories where \texttt{NIGHT} is the \texttt{YEARMMDD} date of
sunset,\footnote{The DESI ``night'' rolls over at KPNO noon, not midnight,
so that data from a single observing night are grouped together even though
they span two calendar days.} and \texttt{EXPID} is the zero-padded 8-digit
monotonically increasing exposure identification number.  These directories
contain the original fiber assignment observing request,
raw data from the spectrographs, guiders, fiber view camera, and sky monitors,
and fits to those data performed as part of fiber positioning and field acquisition. For discussion of the instrument components, see \cite{desi-collaboration22a}, and for discussion of the survey operations, see \cite{schlafly23a}.

Intermediate data processing files in subdirectories of \texttt{spectro/redux/fuji/}
include preprocessed spectrograph CCD data (\texttt{preproc/});
nightly biases, and arc- and flat-lamp calibrations (\texttt{calibnight/});
and intermediate spectra steps (\texttt{exposures/})
\eg~sky-subtracted spectra that are not flux-calibrated (\texttt{sframe-$\ast$.fits.gz}).

For completeness, the EDR contains all inputs used by the spectroscopic processing pipeline, including
CCD calibration files in \texttt{spectro/desi\_spectro\_calib/} and \texttt{spectro/desi\_spectro\_dark/};
stellar templates used to fit standard stars in \texttt{spectro/templates/basis\_templates/};
and survey progress bookkeeping files in \texttt{survey/ops/}
used to track if a tile is ``done'' and should be included in a release.

\subsection{Known Issues}
\label{sec:known_issues}

While working with the internal pre-release of EDR, the DESI collaboration has identified several issues with the data produced by the pipeline, which we report here for completeness.

\begin{itemize}

\item Redrock templates do not include Active Galactic Nucleus (AGN)-like galaxies with a mixture of broad and narrow lines.  As a result, these types of galaxies are often fit equally well (or equally poorly) with either \texttt{GALAXY} or \texttt{QSO} templates at the same redshift, which can also trigger \texttt{ZWARN} bit 2 (value $2^2=4$) for \texttt{LOW_DELTACHI2} since the $\chi^2$ difference between the two fits is small, indicating an ambiguous answer.

\item There are cases where Redrock is overconfident and reports \texttt{ZWARN==0}, \ie~no known problems, even though the fit is incorrect.  This can include unphysical fits due to the over-flexibility of PCA template linear combinations. This is particularly true for sky fibers which have a higher fraction of \texttt{ZWARN==0} than would be expected from purely random fluctuations.  Users should be especially cautious in any search for serendipitous targets in nominally blank sky fibers.

\item The Redrock galaxy fits extend to redshift $z=1.7$, though the range $1.6 < z < 1.63$ is only constrained by the \oii doublet in the midst of significant sky background while $1.63 < z < 1.7$ has no major emission line coverage.
Thus $1.6 < z < 1.7$ is particularly susceptible to unphysical fits.  This was the motivation for the LSS catalogs to only consider galaxies with $z<1.6$ (see \S\ref{sec:lsscat}).

\item Negative \texttt{TARGETID}s indicate positioners that were non-functional and thus were not pointing at a known science target. Although these are unique within a given \texttt{TILEID}, they are not unique values across different \texttt{TILEID}s. Since these are not science targets, most users can discard any negative \texttt{TARGETID} targets.  This has been fixed for Main Survey
non-functional positioners in future data releases, where negative \texttt{TARGETID}s will be unique.

\item For most of the tiles in Target Selection Validation, proper-motion corrections were applied in fiberassign when the tile was designed.\footnote{The design date can differ from when a tile was observed.} A consequence is that the (\texttt{TARGET\_RA}, \texttt{TARGET\_DEC}, and \texttt{REF\_EPOCH}) values are altered to have a \texttt{REF\_EPOCH} of the date that the tile was designed, which makes them differ from the input photometric column values. The information is correct and consistent with the photometry, however.

\item In the coadded \texttt{FIBERMAP} tables, 0.03\% of targets incorrectly have \texttt{COADD\_FIBERSTATUS==0}
even though all of their data are masked.  These result in \texttt{ZWARN} $\neq 0$ in the redshift fits, but quality cuts based solely upon \texttt{COADD\_FIBERSTATUS} have a tiny amount of contamination.

\item In the coadded \texttt{FIBERMAP} tables, \texttt{MEAN\_FIBER\_RA} and \texttt{MEAN\_FIBER\_DEC} record the
average as-observed position of the fibers (in comparison to the intended positions recorded in
\texttt{TARGET\_RA} and \texttt{TARGET\_DEC} plus proper motions \texttt{PMRA}, \texttt{PMDEC}, \texttt{REF\_EPOCH}).
The \texttt{FIBERMAP} coordinate coaddition incorrectly included exposures that had been excluded from the spectral coaddition, which can result in incorrect \texttt{MEAN\_FIBER\_RA/DEC} values.  The same issue applies to the standard deviations recorded in \texttt{STD\_FIBER\_RA/DEC}.
As a result, \texttt{TARGET\_RA/DEC} are more reliable than \texttt{MEAN\_FIBER\_RA/DEC}, while noting that
the actual positioning can vary by O(0.1$^{\prime\prime}$), which is still small compared to the $\sim 1.5^{\prime\prime}$ diameter fibers.

%%% SB/AK/AR: too detailed
%%% \item During SV, the \texttt{SUBPRIORITY} column was overwritten in the \texttt{fiberassign} files. These were propagated to the \texttt{FIBERMAP} extension of many data products; see \S5.2 of \citep{myers23a} for more details.

\end{itemize}

Additional known issues and clarifications will be documented at \url{https://data.desi.lbl.gov/doc/releases/edr} when needed.

\section{One-Percent Survey LSS Catalogs}\label{sec:lsscat}

DESI creates large-scale structure (LSS) catalogs from its data in order to facilitate clustering measurements. The overall process is similar to that applied to SDSS (most recently eBOSS; \citealt{ebossLSS}). We determine the area on the sky where good observations were possible for each tracer, applying criteria 
on the DESI data to select reliable redshifts, and provide weights that correct for variations in observing completeness. The end results are data and matched random catalogs, split into the various relevant DESI tracer classes, that can be passed directly to any common software for calculating redshift-space clustering measurements.

These catalogs for the One-Percent Survey are ideal for studying small-scale clustering, as the tiling strategy makes them highly complete for all tracer types \citep{DESI23a}. Some studies using the results derived from these catalogs include \cite{Gao2023,pearl2023,Prada2023,Rocher2023,Yu2023,Yuan2023}. The catalog files are available in the EDR at \texttt{vac/edr/lss/v2.0}; see \S\ref{sec:access} for data access details.
% original had a "abacusELG" citation to Rocher+ 2023 in prep, but SB thinks that is the same as Rocher2023 refering to "The Halo Occupation Distribution of DESI 1% ELG sample".

\subsection{Gathering Assignment and Observation Information}
\label{sec:lss_inputs}

%In this section we describe

The DESI LSS catalogs begin by gathering the information that describes where and what on the sky DESI could observe. Two fundamental pieces of this information were generated by the DESI targeting team \citep{myers23a}. These are: 1) the data chosen (`targeted') for spectroscopic follow-up, including photometric properties of the targets and meta-data related to their observation (hereafter simply `data') and 2) a uniform random distribution of points on the sky occupying the area covered by Legacy Survey imaging that also includes the meta-data related to the photometric observations at the given location (hereafter `randoms'; see \S4.5 of \citealt{myers23a}). Both the data and randoms were given a unique identifier, \texttt{TARGETID}, by the DESI targeting team and this identifier is used to match between relevant data files in all relevant cases. We use 18 random files (more are available), each with a density 2500/deg$^2$. The 2500/deg$^2$ is convenient, as it allows for quick determination of footprint area after various cuts. 

With these data and randoms, we first track the locations on the sky where DESI observations could have happened. We do this using the outputs of the DESI fiberassign software \citep{raichoor23a}. As described in \S\ref{sec:fiberassign-files}, prior to the observation of each DESI tile, targets are processed through fiberassign in order to assign targets to particular fibers (with each unique fiber corresponding to a unique robotic positioner and a unique location in the spectrograph CCDs). In addition to the particular assignment, the information on all potential assignments is also stored for each tile. Further, all settings used when running the fiberassign software are stored so that the assignments can be reproduced. This means that we can also run the randoms through fiberassign, with the matched settings. The positions of the randoms in the potential assignments thus provide a superset of the geometric area observable by DESI on every tile. We thus concatenate the potential assignment information across all tiles for both data and randoms.\footnote{They are available in \texttt{vac/edr/lss/v2.0/potential_assignments}.}

The total set of concatenated potential assignment information includes many duplicated targets. In the One-Percent Survey, each location on the sky was potentially covered 13 times. Further, many sky locations are accessible to more than one robotic positioner. Initially, we keep all of the information on repeated potential assignments. Downstream, many will be vetoed, \eg, due to hardware performance or low data quality. At this stage, we also match to the information on redshifts in the cumulative tile-based redshift catalog provided in the EDR (see \S\ref{sec:spectra-files}). For the data, we match not just to the target, but to the particular tile and fiber on which it was observed.\footnote{The particular columns used for the match are \texttt{TARGETID},\texttt{TILEID},\texttt{LOCATION}; \texttt{LOCATION} refers to the robotic positioner, which corresponds to a unique fiber.} Many rows for the data will have no match and the corresponding entries will simply be null. For the randoms, we match in order to obtain the metadata related to the particular spectrum. We, therefore, match only to the tile and fiber and will have a match for all random targets.\footnote{The files with matches to spectroscopic information are available 
at \texttt{vac/edr/lss/v2.0/inputs_wspec}.}

%The \texttt{TILES} information provides the information on unique tile groupings, analogous to the `sectors' defined in SDSS. We determine the completeness within these groups as \texttt{COMP\_TILE}. 
The information on the number of overlapping tiles allows us to divide the observed area by coverage.
The covered area listed in Table~\ref{tab:survey_nights_tiles_exp} is simply the unique area that is overlapped by any tile in that survey, but it does not include the detailed accounting for focal plane geometry, disabled/broken hardware, or higher-priority targets blocking lower-priority targets from being observed.  Using data and randoms and repeated realizations of fiber assignment provides a much more accurate and geometrically detailed measurement of the true coverage.

\begin{figure}
    \centering 
    \includegraphics[width=1.01\columnwidth]{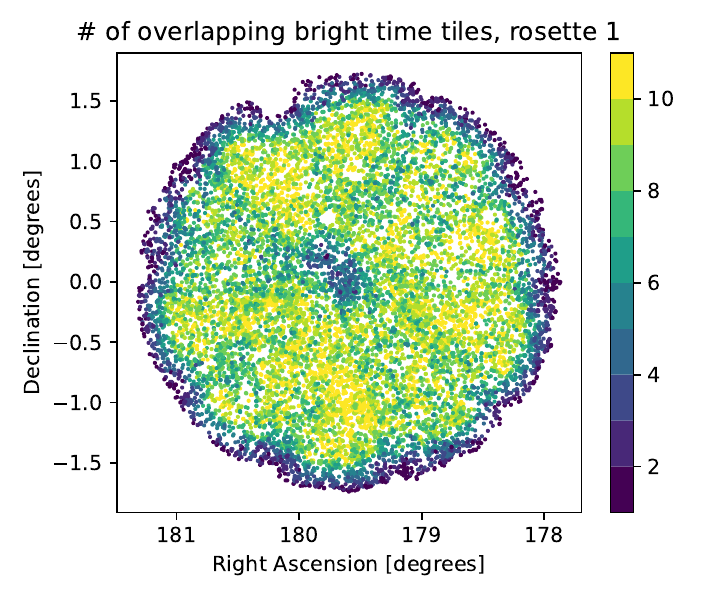}
    \caption{The number of overlapping bright time tiles at each location of a BGS target in `rosette' number 1 of the DESI One-Percent Survey (SV3 R1 in Table \ref{tab:SV_fields}).}
    \label{fig:BGSntile}
\end{figure}

Fig.~\ref{fig:BGSntile} displays this information for rosette number 1 (SV3 R1 in Table~\ref{tab:SV_fields}) of the bright time program. Each point is a BGS target at a location covered at least once by the One-Percent Survey. One can see that some areas were covered up to 11 times. For dark time observations, areas were covered up to 13 times. This dense coverage provides highly complete samples.

%By tracking this potential assignment information
%For the data, these are the targets and their photometric properties. For convenience, we first compile the One-Percent Survey target files (which are divided by healpix in order to keep the memory load of each small enough) into all of those within the greater area of the SV3 tiles, applying the process separately for bright and dark time. %\texttt{(CATROOT)/SV3/LSS/(bright\_or\_dark)\_targets.fits}

%For the data, we also use the information on assignments generated by \textsc{fiberassign} (one file per observed tile) and the information from the redshift catalogs {\bf described somewhere in this paper}. The process is described below.

%These 

%In order to match the footprint of observed data, for every observed tile, we run \textsc{fiberassign} for each random file, using the same settings as applied to the data. These settings are read from the headers of the data \textsc{fiberassign} files. These are run treating all random points the same. The only information that is used downstream is that in the \texttt{FAVAIL} HDU, which contains the information on random points that were reachable by positioners. For every observed tile, we thus have a \textsc{fiberassign} file for both data and randoms.%The outputs are stored like \\ \texttt{(CATROOT)/(survey)/LSS/random(random\_number)/fba-(0padded\_tileid).fits}
%in \texttt{(DESIROOT)/target/fiberassign/tiles/trunk/(0padded\_tileid[:3])/} and named like \\ \texttt{fiberassign-(0padded\_tileid).fits.gz}. Link to code
In order to jointly model the completeness of all targets with respect to each other, we simulate the observation of targets in the One-Percent Survey using multiple realizations of their assignment histories. The process and its results are described in more detail in \cite{lasker2023a}. Briefly, all DESI targets were initially assigned a random \texttt{SUBPRIORITY}, which is used to determine which target gets assigned a fiber when available targets have the same overall priority (see \citealt{raichoor23a,myers23a}). We create 128 alternative assignment histories by randomly shuffling the subpriorities 128 times.\footnote{Each of the alternative assignment histories is available in \texttt{vac/edr/lss/v2.0/altmtl}.} For each of the 128 realizations, the One-Percent Survey observations are simulated by following the same order of tile observations and targeting feedback loop.\footnote{Targets with good observations have their priority adjusted so that unobserved targets are more likely to be assigned, see \S5 of \cite{schlafly23a}.} From these simulations, we package the results as a bit value for each target that encodes the realization numbers it was observed in. From these bit values, one can determine the individual and pairwise probabilities required to obtain unbiased clustering statistics \citep{2017Bianchi}. 

\subsection{LSS Catalogs}

The LSS catalogs are cut (subselected down) to unique \texttt{TARGETID} per supported tracer type. They come in two flavors. The `full' catalogs contain an entry for all reachable targets, whether or not they were observed\footnote{They are available in \texttt{vac/edr/lss/v2.0/LSScats/full}.}. They also include all columns believed to possibly be relevant. The `clustering' catalogs cut to good spectroscopic observations and the redshift range intended for clustering analysis\footnote{They are available in \texttt{vac/edr/lss/v2.0/LSScats/clustering}.}. They include weights to account for variations in the selection function and only include the columns required to calculate clustering statistics.

%The outputs are all in \texttt{LSSdir} = \texttt{(CATROOT/(suvey)/LSS/(spec\_release)/LSScats/(version)}

Catalogs were created for the four extragalactic DESI target types:
BGS, LRG, ELG, and QSO. For all except QSO, catalogs are produced for additional sub-type definitions. In the cases where the sub-type corresponds to a bitname from SV3 targeting (see Table \ref{table:sv3bits} and surrounding text), we use that name. The additional LRG selection, named  
LRG\_main, keeps only targets that satisfy the Main Survey selection (see \citealt{zhou22a} for details). The ELG sample is cut in three additional ways. First, ELG\_HIP contains only the $\sim$75\% of the ELG sample assigned higher-priority (see \citealt{raichoor22a} for more details). Then, for each of ELG and ELG\_HIP, we also remove QSO targets. QSO targets have the highest priority and it may, therefore, be useful to treat any targets that satisfy both the QSO and ELG selections within the QSO analysis. Finally, there are two BGS samples: BGS\_ANY and BGS\_BRIGHT. BGS\_ANY is the combination of both the BGS\_BRIGHT and BGS\_FAINT BGS selections. See \cite{hahn22a} for more details. A summary of the statistics for all tracer types is given in Table~\ref{tab:sv3targets}.

\begin{table*}
\centering
\caption{\label{tab:sv3targets}Statistics for each of the DESI tracer types for which One-Percent Survey LSS catalogs were created. We list the number of good redshifts included, the redshift range we included them from, the sky area occupied, and the observational completeness within that area. The area is slightly different for different tracer types due to priority vetoes (\eg, a QSO target can remove sky area from lower-priority samples). The completeness listed is the number of targets observed divided by the number of targets within the entire observable area. The completeness can be increased by cutting on the rosette radius (see text and Fig. \ref{fig:comp_rosette}).}
\begin{tabular}{l|cccc}
\hline\hline
Tracer & \# of good z  & $z$ range & Area [deg$^2$] & Completeness  \\\hline
BGS\_ANY & 241746 & $0.01 < z < 0.6$ & 174 & 94\% \\
BGS\_BRIGHT & 143853 & $0.01 < z < 0.6$ & 174 & 96\%\\
LRG  & 112649 & $0.4 < z < 1.1$ & 167 & 95\% \\
LRG\_main  & 86040 & $0.4 < z < 1.1$ & 167 & 95\% \\
ELG  & 267345 & $0.6 < z < 1.6$ & 169 & 86\% \\
ELGnotqso  & 259317 & $0.6 < z < 1.6$ & 164 & 88\% \\
ELG\_HIP  & 209833 & $0.6 < z < 1.6$ & 169 & 89\% \\
ELG\_HIPnotqso  & 202734 & $0.6 < z < 1.6$ & 164 & 90\% \\
QSO & 35566 & $0.6 < z < 3.5$ & 175 & 98\% \\\hline
\end{tabular}
\end{table*}

\subsubsection{Full Catalogs \label{sec:lss_fullcat}}
Starting from the compilation of all potential observations, the first step in creating the full catalogs for the data is to cut to targets of a given target type. We then cut to unique targets. We cannot do so randomly, as, \eg, we must keep the cases that have been observed. We define the following boolean quantities:

\textbullet~ $H_{\rm good}$: The particular fiber on the given tile was observed with good hardware. %\texttt{TILELOCID} (but not necessarily the particular \texttt{TARGETID}) ; denoted $G$ in below equation.

%\texttt{GOODHARDLOC} 

\textbullet~ $L_{\rm fa}$: The particular target, fiber, and tile was assigned and observed.
%\texttt{LOCATION\_ASSIGNED} The particular \texttt{TARGETID},\texttt{TILELOCID} was assigned for observation; denoted $L$ in below equation.

\textbullet~ $T_{\rm fa}$: The particular fiber and tile (but not necessarily target) was assigned and observed.
%\texttt{TILELOCID\_ASSIGNED} The \texttt{TILELOCID} was assigned for observation (but not necessarily to the given \texttt{TARGETID}; denoted $T$ in below equation.

\textbullet~ $S_{\rm good}$: The particular fiber and tile (but not necessarily target) was determined by the spectroscopic pipeline to have a template squared signal-to-noise ratio (\texttt{TSNR2}, referred to here as $S_{\rm ratio}$) above the vetoing threshold (defined below).
%\texttt{TSNR2\_(tracer)} value above the vetoing threshold; denoted $S$ in below equation.
%\texttt{GOODTSNR} The particular \texttt{TILELOCID} (but not necessarily the particular \texttt{TARGETID}) was determined by the spectroscopic pipeline to have a \texttt{TSNR2\_(tracer)} value above the vetoing threshold; denoted $S$ in below equation.

We clip $S_{\rm ratio}$ to be within the range (0,200) and then these quantities are combined to create a value to sort by
\begin{equation}
    v_{\rm sort} = L_{\rm fa} S_{\rm good} H_{\rm good}(1+S_{\rm ratio})+T_{\rm fa} H_{\rm good}+H_{\rm good}.
\end{equation}
We sorted by this $v_{\rm sort}$ in ascending order and then cut to unique targets by selecting the last entry for each unique target. After cutting to unique targets, we join to the redshift determined in the HEALPix-based redshift catalog and use \texttt{Z\_HP} as the column name.

%We also add:

%\textbullet~ \texttt{ROSETTE\_NUMBER} The number of the `rosette' where tiles are grouped.

%$\textbullet~ \texttt{ROSETTE\_R} The angular distance from the center of the rosette, in degrees.

For ELG catalogs, we join the information on the \oii emission line fits, produced with the spectroscopic release using the HEALPix-based coadd. We combine the information on \oii flux and its inverse variance to obtain the signal-to-noise ratio of the \oii flux for each observed spectrum, which we denote $S_{\rm \otwo}$. Following \cite{raichoor22a}, this information is combined with the $\Delta\chi^2$ obtained from the redshift fitting pipeline between the best and next-best-fit redshifts (the quantity \texttt{DELTACHI2} in \S\ref{sec:redrock}) to determine a criterion, ${\rm \otwo}_{\rm crit}$, for selecting reliable redshifts:

%\begin{split}
%\begin{multiline}
\begin{equation}
%\begin{dmath}
{\rm \otwo}_{\rm crit} =  {\rm log10}\left(S_{\rm \otwo}\right)\\
 +0.2{\rm log10}(\Delta\chi^2).
%\end{dmath}
\label{eq:elg_zcrit}
\end{equation}
%\end{multiline}
%\end{split}

For quasars, we join to the HEALPix-based quasar catalog, which contains extra information related to classifying the spectra as \texttt{QSO} or not and improved redshift estimates. These catalogs will be released as a VAC and will be fully documented in \cite{Canning2023}. The fiducial pipeline estimates of the redshifts are replaced by those from the quasar catalog.\footnote{We replace the original \texttt{Z\_HP} column with that from the quasar catalog and rename it \texttt{Z\_RR} (to indicate `Redrock'). \texttt{ZERR} from the quasar catalog is renamed as \texttt{ZERR\_QF} and \texttt{ZERR} remains the \texttt{ZERR} from Redrock.}
%From it, we add the columns 'Z', 'ZERR', and 'Z\_QN'. We replace the original 'Z' column with that from the QSO catalog and rename it 'Z\_RR' (to indicate 'redrock'). 'ZERR' from the QSO catalog is renamed as 'ZERR\_QF' and 'ZERR' remains the 'ZERR' from redrock.
The flavor of the quasar catalog that we use is the one that only contains entries for objects believed to be quasars.\footnote{We use the file: \\ \texttt{QSO\_cat\_fuji\_sv3\_dark\_healpix\_only\_qso\_targets.fits}. The quasar catalog will also provide a flavor that includes the diagnostic information for all targets.}  Thus, there will be many more rows with null entries for the quasar information than for those with Redrock information. 

%The tables are output to fits files, named like \texttt{(LSSdir)/(tracer)\_full\_noveto.dat.fits}.

For randoms, we must also cut to unique \texttt{TARGETID} and we do so separately for each tracer type, as the tracer information is included in the sort. We also apply the imaging mask bits that were applied to the target samples. These are Legacy Survey bits\footnote{\url{https://www.legacysurvey.org/dr9/bitmasks/\#maskbits}} 1 \revisionforreviewer{(\texttt{BRIGHT}) and 13 (\texttt{CLUSTER}) applied to bright time targets to flag objects with imaging pixels within half of the locus of a radius-magnitude relation of a `bright' star or in a globular cluster respectively. In addition, bit 12 (\texttt{GALAXY}) is included for dark time targets to flag targets in a large galaxy in the Siena Galaxy Atlas \citep[SGA;][]{moustakas23a}.}\footnote{\url{https://github.com/desihub/desitarget/blob/1.1.1/py/desitarget/geomask.py\#L137}} In addition to the boolean quantities used to sort the data, we use the following:

%\textbullet~ \texttt{GOODHARDLOC} Same as for the data, except that all observed fibers are included (ignoring the target type).

%\textbullet~ \texttt{GOODTSNR} Again, same as for the data, except that all observed fibers are included (ignoring the target type).

\textbullet~ $P_{\rm good}$: The \texttt{PRIORITY} of the target that was assigned at the given tile and fiber was less than or equal to the \texttt{PRIORITY} of the given target class.  
%\texttt{GOODPRI} The \texttt{priority} of the target that was assigned at the given \texttt{TILELOCID} was less than or equal to the \texttt{PRIORITY} of the given target class; denoted $P$ in the equation below.

\textbullet~ $Z_{\rm poss}$: The tile and fiber was either assigned to a target of the given target class or no unassigned targets of the given target class were reachable by the fiber on this tile.
%\texttt{ZPOSSLOC} The \texttt{TILELOCID} was either assigned to a target of the given target class or no unassigned targets of the given target class were reachable at this \texttt{TILELOCID}; denoted $ZP$ below.

The randoms are then sorted by
\begin{equation}
    v_{\rm sort} = S_{\rm ratio} H_{\rm good} P_{\rm good} Z_{\rm poss}
\end{equation}
and we cut to unique random points by selecting the highest $v_{\rm sort}$ value for each. %The \texttt{ROSETTE\_NUMBER} and \texttt{ROSETTE\_R} columns are added like for the data. %The tables are output to fits files, named like \texttt{(LSSdir)/(tracer)\_(random\_number)\_full\_noveto.ran.fits}.

The final step for the `full' catalogs is to apply vetoes.\footnote{The catalogs that have `noveto' in the file name are the catalogs produced prior to this veto process.}  All of the boolean columns defined above for data and randoms must be \texttt{True} to pass the veto. Note that the combination of the $P_{\rm good}$ and $Z_{\rm poss}$ criteria act as a priority veto mask. Thus, the area occupied by the lowest priority targets (as traced by the number of randoms) will be less than the highest priority ones. 

The sky area occupied for each target class in the One-Percent Survey is given in Table \ref{tab:sv3targets}. The difference between the highest (QSO) and lowest (ELGnotqso) priority targets is only 7\% due to the survey strategy to achieve high coverage. The ELG catalogs with QSO targets removed have a smaller area than those including the QSO targets because the priority used for the $P_{\rm good}$ determination is that of QSO when they are included but that of ELG\_HIP targets when they are not. Correspondingly, there is a small increase in the completeness of the ELG samples with the QSO targets removed, as the area removed was at sky locations where only QSO targets were observable.

Finally, additional vetoes are applied based on the imaging data. For LRGs, we apply a custom mask, described in \cite{zhou22a}. For the other tracers, Legacy Survey bit 11 \revisionforreviewer{(\texttt{MEDIUM}) is always used to flag objects with imaging pixels within the locus of a radius-magnitude relation of a `medium' brightness star. For QSO, bits 8 (\texttt{WISEM1}) and 9 (\texttt{WISEM1}) are also applied. These refer to targets with pixels in the imaging that have a \texttt{WISEMASK_W1} or \texttt{WISEMASK_W2} bright star mask flag set respectively.}

\begin{figure}
     \centering 
     \includegraphics[width=1.01\columnwidth]{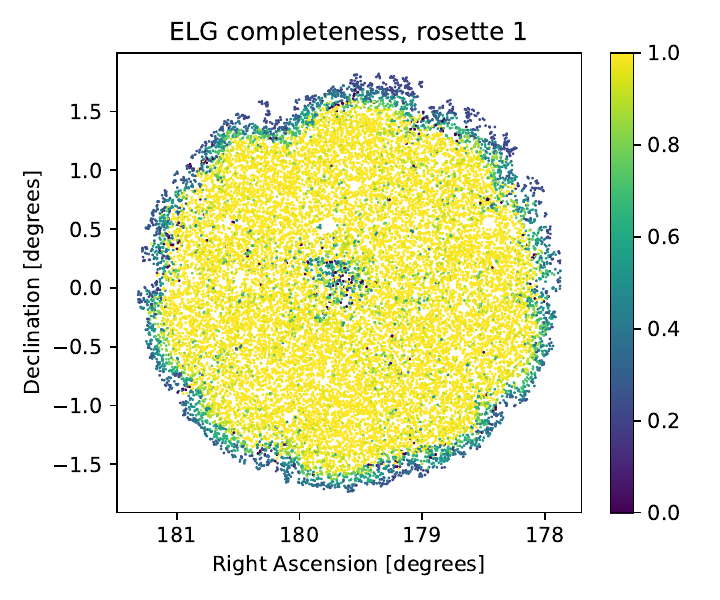}
     \caption{The observational completeness within each tile grouping of ELG targets on `rosette' number 1 (SV3 R1 in Table \ref{tab:SV_fields}) from the DESI One-Percent Survey.}
     \label{fig:comp_rosette}
 \end{figure}

Given the vetoed catalogs, we determine a completeness both per fiber, $C_{\rm fib}$, and per unique tile grouping, $C_{\rm tile}$. Per fiber is simply the inverse of the number of data points at a given tile and fiber.\footnote{In the catalogs, the column is \texttt{FRACZ\_TILELOCID}.} The values of $C_{\rm fib}$ are similar to the $P_{\rm obs}$ value obtained from 128 alternative assignment histories. The completeness per tile grouping is simply the observed number divided by the total number within each tile group, \ie, $C_{\rm tile} = N_{\rm observed}/N_{\rm total}$. As an example, Fig. \ref{fig:comp_rosette} shows the completeness, $C_{\rm tile}$, of ELG targets in rosette number 1. The pattern observed there is typical of all rosettes. The ELG completeness in the plotted rosette is 87\% and across all rosettes it is 86\%. The ELG sample has the lowest completeness, as they had the lowest target priority. The completeness decreases towards the edges and the center, as the total number of overlapping tiles decreases in these regions. The completeness increases to 95\% if one cuts to $0.2 < r_{\rm rosette} < 1.45$, where $r_{\rm rosette}$ is the angular distance from the center of the rosette, in degrees.\footnote{The corresponding column name is \texttt{ROSETTE\_R} in the catalogs.}

\subsubsection{Clustering Catalogs}\label{sec:lss_clusteringcat}

For both data and randoms, the clustering files take the full files and reduce them to a subset of columns necessary for calculating 2-point statistics. The full data file is cut to only those objects with good redshifts and weights to account for selection function variations.

%The files are named like \texttt{(LSSdir)/(tracer)\_(region)\_clustering.dat.fits} for the data and \\ \texttt{(LSSdir)/(tracer)\_(region)\_(random\_number)\_clustering.ran.fits} for the randoms

The catalogs are provided for the `S' (DECaLS) and `N' (BASS/MzLS) photometric regions (see \S4.1.3 of \citealt{myers23a}). Given that the photometry is different because different cameras/filters were used for each, we expect at least slight differences in the selection function between the two regions and thus will always estimate it separately. The area in the S and N regions is nearly identical for the One-Percent Survey. The difference varies from the S region being less than one percent greater for the ELG samples to four percent greater for the LRG sample.\footnote{The N region in the One-Percent Survey has more area affected by stars that are bright in the infrared.} 

Only data with good redshifts are kept. We use the HEALPix-based redshift (see \S\ref{sec:spectra-files}) as the estimate for the redshift.\footnote{Thus, for the `clustering' catalogs, the column `Z\_HP' is changed to `Z'.} Each tracer has a different definition for a `good' redshift, as detailed in the respective target selection papers (BGS, \citealt{hahn22a}; LRG, \citealt{zhou22a}; ELG, \citealt{raichoor22a}; and QSO, \citealt{chaussidon22a}). The criteria were motivated by maximizing the completeness while minimizing the fraction of catastrophic failures expected for main survey observations and used a combination of comparisons of Redrock redshift to a visually inspected redshift or multiple Redrock redshifts from repeated observations of the target. The estimated catastrophic failure fractions are less than 0.5\% for all tracer types after restricting to `good' redshifts. Key quantities used to select good redshifts are the redshift pipeline flag \texttt{ZWARN} and the $\Delta\chi^2$~(\texttt{DELTACHI2}) obtained from the redshift fitting pipeline between the best and next-best-fit redshifts. We also restrict to a given redshift ($z$) range that is different for each tracer. The combined criteria are:

\textbullet~ BGS: \texttt{ZWARN==0}, $\Delta\chi^2>40$, $0.01 < z < 0.6$

\textbullet~ LRG: \texttt{ZWARN==0}, $\Delta\chi^2>15$, $0.4 < z < 1.1$

\textbullet~ ELG: ${\rm \otwo}_{\rm crit}$ $>0.9$, $0.6 < z < 1.6$ (with ${\rm \otwo}_{\rm crit}$ defined by Eq.~\ref{eq:elg_zcrit}).

\textbullet~ QSO: Not already rejected by the quasar catalog, $0.6 < z < 3.5$

The quasar catalog requires that either Redrock or QuasarNET identified the object as a QSO, while the galaxy catalogs (BGS, LRG, ELG) do not explicitly require that the targeted objects are spectrally classified as galaxies.

For the DESI One-Percent Survey, we provide two weights to be used with the clustering catalogs. One, $w_{\rm comp}$, accounts for fiber assignment incompleteness. The other, $w_{\rm FKP}$, optimizes against expected signal-to-noise in 2-point clustering measurements as a function of redshift and is based on \cite{FKP}. We will describe both below. In previous SDSS (most recently, \citealt{ebossLSS}) and future DESI LSS catalogs, weights that account for fluctuations in both target density due to imaging quality and redshift success due to the signal-to-noise of spectroscopic observations are added. We do not include such weights for the DESI One-Percent Survey LSS catalogs. For target density fluctuations, their effects typically manifest on large angular scales and the relatively small area of the One-Percent Survey footprint is not ideal for the regression methods typically used to define them. For the redshift success, the effective exposure time reached during the One-Percent Survey was such that (after selecting the greatest signal-to-noise measurement out of any repeat observations) the variation in success rate is relatively low. Determining these weights for DESI Main Survey data is a primary focus of DESI year 1 analyses.
%and the unique observing strategy that prioritized targets that failed to obtain a high {\bf specifics + citation needed} $\Delta\chi^2$ on their first observation meant that we were unable to determine a scheme that would provide the relative probability of achieving a good redshift for a given target. 

%For the One-Percent Survey LSS catalogs, the \texttt{WEIGHT} column is identical to the \texttt{WEIGHT\_COMP} column, as we do not determine other types of weights to account for variations in the survey selection function (\eg, for imaging systematics or varying spectroscopic success rate). 

The completeness weights,\footnote{Their column name is \texttt{WEIGHT\_COMP}; the column \texttt{WEIGHT} is identical to \texttt{WEIGHT\_COMP} for the One-Percent Survey LSS catalogs, but this will not be the case for future DESI LSS catalogs.} $w_{\rm comp}$, are determined from the 128 realizations of alternative assignment histories. The process of generating these realizations is detailed in \cite{lasker2023a}. Given there is one data realization, we have 129 total realizations. The number of realizations in which a target was assigned is thus $128 P_{\rm obs}$ + 1. The probability of assignment is N${}_{\mathrm{assigned}}$/N${}_{\mathrm{tot}}$, and we wish to use the inverse probability as the weight. Thus, 
\begin{equation}
  w_{\rm comp} = 129/(128 P_{\rm obs}+1).  
\end{equation}
 These weights can be used to obtain unbiased statistics for any one-point measurement or for clustering measurements on projected scales that are large relative to the fiber patrol radius, which is at most 89$^{\prime\prime}$ (it depends on \eg, the focal plane position due to the optics). For unbiased $N$-point clustering statistics in general, one should use the bit values to determine the joint probability for any configuration, and, \eg, for 2-point statistics follow the process outlined in \cite{2018Bianchi}. Comparisons of clustering results applying (or not) various weighting prescriptions to the One-Percent Survey data are presented in \cite{lasker2023a} and \cite{Rocher2023}.

The clustering randoms contain the same rows as the full random files. Redshifts and weights are added to the randoms by randomly sampling the data. In this way, the weighted $dN/dz$ of the data and random should match (and the weights on the random points are there {\it only} for this purpose). Other columns, such as photometry, that vary with redshift are similarly sampled. In all cases, any cuts that are applied to the data sample should also be applied to the random sample. Potential choices include, \eg, cuts on $r_{\rm rosette}$, $N_{\rm tile}$, or redshift. Any number of random files (recall, 18 total are available, each with a projected density of 2500/deg$^2$), or even a sub-selection of a random file, can be used without biasing any potential statistic, with the caveat that using less random points means a higher shot-noise contribution from the randoms.

\begin{figure}
    \centering 
    \includegraphics[width=1.01\columnwidth]{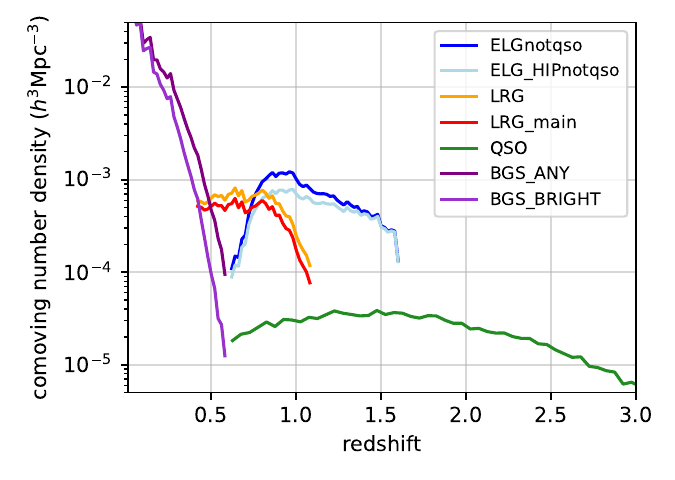}
    \caption{The comoving number density for samples used to create LSS catalogs in the DESI One-Percent Survey. For this display, we have taken the mean of the results in the `N' and `S' regions (see text for details).}
    \label{fig:nz}
\end{figure}

The comoving number density as a function of redshift, $n(z)$, is determined for each tracer by applying the completeness weights, and represents the estimated density for a complete sample. In order to calculate the comoving volume, we use a cosmological model based on the Planck 2018 results\footnote{Specifically, the 2018 {\it Planck} TT,TE,EE+lowE+lensing mean
results, including massive neutrinos.} \citep{Planck2018} and calculate all comoving distances in the units $h^{-1}$Mpc. The $n(z)$ are determined separately for the `N' and `S' regions. Fig. \ref{fig:nz} shows the measured $n(z)$, taking a simple mean of the `N' and `S' results. The $n(z)$ are used to determine the $w_{\rm FKP}$ weights that are included in the clustering catalogs.\footnote{The column name is \texttt{WEIGHT\_FKP}.} We use
\begin{equation}
w_{\rm FKP}(z) = \frac{1}{1+n(z)P_0}.
\end{equation}
We use a value of $P_0$ that is different for each tracer type and is approximately equal to the power spectrum amplitude at $k=0.15 h$Mpc$^{-1}$; the values are 10$^4$, 7000, 6000, and 4000 Mpc$^3h^{-3}$ for LRG, BGS, QSO, and ELG.

\subsubsection{BGS $k$- and $e$-Corrections}

The BGS sample is approximately flux-limited, thus its galaxies will have an especially large range in their intrinsic luminosity. In order to compare the clustering of BGS galaxies at different redshifts and luminosities, we provide `$k$' and `$e$' corrections. Typically, absolute magnitudes are corrected by $k$-corrections to account for bandshifting effects, specifically that the observed flux distributions in a given passband will be different in the rest frames of galaxies at different redshifts. We thus provide $r$-band absolute magnitudes, $M_r$, with the BGS clustering catalogs,\footnote{The column name is \texttt{ABSMAG\_R} and is for the $z=0.1$ reference-frame.} defined via
\begin{equation}
    M_r - 5 {\rm log}_{10}(h) = m_r - 5{\rm log}_{10} (d_L (z)) - k_r(z).
    \label{eq:kmr}
\end{equation}
\noindent Here, the subscript $r$ represents the $r$-band, $k(z)$ represents the $k$-correction of the galaxy, and $d_L(z)$ is the luminosity distance to the redshift $z$, determined using the same cosmology defined in the previous subsection (see \citealt{hogg2002k} for a good overview of $k$-corrections). Optionally and in addition, an $e$-correction may be applied in order to account for the intrinsic luminosity evolution of a galaxy over time. Further, we derive the reference-frame $g-r$ color and thus also the $g$-band $k$ correction. Results are provided using both $z=0$ and $z=0.1$ as the reference-frame.\footnote{Denoted via \texttt{\_0P0} and \texttt{\_0P1} in the column name for $z=0$ and $z=0.1$, respectively.}

\begin{figure}
    \centering 
    \includegraphics[width=1.01\columnwidth]{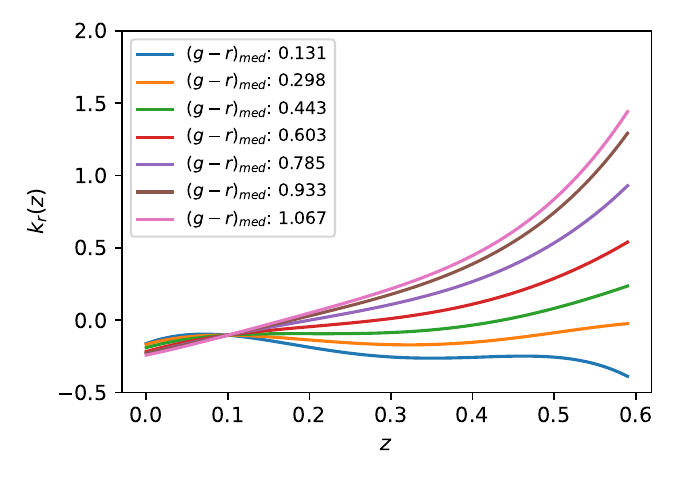}
    \caption{The seven $r$-band $k$-correction polynomials for median rest-frame $(g-r)_0$ in each color bin.}
    \label{fig:kcorr}
\end{figure}

The methods we use to determine the BGS EDR $k+e$ corrections are detailed in \cite{Moore2023}.\footnote{The code is at \url{https://github.com/SgmAstro/DESI}.} To begin with, we make use of the Galaxy and Mass Assembly (GAMA) DR4 dataset to create estimates of the $k$- and $e$-corrections \citep{Driver_2022}. Each galaxy has an individual $k$-correction polynomial, calculated using KCORRECT v4.2 (see \citealt{BlantonRoweis2007} and \citealt{Loveday2012} for further details). As such, each galaxy has an individual $k(z)$, given its observed $g-r$. These $k(z)$ values are divided into seven equal-width $(g\hspace{1 pt}-\hspace{1 pt}r)_0$ color bins. Within each of these color bins, a fourth-order polynomial is fitted to the data points corresponding to the median $(g-r)_0$ color of the bin (see Fig. \ref{fig:kcorr}). The polynomial is based on the functional form shown in Equation \ref{eq:kpoly}.

\begin{equation}
    k(z) = \sum_{i=0}^{4} a_i (z - z_{\rm ref})^{4-i}
    \label{eq:kpoly}
\end{equation}

The coefficients of the polynomials for $z_{\rm ref} = 0.1$ are shown in Table \ref{tab:krcoeffs}. Moreover, a linear interpolation between these color polynomials is done such that the $k$-correction of a galaxy is found based on its rest-frame $(g-r)_0$ color and its redshift.  The rest-frame $(g-r)_0$ colors for the DESI galaxies are found using an iterative root-finding method (Brent's method as the default). Note that for all $k$-corrections, 

\begin{equation}
    k(z_{\rm ref}) = -2.5 {\rm log}_{10}(1+z_{\rm ref})
    \label{eq:kzref}
\end{equation}

\noindent is true. As such, the zeroth-power coefficient ($a_4$) enforces this condition for all individual $k$-correction polynomials, explaining why they are all the same value for each color bin.

\begin{table}%[]
\begin{center}
\caption{\label{tab:krcoeffs}{Coefficient table for the $r$-band $k$-correction polynomials.}}
\begin{tabular}{@{}llllllll@{}}
\hline
\hline
min  & max  & median & $a_0$    & $a_1$   & $a_2$   & $a_3$   & $a_4$   \\ \hline

-100 & 0.18 & 0.131  & -45.328 & 35.277 & -6.604 & -0.481 & -0.104 \\
0.18 & 0.35 & 0.298  & -20.077 & 20.145 & -4.620 & -0.048 & -0.104 \\
0.35 & 0.52 & 0.443  & -10.982 & 14.357 & -3.676 & 0.339  & -0.104 \\
0.52 & 0.69 & 0.603  & -3.428  & 9.478  & -2.703 & 0.765  & -0.104 \\
0.69 & 0.86 & 0.785  & 6.717   & 3.250  & -1.176 & 1.113  & -0.104 \\
0.86 & 1.03 & 0.933  & 16.761  & -2.514 & 0.351  & 1.307  & -0.104 \\
1.03 & 100  & 1.067  & 20.302 & -4.189  & 0.562  & 1.494  & -0.104 \\ \hline
\end{tabular}
\end{center}
\end{table}

Finally, the $e$-corrections we provide are determined using the same functional form as \cite{McNaught_Roberts_2014}: 
\begin{equation}
    E(z) = -Q_{0} (z-z_{\rm ref}).
    \label{eq:ecorr}
\end{equation}
Specifically, we assume that the density evolution ($P$) is zero and the luminosity evolution ($Q$) is the only factor to be considered. We provide results for the $r$-band, for which we use $Q_0 = 0.97$ following the result found empirically in \cite{McNaught_Roberts_2014}. One can subtract the resulting $E(z)$ from $M_r$ in order to obtain, \eg, BGS samples that have minimal evolution in number density with redshift.

\section{Data Access}\label{sec:access}

DESI data access is currently very file-oriented, reflecting the manner in
which most DESI collaborators access and use the data.
In addition to these files, value-added data services and products are under development including a database and a suite of tutorials. The latter are provided on a best-effort basis to the community for convenience. We describe these data access methods as well as the data license and acknowledgments below. Documentation of DESI data access is further maintained at \url{https://data.desi.lbl.gov/doc/access/}.

\subsection{File Access}

For researchers who are members of other collaborations with NERSC account access,
files are directly available at
\texttt{/global/cfs/cdirs/desi/public/edr/}, without requiring DESI membership.
The exact same directory structure can be inspected and individual files
downloaded via \url{https://data.desi.lbl.gov/public/edr} without requiring
a NERSC account.  Efficient bulk download of larger amounts of data is
available using the Globus\footnote{\url{https://globus.org}}
endpoint ``DESI Public Data'', also without requiring a NERSC account.
All 3 of these methods access the same files on disk at NERSC.
Previous descriptions of file and directory locations in this paper started
at the \texttt{public/edr/} level, regardless of whether that is prefixed by
direct file access, https, Globus, or possibly some future data access method.

The entire EDR is over 80 TB, so users are encouraged to be selective in
downloading only what they really need for an analysis.
We anticipate that most users will start with one of the redshift catalogs
in \texttt{vac/edr/lss/} or \texttt{spectro/redux/fuji/zcatalog/},
sub-select to the objects of interest, and then proceed to download only
the files containing spectra of those targets.

\subsection{Other Interfaces}

Catalog-level data (target photometry, fiber assignments, exposure metadata, spectral classifications, and redshifts,
but not the spectra themselves nor most of the value added catalogs)
are available in a Postgres database with tables and access credentials described at \url{https://data.desi.lbl.gov/doc/access/database}.
Users with NERSC accounts can directly query this database with SQL, or use \texttt{SQLAlchemy} wrapper objects provided with pre-installed
\texttt{specprodDB} Python code.
% For users without NERSC credentials, a public copy of the database is expected to be made available from  the Astro Data Lab as described at \url{https://datalab.noirlab.edu/desi}. This platform includes anonymous public access via a web query interface and authenticated access via a JupyterLab server. Additionally, the Astro Data Lab plans to serve a subset of DESI spectra via the SPectra Analysis and Retrievable Catalog Lab (\texttt{SPARCL}\footnote{\url{https://astrosparcl.datalab.noirlab.edu}}),
% which consists of a spectral database with a programmatic interface. The subset is limited to healpix-coadded spectra which have been combined across cameras. 
For users without NERSC credentials, a public copy of the database is available from the Astro Data Lab as described at \url{https://datalab.noirlab.edu/desi}. This platform includes anonymous public access via a web query interface and authenticated access via a JupyterLab server. Additionally, the Astro Data Lab serves a subset of DESI spectra via the SPectra Analysis and Retrievable Catalog Lab (\texttt{SPARCL}\footnote{\url{https://astrosparcl.datalab.noirlab.edu}}),
which consists of a spectral database with a programmatic interface. The subset is limited to healpix-coadded spectra which have been combined across cameras.
Other types of spectra and files are only available at the primary file-based archive at NERSC as described previously.

Other access methods may be provided in the future, \eg,
webpages for interactively browsing the data, or interfaces for
downloading individual spectra or custom collections of spectra.
If and when these become available, they will be documented at
\url{https://data.desi.lbl.gov/doc/access/}.

\subsection{Tutorials}

There is an internal effort within the DESI collaboration to design notebooks to help introduce
specific data products or ways of accessing specific types of data.
These are aggregated in a GitHub repository:
\url{https://github.com/desihub/tutorials}.
The tutorials are divided into thematic and topical sub-directories.
Of most relevance is the \texttt{getting_started} sub-directory, which
includes introductions to the types of files found in the EDR, and a notebook
on working with the \texttt{zcatalog} files. 

\subsection{Data License and Acknowledgments}

DESI Data are released under the Creative Commons Attribution 4.0
International License.\footnote{\url{https://creativecommons.org/licenses/by/4.0/}}
This allows users to share, copy, redistribute, adapt, transform, and build upon
the DESI data for any purpose, including commercially, as long as attribution
is given by citing this paper and including the acknowledgments text
listed at \url{https://data.desi.lbl.gov/doc/acknowledgments/}.

\section{Conclusion}\label{sec:conclusion}

This paper describes the release of the first science data taken with the DESI instrument.
This new sample consists of all commissioning and Survey Validation data taken between December 14, 2020 and June 10, 2021; with the majority of the data acquisition ending on May 13, 2021.
These observations were taken to validate the survey design and observing strategy for the DESI Main Survey that commenced on May 14, 2021.
The release includes deep spectra with robust visual inspection classification of 769 stars, 14,735 galaxies, and 3,121 quasars from extended selection algorithms that were tested during the Survey Validation period.
The release includes highly complete, good-quality, spectroscopic samples of 306,052 stars; 721,026 galaxies; and 44,151 quasars obtained over an area of roughly 170 deg$^2$ in the One-Percent Survey.
In total, after accounting for additional spectra from secondary programs and all other DESI surveys, this release includes good calibrated spectra and catalog information for 496,128 stars; 1,125,635 galaxies; and 90,241 quasars that were spectroscopically classified and free of any known hardware, observational, or redshift fitting issues.

These data were all reprocessed with the DESI data reduction pipeline and redshift classification algorithms \citep{guy22a,bailey23a}.
Updated versions are expected to make only minor changes to the quality of one-dimensional spectra and redshift classifications.

The early data from DESI can be accessed as described in \S\ref{sec:access}.
Plots and key numbers in this paper were produced from notebooks and code in \url{https://github.com/desihub/edrpaper} using
data from this Early Data Release.
The Digital Object Identifier (DOI) for EDR is
\href{https://doi.org/10.5281/zenodo.7964161}{\texttt{doi:10.5281/zenodo.7964161}},
which includes both the EDR dataset and the data for the plots in this paper.

The redshift range, depth, and variety of spectroscopic
targets make the data in this release an ideal sample for
studies of stellar astrophysics, galaxy and quasar astrophysics, and early studies of the clustering of matter. Representing the range of potential studies, the DESI
collaboration has already used these data to 
measure the 1D and 3D Ly$\alpha$ forest \citep{Karacayli2023,Perez2023,Ravoux2023}, 
model the connection between galaxies and halos \citep{Gao2023,Prada2023,Yu2023,Yuan2023}, 
derive the probabilistic stellar mass functions from hundreds of thousands of BGS galaxies \citep{Hahn2023},
and identify very metal-poor stars in the Milky Way \citep{Prieto2023},
% fit the Fundamental Plane \citep{Said2023} and Tully-Fisher \citep{Douglass2023} Relations, 
among other topics.

The first year of the full five-year DESI survey concluded on June 13, 2022 and the DESI collaboration has finalized the internal data reductions for that sample.
Those data will be publicly released after the completion and publication of the BAO and RSD measurements that motivated the construction of DESI.
Meanwhile, DESI continues successful and efficient operations, planning its next major data sample when a third year of observation is complete.
That three-year sample is expected to include more than 30 million spectroscopically-confirmed galaxies and quasars, with a full ~14,000 deg$^2$ footprint of the BGS and MWS programs.
DESI will make the first-year, the three-year, and eventually its final sample, publicly available following the release of the key cosmological analyses for each corresponding dataset.

\acknowledgments
{This material is based upon work supported by the U.S. Department of Energy (DOE), Office of Science, Office of High-Energy Physics, under Contract No. DE–AC02–05CH11231, and by the National Energy Research Scientific Computing Center, a DOE Office of Science User Facility under the same contract. Additional support for DESI was provided by the U.S. National Science Foundation (NSF), Division of Astronomical Sciences under Contract No. AST-0950945 to the NSF’s National Optical-Infrared Astronomy Research Laboratory; the Science and Technology Facilities Council of the United Kingdom; the Gordon and Betty Moore Foundation; the Heising-Simons Foundation; the French Alternative Energies and Atomic Energy Commission (CEA); the National Council of Science and Technology of Mexico (CONACYT); the Ministry of Science and Innovation of Spain (MICINN), and by the DESI Member Institutions: \url{https://www.desi.lbl.gov/collaborating-institutions}. Any opinions, findings, and conclusions or recommendations expressed in this material are those of the author(s) and do not necessarily reflect the views of the U.S.~National Science Foundation, the U.S.~Department of Energy, or any of the listed funding agencies.

The authors are honored to be permitted to conduct scientific research on Iolkam Du’ag (Kitt Peak), a mountain with particular significance to the Tohono O’odham Nation.
} 

\vspace{0.6in}

\textit{Facilities}:
Mayall (DESI)

\textit{Software}:
astropy~\citep{astropy1, astropy2, astropy3},
healpy~\citep{healpy},
desispec~\citep{guy22a},
desitarget~\citep{myers23a},
Redrock~\citep{bailey23a}.

%%% fitsio (https://github.com/esheldon/fitsio), healpy (Zonca et al. 2019), matplotlib (Caswell et al. 2021), numpy (Harris et al. 2020), photutils (Bradley et al. 2019), pyyaml (https://pyyaml.org/), scipy (Gommers et al. 2021).

% \setcounter{secnumdepth}{1}
% \setcounter{tocdepth}{1}
% \addtocontents{toc}{\protect\setcounter{tocdepth}{0}}% This turns off subsections

\clearpage

% \begin{appendix}
\appendix
\section{Primary Targets}\label{app:primarytargets}

This appendix includes tables for the primary targeting bits, some of which are replicated from
\cite{myers23a}, for convenience and described in \S\ref{sec:sv_target_samples}.

%-------------------
% SV1 targeting bits
%-------------------
\subsection{Target Selection Validation (SV1) Targeting Bitmasks}
\label{sec:sv1_target_bitmasks}

Target Selection Validation (\texttt{SURVEY=sv1}) bitmasks are recorded in fibermap columns
\texttt{SV1\_DESI\_TARGET}, \texttt{SV1\_BGS\_TARGET}, and \texttt{SV1\_MWS\_TARGET}.
Table~\ref{table:sv1dark} lists the \texttt{SV1\_DESI\_TARGET} bits for dark-time targets,
Table~\ref{table:sv1cal} lists the \texttt{SV1\_DESI\_TARGET} bits for general calibration targets such as standard stars and sky locations, Table~\ref{table:sv1bgs} lists the \texttt{SV1\_BGS\_TARGET} bits for the Bright Galaxy Survey, and
Table~\ref{table:sv1mws} lists the \texttt{SV1\_MWS\_TARGET} bits for Milky Way Survey targets.
These target selection bits are also defined programmatically in the open-source
\texttt{desitarget}\footnote{\url{https://github.com/desihub/desitarget}} software package.
A YAML-format file describing the bits is in subdirectory \texttt{py/desitarget/sv1/data/sv1_targetmask.yaml},
with convenience wrapper objects in the Python module \texttt{desitarget.sv1.sv1_targetmask.desi_mask}.
Examples of using these bitmasks with this code are given in \S2 of \cite{myers23a}.

{
\begin{deluxetable*}{ccc}[htb]
\tablecaption{Dark-time targeting bits for SV1.\label{table:sv1dark}}
\tablewidth{0pt}
\tablehead{
\colhead{Bit-name} & 
\colhead{Bit-value} &
\colhead{Description}
}
\startdata
{\tt LRG}                    & 0 & LRG \\
{\tt ELG}                    & 1 & ELG \\
{\tt QSO}                    & 2 & QSO \\
{\tt LRG\_OPT}               & 3 & LRG from baseline version of optical cuts \\
{\tt LRG\_IR}                & 4 & LRG from baseline version of IR cuts \\ 
{\tt LRG\_SV\_OPT}           & 5 & LRG from relaxed version of optical cuts \\
{\tt LRG\_SV\_IR}            & 6 & LRG from relaxed version of IR cuts \\        
{\tt LOWZ\_FILLER}           & 7 & LRG-like low-$z$ filler sample ({\em not} used for SV1) \\
{\tt ELG\_SV\_GTOT}          & 8 & ELG from relaxed version of FDR\tablenotemark{a} cuts and $g$-band magnitude limit \\
{\tt ELG\_SV\_GFIB}          & 9 & ELG from relaxed version of FDR\tablenotemark{a} cuts and $g$-band  fiber-magnitude limit \\
{\tt ELG\_FDR\_GTOT}        & 10 & ELG from FDR\tablenotemark{a} cuts with $g$-band magnitude limit \\
{\tt ELG\_FDR\_GFIB}        & 11 & ELG from FDR\tablenotemark{a} cuts with $g$-band fiber-magnitude limit \\
{\tt QSO\_COLOR\_4PASS}     & 12 & Low-$z$ (tracer) QSO using color cuts \\
{\tt QSO\_RF\_4PASS}        & 13 & Low-$z$ (tracer) QSO using random forest \\
{\tt QSO\_COLOR\_8PASS}     & 14 & High-$z$ (Lyman-$\alpha$) QSO using color cuts \\
{\tt QSO\_RF\_8PASS}        & 15 & High-$z$ (Lyman-$\alpha$) QSO using random forest \\
{\tt QSO\_HZ\_F}            & 16 & Faint, high-redshift QSO \\
{\tt QSO\_Z5}               & 17 & $z\sim5$ QSO \\
{\tt LRG\_OPT\_NORTH}       & 18 & LRG from baseline version of optical cuts tuned for Bok/Mosaic \\
{\tt LRG\_IR\_NORTH}        & 19 & LRG from baseline version of IR cuts tuned for Bok/Mosaic \\
{\tt LRG\_SV\_OPT\_NORTH}   & 20 & LRG from relaxed version of optical cuts tuned for Bok/Mosaic \\
{\tt LRG\_SV\_IR\_NORTH}    & 21 & LRG from relaxed version of IR cuts tuned for Bok/Mosaic \\
{\tt LOWZ\_FILLER\_NORTH}   & 22 & LRG-like low-$z$ filler sample tuned for Bok/Mosaic ({\em not} used for SV1) \\
{\tt LRG\_OPT\_SOUTH}       & 23 & LRG from baseline version of optical cuts tuned for DECam \\
{\tt LRG\_IR\_SOUTH}        & 24 & LRG from baseline version of IR cuts tuned for DECam \\
{\tt LRG\_SV\_OPT\_SOUTH}   & 25 & LRG from relaxed version of optical cuts tuned for DECam \\
{\tt LRG\_SV\_IR\_SOUTH}    & 26 & LRG from relaxed version of IR cuts tuned for DECam \\
{\tt LOWZ\_FILLER\_SOUTH}   & 27 & LRG-like low-$z$ filler sample tuned for DECam ({\em not} used for SV1) \\
{\tt ELG\_SV\_GTOT\_NORTH}  & 28 & ELG from relaxed version of FDR\tablenotemark{a} cuts and $g$-band limit for Bok/Mosaic \\ 
{\tt ELG\_SV\_GFIB\_NORTH}  & 29 & As for {\tt ELG\_SV\_GTOT\_NORTH} but using a fiber-magnitude limit in $g$-band \\ 
{\tt ELG\_FDR\_GTOT\_NORTH} & 30 & ELG from FDR\tablenotemark{a} cuts with $g$-band magnitude limit for Bok/Mosaic \\ 
{\tt ELG\_FDR\_GFIB\_NORTH} & 31 & ELG from FDR\tablenotemark{a} cuts with $g$-band fiber-magnitude limit for Bok/Mosaic \\
{\tt ELG\_SV\_GTOT\_SOUTH}  & 38 & ELG from relaxed version of FDR\tablenotemark{a} cuts and $g$-band magnitude limit for DECam \\
{\tt ELG\_SV\_GFIB\_SOUTH}  & 39 & As for {\tt ELG\_SV\_GTOT\_SOUTH} but using a fiber-magnitude limit in $g$-band  \\
{\tt ELG\_FDR\_GTOT\_SOUTH} & 40 & ELG from FDR\tablenotemark{a} cuts with $g$-band magnitude limit for DECam \\
{\tt ELG\_FDR\_GFIB\_SOUTH} & 41 & ELG from FDR\tablenotemark{a} cuts with $g$-band fiber-magnitude limit for DECam \\
{\tt QSO\_COLOR\_4PASS\_NORTH} & 42 & Low-$z$ (tracer) QSO using color cuts tuned for Bok/Mosaic \\
{\tt QSO\_RF\_4PASS\_NORTH}    & 43 & Low-$z$ (tracer) QSO using random forest tuned for Bok/Mosaic \\
{\tt QSO\_COLOR\_8PASS\_NORTH} & 44 & High-$z$ (Lyman-$\alpha$) QSO using color cuts tuned for Bok/Mosaic \\
{\tt QSO\_RF\_8PASS\_NORTH}    & 45 & High-$z$ (Lyman-$\alpha$) QSO using random forest tuned for Bok/Mosaic \\
{\tt QSO\_HZ\_F\_NORTH}        & 46 & QSO at high-redshift and faint tuned for Bok/Mosaic \\ 
{\tt QSO\_Z5\_NORTH}           & 47 & $z\sim5$ QSO tuned for Bok/Mosaic \\
{\tt QSO\_COLOR\_4PASS\_SOUTH} & 48 & Low-$z$ (tracer) QSO using color cuts tuned for DECam \\ 
{\tt QSO\_RF\_4PASS\_SOUTH}    & 53 & Low-$z$ (tracer) QSO using random forest tuned for DECam \\ 
{\tt QSO\_COLOR\_8PASS\_SOUTH} & 54 & High-$z$ (Lyman-Alpha) QSO using color cuts tuned for DECam \\
{\tt QSO\_RF\_8PASS\_SOUTH}    & 55 & High-$z$ (Lyman-Alpha) QSO using random forest tuned for DECam \\
{\tt QSO\_HZ\_F\_SOUTH}        & 56 & Faint, high-redshift QSO tuned for DECam \\
{\tt QSO\_Z5\_SOUTH}           & 57 & $z\sim5$ QSO tuned for DECam \\
\enddata
\begin{center}
\tablenotetext{}{Bits are stored in the {\tt sv1\_desi\_mask} and accessed via the {\tt SV1\_DESI\_TARGET} column \citep[see][for more details]{myers23a}.}
\tablenotetext{a}{``FDR'' refers to the DESI Final Design Report \citep[see][]{desi16a}.}
\end{center}
\end{deluxetable*}

\begin{deluxetable*}{ccc}[htb]
\tablecaption{SV1 bits for calibration, object-avoidance, and to indicate non-dark-time programs\label{table:sv1cal}}
\tablewidth{0pt}
\tablehead{
\colhead{Bit-name} & 
\colhead{Bit-value} &
\colhead{Description}
}
\startdata
{\tt SKY}                   & 32 &   Blank sky locations \\
{\tt STD\_FAINT}            & 33 &   Standard stars for dark/gray conditions \\
{\tt STD\_WD}               & 34 &   White Dwarf standard stars \\
{\tt STD\_BRIGHT}           & 35 &   Standard stars for bright conditions \\
{\tt BAD\_SKY}              & 36 &   Blank sky locations that are imperfect but still useable \\
{\tt SUPP\_SKY}             & 37 &   SKY is based on \textit{Gaia}-avoidance ({\tt SKY} will be set, too) \\
{\tt NO\_TARGET}               & 49 & No known target at this location \\
{\tt BRIGHT\_OBJECT}           & 50 & Known bright object to avoid \\
{\tt IN\_BRIGHT\_OBJECT}       & 51 & Too near a bright object; {\em do not observe} \\
{\tt NEAR\_BRIGHT\_OBJECT}     & 52 & Near a bright object but ok to observe \\
{\tt BGS\_ANY}       & 60 & Any BGS bit is set (see Table~\ref{table:sv1bgs}) \\
{\tt MWS\_ANY}       & 61 & Any MWS bit is set (see Table~\ref{table:sv1mws}) \\
{\tt SCND\_ANY}      & 62 & Any secondary bit is set (see Table~\ref{table:sv1sec})  \\
\enddata
\begin{center}
\tablenotetext{}{Bits are stored in the {\tt sv1\_desi\_mask} and accessed via the {\tt SV1\_DESI\_TARGET} column \citep[see][for more details]{myers23a}. Additional standard star targets based purely on \textit{Gaia} are included in Table~\ref{table:sv1mws}.}
\end{center}
\end{deluxetable*}

\begin{deluxetable*}{ccc}[htb]
\tablecaption{Bright Galaxy Survey (BGS) targeting bits for SV1\label{table:sv1bgs}}
\tablewidth{0pt}
\tablehead{
\colhead{Bit-name} & 
\colhead{Bit-value} &
\colhead{Description}
}
\startdata
{\tt BGS\_FAINT}              & 0  & BGS faint targets                              \\ 
{\tt BGS\_BRIGHT}             & 1  & BGS bright targets                             \\
{\tt BGS\_FAINT\_EXT}         & 2  & BGS faint extended targets                     \\
{\tt BGS\_LOWQ}               & 3  & BGS low-quality targets                        \\
{\tt BGS\_FIBMAG}             & 4  & BGS fiber-magnitude targets                    \\
{\tt BGS\_FAINT\_NORTH}       & 8  & BGS faint cuts tuned for Bok/Mosaic            \\
{\tt BGS\_BRIGHT\_NORTH}      & 9  & BGS bright cuts tuned for Bok/Mosaic           \\
{\tt BGS\_FAINT\_EXT\_NORTH}  & 10 & BGS faint extended cuts tuned for Bok/Mosaic   \\
{\tt BGS\_LOWQ\_NORTH}        & 11 & BGS low-quality cuts tuned for Bok/Mosaic      \\
{\tt BGS\_FIBMAG\_NORTH}      & 12 & BGS low-quality cuts tuned for Bok/Mosaic      \\
{\tt BGS\_FAINT\_SOUTH}       & 16 & BGS faint cuts tuned for DECam                 \\
{\tt BGS\_BRIGHT\_SOUTH}      & 17 & BGS bright cuts tuned for DECam                \\
{\tt BGS\_FAINT\_EXT\_SOUTH}  & 18 & BGS faint extended cuts tuned for DECam        \\
{\tt BGS\_LOWQ\_SOUTH}        & 19 & BGS low-quality cuts tuned for DECam           \\
{\tt BGS\_FIBMAG\_SOUTH}      & 20 & BGS fiber-magnitude cuts tuned for DECam       \\
{\tt BGS\_KNOWN\_ANY}         & 40 & Known target from another survey               \\
{\tt BGS\_KNOWN\_COLLIDED}    & 41 & BGS known SDSS/BOSS fiber-collided             \\
{\tt BGS\_KNOWN\_SDSS}        & 42 & BGS known SDSS targets                         \\
{\tt BGS\_KNOWN\_BOSS}        & 43 & BGS known BOSS targets                         \\
\enddata
\begin{center}
\tablenotetext{}{Bits are stored in the {\tt sv1\_bgs\_mask} and accessed via the {\tt SV1\_BGS\_TARGET} column \citep[see][for more details]{myers23a}.}
\end{center}
\end{deluxetable*}

\begin{deluxetable*}{ccc}[htb]
\tablecaption{Milky Way Survey (MWS) targeting bits for SV1\label{table:sv1mws}}
\tablewidth{0pt}
\tablehead{
\colhead{Bit-name} & 
\colhead{Bit-value} &
\colhead{Description}
}
\startdata
{\tt MWS\_MAIN\_BROAD}        & 0  & MWS magnitude-limited bulk sample \\ 
{\tt MWS\_WD}                 & 1  & MWS White Dwarf target            \\
{\tt MWS\_NEARBY}             & 2  & MWS volume-complete ($\sim$100\,pc) sample \\
{\tt MWS\_MAIN\_BROAD\_NORTH} & 4  & MWS targets from Bok/Mosaic       \\
{\tt MWS\_MAIN\_BROAD\_SOUTH} & 5  & MWS targets from DECam            \\
{\tt MWS\_BHB}                & 6  & MWS Blue Horizontal Branch targets                          \\
{\tt MWS\_MAIN\_FAINT}        & 14 & MWS magnitude-limited sample                   \\
{\tt MWS\_MAIN\_FAINT\_NORTH} & 15 & MWS magnitude-limited sample from Bok/Mosaic   \\
{\tt MWS\_MAIN\_FAINT\_SOUTH} & 16 & MWS magnitude-limited sample from DECam        \\
{\tt GAIA\_STD\_FAINT}        & 33 & {\it Gaia}-based standard stars for dark/gray conditions   \\
{\tt GAIA\_STD\_WD}           & 34 & {\it Gaia}-based White Dwarf stars for use as standards              \\
{\tt GAIA\_STD\_BRIGHT}       & 35 & {\it Gaia}-based standard stars for bright conditions                \\
{\tt BACKUP\_BRIGHT}          & 60 & Bright backup {\it Gaia} targets                     \\
{\tt BACKUP\_FAINT}           & 61 & Fainter backup {\it Gaia} targets                    \\
{\tt BACKUP\_VERY\_FAINT}     & 62 & Even fainter backup {\it Gaia} targets               \\
\enddata
\begin{center}
\tablenotetext{}{Bits are stored in the {\tt sv1\_mws\_mask} and accessed via the {\tt SV1\_MWS\_TARGET} column \citep[see][for more details]{myers23a}. \revisionforreviewer{\textit{Gaia} DR2 was used for target selection of \textit{Gaia} targets during Survey Validation \citep{GaiaDR2}.}}
\end{center}
\end{deluxetable*}

%-------------------
% SV2 targeting bits
%-------------------

\subsection{Operations Development (SV2) Targeting Bitmasks}
\label{sec:sv2_target_bitmasks}

The relevant targeting bits for Operations Development (SV2) are outlined in Table~\ref{table:sv2bits}. Note that some targeting bits were retained moving from Target Selection Validation to SV2 and, therefore, only {\em new} bits and {\em changes} to existing bits are included in Table~\ref{table:sv2bits}. The bit-mask used to track SV2 sub-programs is called {\tt sv2\_desi\_mask}, and the bit-values for each SV2 target can be accessed via the {\tt SV2\_DESI\_TARGET}, {\tt SV2\_BGS\_TARGET}, and {\tt SV2\_MWS\_TARGET}
column in data files \citep[again, for more details, see \S2.4 of][]{myers23a}.

\begin{deluxetable*}{ccc}[htb]
\tablecaption{Additional and updated targeting bits for Operations Developement (SV2).\label{table:sv2bits}}
\tablewidth{0pt}
\tablehead{
\colhead{Bit-name} & 
\colhead{Bit-value} &
\colhead{Description}
}
\startdata
\multicolumn{3}{l}{{\em Dark-time targeting bits} in \texttt{SV2\_DESI\_TARGET}} \\
{\tt QSO\_HIZ}    &    4 & QSO selected using high-redshift Random Forest \\ 
{\tt LRG\_NORTH}  &    8 & LRG cuts tuned for Bok/Mosaic imaging data     \\
{\tt ELG\_NORTH}  &    9 & ELG cuts tuned for Bok/Mosaic imaging data     \\
{\tt QSO\_NORTH}  &   10 & QSO cuts tuned for Bok/Mosaic imaging data     \\
{\tt LRG\_SOUTH}  &   16 & LRG cuts tuned for DECam imaging data          \\
{\tt ELG\_SOUTH}  &   17 & ELG cuts tuned for DECam imaging data          \\
{\tt QSO\_SOUTH}  &   18 & QSO cuts tuned for DECam imaging data          \\
\hline
\multicolumn{3}{l}{{\em BGS targeting bits} in \texttt{SV2\_BGS\_TARGET}} \\
{\tt BGS\_WISE}       & 2 & BGS targets selected using {\em WISE} imaging   \\        {\tt BGS\_FAINT\_HIP} & 3 & BGS faint targets prioritized like {\tt BGS\_BRIGHT} targets \\ 
{\tt BGS\_WISE\_NORTH} & 10 & {\tt BGS\_WISE} cuts tuned for Bok/Mosaic imaging \\ 
{\tt BGS\_WISE\_SOUTH} & 18 & {\tt BGS\_WISE} cuts tuned for DECam imaging      \\
\hline
\multicolumn{3}{l}{{\em MWS targeting bits} in \texttt{SV2\_MWS\_TARGET}} \\
{\tt MWS\_BROAD}             & 0 & MWS magnitude-limited bulk sample      \\
{\tt MWS\_BROAD\_NORTH}      & 4 & MWS cuts tuned for Bok/Mosaic imaging  \\
{\tt MWS\_BROAD\_SOUTH}      & 5 & MWS cuts tuned for DECam imaging       \\
{\tt MWS\_MAIN\_BLUE}        & 8 & MWS magnitude-limited blue sample \\
{\tt MWS\_MAIN\_BLUE\_NORTH} & 9 & MWS magnitude-limited blue sample tuned for Bok/Mosaic imaging \\
{\tt MWS\_MAIN\_BLUE\_SOUTH} & 10 & MWS magnitude-limited blue sample tuned for DECam imaging \\
{\tt MWS\_MAIN\_RED}         & 11 & MWS magnitude-limited red sample \\
{\tt MWS\_MAIN\_RED\_NORTH}  & 12 & MWS magnitude-limited red sample tuned for Bok/Mosaic imaging \\
{\tt MWS\_MAIN\_RED\_SOUTH}  & 13 & MWS magnitude-limited red sample tuned for DECam imaging \\
\enddata
\begin{center}
\tablenotetext{}{Bits are stored in the {\tt sv2\_desi\_mask} and accessed via the {\tt SV2\_DESI\_TARGET} column \citep[see][for more details]{myers23a}. Many bits from Tables~\ref{table:sv1dark}, \ref{table:sv1cal}, \ref{table:sv1bgs} and \ref{table:sv1mws} were reused for SV2, and only new or different bits are included in this table. For example, calibration bits listed in Table~\ref{table:sv1cal} remained the same moving from SV1 to SV2. Where the name and description of a bit-value has changed in this table compared to Tables~\ref{table:sv1dark}, \ref{table:sv1cal}, \ref{table:sv1bgs} or \ref{table:sv1mws}, the bit was deprecated and updated for SV2.}
\end{center}
\end{deluxetable*}

%-------------------
% SV3 targeting bits
%-------------------

\subsection{One-Percent Survey (SV3) Targeting Bitmasks}
\label{sec:sv3_target_bitmasks}

The relevant target selection bitmasks for the One-Percent Survey (SV3) are outlined in Table~\ref{table:sv3bits}. Many targeting bits were retained moving from earlier phases of SV to the One-Percent Survey and, therefore, only {\em new} bits and {\em changes} to existing bits (when compared to Tables~\ref{table:sv1dark}, \ref{table:sv1cal}, \ref{table:sv1bgs}, \ref{table:sv1mws}, and \ref{table:sv2bits}) are included in Table~\ref{table:sv3bits}. The bit-mask used to track sub-programs for the One-Percent Survey is called {\tt sv3\_desi\_mask}, and the bit-values for each target can be accessed via the {\tt SV3\_DESI\_TARGET}, {\tt SV3\_BGS\_TARGET}, and
{\tt SV3\_MWS\_TARGET} columns in data files \citep[again, for more details, see \S2.4 of][]{myers23a}. 

\begin{deluxetable*}{ccc}[htb]
\tablecaption{Additional and updated targeting bits for SV3.\label{table:sv3bits}}
\tablewidth{0pt}
\tablehead{
\colhead{Bit-name} & 
\colhead{Bit-value} &
\colhead{Description}
}
\startdata
\sidehead{\em Dark-time targeting bits}
{\tt LRG\_LOWDENS} & 3 & LRG cuts to produce a lower-than-nominal ($\sim600\,{\rm deg}^{-2}$) target density \\
{\tt ELG\_LOP}     & 5 & ELG scheduled for observations at lower priority  \\
{\tt ELG\_HIP}     & 6 & ELG scheduled for observations at higher priority \\
{\tt ELG\_LOP\_NORTH}     & 11 & ELG at lower priority tuned for Bok/Mosaic imaging \\
{\tt ELG\_HIP\_NORTH}     & 12 & ELG at higher priority tuned for Bok/Mosaic imaging \\
{\tt LRG\_LOWDENS\_NORTH} & 13 & Lower-density LRG cuts tuned for Bok/Mosaic imaging \\
{\tt ELG\_LOP\_SOUTH}     & 19 & ELG at lower priority tuned for DECam imaging \\
{\tt ELG\_HIP\_SOUTH}     & 20 & ELG at higher priority tuned for DECam imaging \\
{\tt LRG\_LOWDENS\_SOUTH} & 21 & Lower-density LRG cuts tuned for DECam imaging \\
\enddata
\begin{center}
\tablenotetext{}{Bits are stored in the {\tt sv3\_desi\_mask} and accessed via the {\tt SV3\_DESI\_TARGET} column \citep[see][for more details]{myers23a}. Many bits from Tables~\ref{table:sv1dark}, \ref{table:sv1cal}, \ref{table:sv1bgs}, \ref{table:sv1mws} and \ref{table:sv2bits} were reused for SV3, and only new or different bits are included in this table. For example, calibration, BGS and MWS bits were not altered moving from SV2 to SV3. Where the name and description of a bit-value has changed in this table compared to Tables~\ref{table:sv1dark}, \ref{table:sv1cal}, \ref{table:sv1bgs}, \ref{table:sv1mws} or \ref{table:sv2bits}, the bit was deprecated and updated for SV3.}
\end{center}
\end{deluxetable*}

}

\section{Secondary Targets}\label{app:secondarytargets}

%---------------------------------------------------------------------------------------
In addition to its primary science goals, the DESI survey incorporates a range of ``secondary'' targets to pursue bespoke research. In this appendix, we describe the secondary target campaigns included in the EDR and outline how the bit-values in their {\tt scnd\_mask} and {\tt SCND\_TARGET} column \citep[see \S2.4 of][]{myers23a} can be linked back to the relevant program. 

DESI pursued secondary targets during both its ``Target Selection Validation'' and ``One-Percent Survey'' phases --- which we refer to here as ``SV1'' and ``SV3,'' respectively, for consistency with how the bit-masks are named. In Table~\ref{table:sv1sec} we list the bit-names and bit-values for secondary targets that were scheduled during SV1. In Table~\ref{table:sv3sec} we indicate how these bits {\em changed} for SV3.\footnote{Many bit-values were also deprecated moving from SV1 to SV3. So, some bits were set for {\em zero} targets for SV3} The target classes listed in Table~\ref{table:sv1sec} were either assigned to fill ``spare'' fibers on regular SV1 tiles, or were assigned to their own ``dedicated'' campaign on custom tiles listed in Table~\ref{tab:secondary_tiles}. Targets that were intended for dedicated observations are marked with a * in 
Table~\ref{table:sv1sec}. 
\revisionforreviewer{Almost all of the programs described in this Appendix continue to be observed as part of the DESI Main Survey, although a few programs were changed to optimize targeting and their bit-names were updated. The exceptions are: \texttt{MWS_DDOGIANTS} and \texttt{VETO}, which were never actually observed; \texttt{GW190412} and \texttt{IC134191}, which were associated with specific fields of transients during SV; \texttt{MWS_CALIB} and \texttt{BACKUP_CALIB}, which were used specifically for calibrating targets for the SV1 stellar survey described in \cite{cooper22a}; and dedicated programs marked with a * in 
Table~\ref{table:sv1sec}.}

In the rest of this appendix, we outline each secondary target class, moving through the bit-names in Tables~\ref{table:sv1sec} and \ref{table:sv3sec}. Further details of the selection of each type of secondary target are available at the associated {\tt docs} link for SV1\footnote{\url{https://data.desi.lbl.gov/public/ets/target/secondary/sv1/}} or SV3.\footnote{\url{https://data.desi.lbl.gov/public/ets/target/secondary/sv3/}} During SV1, secondary targets were permitted multiple observations. But, during SV3, most secondary targets were limited to a single observation, unless otherwise detailed below.

\begin{deluxetable*}{cccccc}[t]
\tablecaption{Secondary targeting bits for SV1}\label{table:sv1sec}
\tablewidth{0pt}
\tablehead{
\colhead{Bit-name} & 
\colhead{Bit-value} &
\colhead{Bit-name} & 
\colhead{Bit-value} &
\colhead{Bit-name} & 
\colhead{Bit-value}
}
\startdata
{\tt VETO} &                                   0    &  {\tt DC3R2\_GAMA} &                    20  &  {\tt BRIGHT\_HPM} &                                40  \\
{\tt UDG} &                                    1    &  {\tt UNWISE\_BLUE}\tablenotemark{*} &  21  &  {\tt WD\_BINARIES\_BRIGHT} &                       41  \\
{\tt FIRST\_MALS} &                            2    &  {\tt UNWISE\_GREEN}\tablenotemark{*} & 22  &  {\tt WD\_BINARIES\_DARK} &                         42  \\
{\tt WD\_BINARIES} &                           3    &  {\tt HETDEX\_MAIN}\tablenotemark{*} &  23  &  {\tt DESILBG}\tablenotemark{*}  &                  43  \\
{\tt LBG\_TOMOG}\tablenotemark{*} &            4    &  {\tt HETDEX\_HP}\tablenotemark{*} &    24  &  {\tt LBG\_TOMOG\_XMM}\tablenotemark{*} &           44  \\
{\tt QSO\_RED} &                               5    &  {\tt PSF\_OUT\_BRIGHT} &               25  &  {\tt LBG\_TOMOG\_COSMOS}\tablenotemark{*} &        45  \\
{\tt M31\_KNOWN}\tablenotemark{*} &            6    &  {\tt PSF\_OUT\_DARK} &                 26  &  {\tt LBG\_TOMOG\_W3}\tablenotemark{*} &            46  \\
{\tt M31\_QSO}\tablenotemark{*} &              7    &  {\tt HPM\_SOUM} &                      27  &  {\tt UNWISE\_GREEN\_II\_3700}\tablenotemark{*} &   47  \\
{\tt M31\_STAR}\tablenotemark{*} &             8    &  {\tt SN\_HOSTS} &                      28  &  {\tt UNWISE\_GREEN\_II\_3800}\tablenotemark{*} &   48  \\
{\tt MWS\_DDOGIANTS} &                         9    &  {\tt GAL\_CLUS\_BCG} &                 29  &  {\tt UNWISE\_GREEN\_II\_3900}\tablenotemark{*} &   49  \\
{\tt MWS\_CLUS\_GAL\_DEEP}\tablenotemark{*} & 10    &  {\tt GAL\_CLUS\_2ND} &                 30  &  {\tt UNWISE\_GREEN\_II\_4000}\tablenotemark{*} &   50  \\
{\tt LOW\_MASS\_AGN} &                        11    &  {\tt GAL\_CLUS\_SAT} &                 31  &  {\tt UNWISE\_BLUE\_FAINT\_II}\tablenotemark{*} &   51  \\
{\tt FAINT\_HPM} &                            12    &  {\tt HSC\_HIZ\_SNE}\tablenotemark{*} & 32  &  {\tt UNWISE\_BLUE\_BRIGHT\_II}\tablenotemark{*} &  52  \\
{\tt GW190412} &                              13    &  {\tt ISM\_CGM\_QGP}\tablenotemark{*} & 33  &  {\tt DESILBG\_TMG\_FINAL}\tablenotemark{*} &       53  \\
{\tt IC134191} &                              14    &  {\tt STRONG\_LENS} &                   34  &  {\tt DESILBG\_G\_FINAL}\tablenotemark{*} &         54  \\
{\tt PV\_BRIGHT} &                            15    &  {\tt WISE\_VAR\_QSO} &                 35  &  {\tt DESILBG\_BXU\_FINAL}\tablenotemark{*} &       55  \\
{\tt PV\_DARK} &                              16    &  {\tt MWS\_CALIB} &                     36  &  {\tt LBG\_TOMOG\_COSMOS\_FINAL}\tablenotemark{*} & 56  \\
{\tt LOW\_Z} &                                17    &  {\tt BACKUP\_CALIB} &                  37  &  {\tt BRIGHT\_TOO} &                                60  \\
{\tt BHB} &                                   18    &  {\tt MWS\_MAIN\_CLUSTER\_SV} &         38  &  {\tt DARK\_TOO} &                                  61  \\
{\tt SPCV} &                                  19    &  {\tt MWS\_RRLYR} &                     39  &                 &                                       \\
\enddata
\begin{center}
\tablenotetext{}{These bits are taken from the {\tt desitarget} code\footnote{See \url{https://github.com/desihub/desitarget/blob/2.5.0/py/desitarget/sv1/data/sv1_targetmask.yaml\#L155-L226}.} and are described in the body of the Appendix.}
\tablenotetext{*}{A dedicated target class intended to be observed on custom tiles.}
\end{center}
\end{deluxetable*}

\subsection{{\tt VETO}}

This targeting bit was reserved to designate targets as unnecessary or problematic. In practice, {\tt VETO} was never used, and, for later DESI programs, flagging targets as bad is done in the MTL ledgers described in \citet{schlafly23a} instead.

% ADM Dennis Zaritsky and Arjun Dey.  CHECKED by DZ
\subsection{{\tt UDG}}

This program was designed to obtain redshifts for, and hence distances to, a sample drawn from the few thousand known ultra-diffuse galaxies (UDGs) across the entire DESI footprint \citep[$\sim0.5\,{\rm deg}^{-2}$;][]{smudges1,smudges3}. The targeted UDGs were field galaxies selected in sparse environments using imaging from the Legacy Survey. Such galaxies are expected to be among the least efficient large galaxies hitherto known. Their distances are essential for understanding their environments, and inferring their sizes and luminosities. Because the bluest UDGs in the field are expected to have emission lines that DESI could detect \citep[see, e.g.][]{udg}, the sample was limited to $g -r < 0.3$.

% ADM R. Srianand, P. Petitjean, N. Gupta (IUCAA, India) and C. Yèche
\subsection{{\tt FIRST\_MALS}}

This project followed up sources from the MeerKAT Absorption Line Survey \citep[MALS, see, e.g.][]{firstmals}, an L- and UHF-band survey targeting $\sim150{,}000$ radio-loud AGN. In the DESI footprint $\sim 60\%$ ($\sim6\,{\rm deg}^{-2}$) of these AGN are expected to have optical counterparts to $r < 23$. The project sought to obtain optical spectra for sources that have 21-cm and OH absorption spectra from MALS. The main goals were to help to characterize the redshift distribution of MALS AGN and intervening and associated absorption systems, to quantify biases due to dust in optically selected AGN samples, and to facilitate photometric redshift estimates for larger radio-selected samples.

% ADM Boris Gansicke, Arjun Dey, Jay Farihi, Joan Najita, MWS Group
\subsection{{\tt WD\_BINARIES}}

This targeting bit was only briefly used for observations associated with version {\tt 0.48.0} of the {\tt desitarget} code before being replaced by {\tt WD\_BINARIES\_BRIGHT} and {\tt WD\_BINARIES\_DARK} (see \S\ref{sec:wdbins} for a more detailed description).

% ADM Corentin Ravoux, Christophe Yeche, Nathalie Palanque-Delabrouille, Eric Armengaud, Alex Krolewski, Michael J. Wilson, David Schlegel, Martin White, Kyle Dawson, Arjun Dey, Simone Ferraro, Zheng Cai
\subsection{{\tt LBG\_TOMOG}}
\label{sec:lbgtomogearly}
This campaign requested $\sim$10 hours of dedicated dark time on a single DESI tile in the COSMOS or XMM-LSS fields of the CFHT Large Area $U$-band Deep Survey \citep[CLAUDS;][]{clauds} to target $\sim$4{,}500 Lyman Break Galaxies (LBGs) and quasars in the redshift range $2 < z < 3.5$. A major goal was to map the Lyman-$\alpha$ Forest in detail by creating a 3D tomographic map of neutral hydrogen absorption \citep[see, e.g.,][]{lbgtomog}. Additional goals included finding high-redshift protoclusters and voids. Quasars were targeted to $r~\leqsim\,23.5$ using the standard DESI targeting approach \citep{chaussidon22a} and LBGs were selected to $r~\leqsim\,24.5$ using a $U$-dropout method applied to the CLAUDS imaging catalogs. The {\tt LBG\_TOMOG} bit was used in files associated with versions {\tt 0.48.0}, {\tt 0.49.0}, {\tt 0.50.0} and {\tt 0.51.0} of the {\tt desitarget} code but was gradually deprecated by the {\tt LBG\_TOMOG\_XMM}, {\tt LBG\_TOMOG\_COSMOS}, {\tt LBG\_TOMOG\_W3} and {\tt LBG\_TOMOG\_COSMOS\_FINAL} bits (see \S\ref{sec:lbgtomog}).

% ADM V. A. Fawcett, D. M. Alexander & D. J. Rosario CHECKED
\subsection{{\tt QSO\_RED}}

The {\tt QSO\_RED} sample targeted mildly dusty quasars that are too red to meet standard DESI criteria. The project sought to ascertain whether obscuration in quasars is explained by viewing the broad-line region through a dusty torus at a grazing angle or by an early, dusty phase in the lifetime of quasars \citep[see, e.g.,][]{qsored}. The {\tt QSO\_RED} targets were selected from point-sources in Legacy Surveys imaging using the {\em WISE} $W1W2W3$ color wedge of \citet{mateos12}. Targets consistent with the ``bluer'' colors of existing DESI quasar targets \citep{chaussidon22a} were removed, producing a sample of $\sim$41{,}000 {\tt QSO\_RED} targets over the DESI footprint ($\sim3\,{\rm deg}^{-2}$). As with all quasar-like target classes throughout the DESI survey \citep[see \S5 of][]{schlafly23a}, {\tt QSO\_RED} sources were scheduled for 4 total observations (starting with SV3).

% ADM Arjun Dey, David Aguado, Carlos Allende Prieto, Andrew Cooper, Katia Cunha, Raja Guhathakurta, Sergey Koposov, Joan Najita, et al.
\subsection{{\tt M31\_KNOWN}, {\tt M31\_QSO}, {\tt M31\_STAR}}

The DESI Andromeda Region Kinematic (``DARK'') survey comprised three complementary programs aimed at studying the dynamics of our nearest large neighbor galaxy. The {\tt M31\_KNOWN} bit covered previously identified, bright targets --- such as objects from the SPLASH survey \citep{splash}, globular clusters, HII regions, planetary nebulae and variable sources. The {\tt M31\_QSO} bit flagged quasar targets behind M31 selected using {\em Gaia} and {\em WISE}. The {\tt M31\_STAR} bit indicated sources selected from a combination of PAndAS \citep{pandas}, {\em Gaia} and {\em WISE}. The dedicated observations of, and scientific results from, the DARK survey are detailed in \citet{dark}.

\subsection{{\tt MWS\_DDOGIANTS}}

The {\tt MWS\_DDOGIANTS} targeting bit was ultimately never used by DESI.

% ADM Ting Li, Carlos Allende Prieto, Andrew Cooper, Sergey Koposov, David Aguado, Katia Cunha, Boris Gansicke, Monica Valluri, Miguel A. Sanchez-Conde, Arjun Dey, Yuan-Sen Ting, Risa Wechsler, Joan Najita
\subsection{{\tt MWS\_CLUS\_GAL\_DEEP}}

This campaign requested dedicated dark-time tiles to obtain spectra of open clusters, globular clusters (GCs), and dwarf spheroidal galaxies in the outskirts of our Galaxy. A major goal was to combine radial velocities with {\em Gaia} astrometry for faint ($19 < r < 21$) stars to constrain cluster membership and measure chemical abundances. Ultimately, the program aimed to characterize the initial mass function for clusters, the kinematics of stellar streams associated with GCs and density profiles for dwarf spheroidals. Targets were selected using Legacy Surveys imaging and {\em Gaia} astrometry at a density of a few thousand per DESI tile.

% ADM Ragadeepika Pucha, Stéphanie Juneau, Mar Mezcua, Arjun Dey
% CHECKED by Raga Pucha
\subsection{{\tt LOW\_MASS\_AGN}}

The {\tt LOW\_MASS\_AGN} program targeted faint ($r > 20$), low-redshift AGN in dwarf galaxies, selected using optical and infrared photometry from the eighth data release of the Legacy Surveys (LS DR8). The targets were pre-selected to be at low redshift ($0.02~\leq\,z_{\rm phot}~\leq\,0.3$) based on photometric redshifts from \citet{zhou21}, and to have faint $z$-band absolute magnitude ($M_z~\geqsim\,21$), which was adopted as a proxy for stellar mass. Candidates were then targeted as AGN on the basis of their $z - W1$, $W1 - W2$, and $W2 - W3$ colors, resulting in a sample with a density of $\sim20\,{\rm deg}^{2}$. The main scientific goal of the {\tt LOW\_MASS\_AGN} program was to identify $\sim$100 AGN driven by intermediate mass ($\leqsim\,10^{6}\,M_{\odot}$) black holes \citep[see, e.g.][]{mezcua20}. The program also aimed to test the validity of a new AGN selection criterion similar to that from \citet{hviding2022}, and extend it to low-mass galaxies for which the application of infrared AGN diagnostics is debated \citep[][]{hainline2016, satyapal2018}.

% ADM Aaron Meisner, Arjun Dey, David Nidever, Joan Najita, Monica Valluri, and Boris Gansicke; CHECKED BY MEISNER
\subsection{{\tt FAINT\_HPM}, {\tt BRIGHT\_HPM}}
\label{sec:hpm}
The {\tt HPM} project sought to measure radial velocities and spectroscopic types for $\sim$1{,}000 faint, high-proper-motion stars drawn from {\em Gaia} and the NOIRLab Source Catalog \citep[NSC;][]{nsc, nsc_dr2}. Targets were selected based on high proper motion ($\mu > 100$ mas~yr$^{-1}$), supplemented with reliable parallax measurements from {\em Gaia} and red colors from {\em WISE} and NSC. The assembled sample --- extending to $G\sim21$ ({\em Gaia}) and $r\sim23$ (NSC) --- was designed to find and study ejected white dwarfs (from double-degenerate binaries), ancient white and brown dwarfs, and hypervelocity stars. Fainter candidates were observed in dark time ({\tt FAINT\_HPM}) and brighter candidates in bright time ({\tt BRIGHT\_HPM}).

% ADM Antonella Palmese, Segev BenZvi, Alex Kim, Ellianna Abrahams, James Annis, Tamara Davis, Ofer Lahav, Risa Wechsler
\subsection{{\tt GW190412}, {\tt IC134191}}

These bits were intended to be used for dedicated, rapid-turnaround observations of DESI ``Targets of Opportunity'' (ToOs) in the vicinity of a gravitational wave signal or an IceCube high-energy neutrino event. The archival gravitational wave alert chosen to mimic a real trigger was  GW190412 \citep{GW190412}, and a real-time follow-up was performed for the ``gold" neutrino event 134191\_17593623.\footnote{See, e.g., \url{https://gcn.gsfc.nasa.gov/amon_icecube_gold_bronze_events.html}.} %Although these simulated tests were never actually conducted, 
{\tt GW190412} and {\tt IC134191} targets were assigned in some files associated with versions {\tt 0.48.0}, {\tt 0.49.0} and {\tt 0.50.0} of the {\tt desitarget} code. The follow-up of {\tt GW190412} was performed two years after the gravitational wave event, since gravitational wave detectors were not operating during SV, both as a test and to provide spectroscopic redshifts for a standard siren measurement \citep{2021arXiv211106445P}.
On the other hand, real time follow-up of ToOs by DESI \citep[e.g.][]{palmese21} was achieved by prioritizing tiles near {\tt IC134191} during afternoon planning \citep[see][]{schlafly23a} or by assigning the {\tt BRIGHT\_TOO\_LOP}, {\tt BRIGHT\_TOO\_HIP}, {\tt DARK\_TOO\_LOP} and {\tt DARK\_TOO\_HIP} bits discussed in \S\ref{sec:ToO}.

% ADM Greg Aldering, Segev BenZvi, Chris Blake, Tamara Davis, Kelly Douglass, Cullan Howlett, Alex Kim, Anthony Kremin, John Lucey, David Parkinson, Fei Qin, Khaled Said, Christoph Saulder, Pauline Zarrouk; CHECKED by Douglass
\subsection{{\tt PV\_BRIGHT}, {\tt PV\_DARK}}
\label{sec:pvsv1}

The low-redshift ($z < 0.15$) DESI Peculiar Velocity (PV) Survey was designed to improve constraints on the growth rate of structure \citep[see, e.g.][]{pv,Kim2020}. The survey comprised three samples \citep[see, e.g.,][]{pvtarget}. First, the fundamental plane (``FP'') sample, which included bright ($r < 18$), elliptically shaped, galaxies, helps characterize the fundamental plane. Second, the Tully-Fisher (``TF'') sample, which included locations along the major axes of SGA galaxies \citep{moustakas23a} with spiral-like colors, probes the TF relation. Third, the ``extended'' sample, which covered positions across the surfaces of large ($> 2\times1.4\arcmin$) SGA galaxies, fills in areas that have no primary DESI science targets due to fiber-patrol limitations. The {\tt PV\_BRIGHT} targets were observed in bright time, and included TF and ``extended'' targets. The {\tt PV\_DARK} sample was observed in dark time, and included FP and TF targets.

% ADM Elise Darragh-Ford, Risa Wechsler, Jeremy Tinker, John Moustakas, Ethan Nadler, Antonella Palmese, Greg Aldering, Segev BenZvi, Kelly Douglass, ChangHoon Hahn, and David Weinberg
\subsection{{\tt LOW\_Z}}
\label{sec:lowzsv1}

This campaign used imaging and photometric data from the Legacy Surveys to identify moderately faint ($19 < r < 21$) very-low-redshift ($z < 0.03$) galaxies. The target space was constrained using cuts on surface brightness and color adapted from the SAGA Survey \citep{geha17, mao21} with minor adjustments. A machine learning method (convolutional neural network; see \citealt{wu2022}) was applied to prioritize the selection of most likely nearby galaxy candidates. Science goals included refining the luminosity function of dwarf galaxies and a census of possible gravitational wave hosts in the local universe. The {\tt LOW\_Z} survey was designed as a ``filler'' class of several hundred targets per sq.\ deg.\ scheduled at a very low priority. The campaign is detailed in \citet{lowz}.

% ADM Sergey Koposov, Carlos Allende, Andrew Cooper, Arjun Dey, Joan Najita, Monica Valluri, Ting Li
\subsection{{\tt BHB}}

The {\tt BHB} sample extended the MWS blue horizontal branch program \citep{cooper22a} to fainter ($19 < g < 21$) targets that required dark-time observations. The {\tt BHB} sample was designed to probe stellar populations and kinematics at distances of $\sim$150\,kpc to constrain the dark matter mass distribution in the outermost Galaxy. Targets were color-selected at a (sub-sampled) density of $\sim2\,{\rm deg}^{-2}$ using a combination of $g-r$ and $r-z$ from the Legacy Surveys \citep[similar to, e.g.,][]{bhb} to separate BHB stars from blue stragglers, quasars and white dwarfs. Further cuts on {\em WISE W}1 and {\em Gaia G} were applied to remove residual quasars. A few targets from RR Lyrae catalogs derived by the {\em Gaia} collaboration or \citet{sesar17} were included in the {\tt BHB} sample.

% ADM Ellianna S. Abrahams, Joshua S. Bloom, Segev BenZvi, Antonella Palmese, Peter Nugent & Alex Kim
\subsection{{\tt SPCV}}

This project aimed to catalog --- and obtain multi-epoch spectra of --- short-period cataclysmic variable stars (spCVs). Science goals included characterizing sources in the ``cataclysmic variable period gap'' of $\sim$2--3 hours \citep[e.g.][]{knigge11}, and finding reference spCVs for the {\em LISA} mission to use as verification binaries \citep[e.g.][]{cornish17}. Targets were selected using {\em Gaia} colors combined with variability amplitudes of $>0.25$ in {\em Gaia G} \citep[see][]{spcv}. The resulting sample was limited to $16 < G < 21$, producing $\sim$1300 candidate spCVs spread across the Milky Way.

% ADM Daniel Gruen, Jamie McCullough, Alexandra Amon, Chris Blake, Becky Canning, Francisco Castander, Joe DeRose, Dan Masters, Ramon Miquel, Justin Myles, Jeff Newman, Aaron Roodman, Anze Slosar, Josh Speagle, Michael Wilson
\subsection{\tt DC3R2\_GAMA}

Spectra were obtained to characterize the relation between $ugriZYJHK$ multi-color and redshift for photo-$z$ calibration across $\approx50\%$ of the color space visible to the Vera C. Rubin Observatory's Legacy Survey of Space and Time (LSST, \citealt{ivezic19}) and Euclid \citep{euclid2022_overview}. Targets were selected from public KiDS+VIKING optical-NIR imaging \citep{wright19,kuijken19} in the GAMA 9h, 12h, and 15h equatorial fields at $z_{\rm fiber} < 22.1$ and assigned to the self-organizing map of the C3R2 survey \citep{dc3r2gama1, dc3r2gama2} transformed to KiDS-VIKING color space. From these, a sample of 13270 spectra from 10376 unique targets was observed on dedicated tiles 80971--80975. Targets were prioritized to maximize the ability to constrain the slope of redshift with respect to magnitude at fixed color, which is a main uncertainty of the C3R2 approach for direct calibration of redshift distributions for faint photometric galaxy samples. Some additional {\tt DC3R2\_GAMA} targets were observed using spare fibers during the course of SV. The first results from this campaign are presented in \citet{mccullough23}.

% ADM Alex Krolewski, Simone Ferraro, Zack Slepian, Eddie Schlafly, Martin White, Rongpu Zhou. CHECKED by Alex Krolewski
\subsection{{\tt UNWISE\_BLUE}, {\tt UNWISE\_GREEN}}
\label{sec:unwiseearly}

This dedicated dark-time program was designed to calibrate the redshift distribution of galaxies for CMB lensing tomography measurements. Targets were randomly sub-selected from the ``blue'' ({\tt UNWISE\_BLUE}) and ``green'' ({\tt UNWISE\_GREEN}) samples of \citet{krolewski20}, which were derived from $W1-W2$ color cuts in the unWISE catalog of \citet{unwise}. The 6'' WISE PSF is too broad to reliably center DESI fibers on the galaxy of interest, so targets were additionally matched to the nearest Subaru Hyper Suprime-Cam (HSC) imaging (with a 2.75\arcsec\ radius) and limited to $y < 22.5$ ($y < 24$) for the blue (green) sample, resulting in $\sim$9{,}000 ($\sim$4{,}500) {\tt UNWISE\_BLUE} ({\tt UNWISE\_GREEN}) targets. The main goal of the program was to improve cosmological constraints from Planck-unWISE lensing measurements by reducing uncertainty in the unWISE redshift distribution, a primary contributor to $S_8$ uncertainty \citep[see, e.g.,][]{krolewski21}. This program was refined in later iterations of SV1 --- files associated with versions {\tt 0.51.0} and {\tt 0.52.0} of the {\tt desitarget code} were supplemented with the additional bits listed in \S\ref{sec:unwise}.

% ADM Martin Landriau
\subsection{{\tt HETDEX\_MAIN}, {\tt HETDEX\_HP}}

The HETDEX dedicated dark-time campaign pursued higher-resolution spectra of a few thousand Lyman-$\alpha$ emitters (LAEs) from the Hobby-Eberly Telescope Dark Energy Experiment \citep{hetdex}. HETDEX targets LAEs in the redshift range $1.9~\leqsim\,z~\leqsim\,3.5$ but its relatively meager spectral resolution \citep[$\sim$800;][]{hetdexres} spawns contamination by low-redshift \otwo\ emitters. The DESI observations sought to characterize this contamination, while also preparing for future DESI-like experiments using LAEs. {\tt HETDEX\_MAIN} targeted HETDEX sources with spectral signal-to-noise $> 5.2$, supplemented by a few hundred lower-significance emitters. {\tt HETDEX\_HIP} targets comprised a few dozen faint, high-redshift LAEs to help characterize DESI detection limits for HETDEX sources.

% ADM David Schlegel, Arjun Dey, Doug Finkbeiner, Aaron Meisner, and Eddie Schlafly
\subsection{{\tt PSF\_OUT\_BRIGHT}, {\tt PSF\_OUT\_DARK}}

The PSF outliers campaign targeted point sources ({\tt TYPE=="PSF"}) in the Legacy Surveys that lie more than 10$\sigma$ from the stellar locus in $grz$ \citep[in the spirit of, e.g.,][]{psfout}. The {\tt PSF\_OUT\_BRIGHT} ({\tt PSF\_OUT\_DARK}) targets were intended for bright-time (dark-time) observations and were limited to $15 < r < 19$ ($16 < r < 22$). The program was designed as a ``filler'' survey --- it comprised a little more than one hundred total targets per sq.\ deg.\ and was scheduled at very low priority to mop up spare fibers. The true target density was lower as $\sim80\%$ of outliers from the stellar locus --- such as candidate quasars, white dwarfs and compact galaxies --- were already targeted by DESI.

% ADM Maayane Soumagnac, Peter Nugent CHECKED
\subsection{{\tt HPM\_SOUM}}

The {\tt HPM\_SOUM} survey pursued spectroscopy of high proper motion stars across the DESI footprint. The scientific goals of the program were similar to those for the {\tt FAINT\_HPM} campaign (see \S\ref{sec:hpm}). The target list comprised $\sim$2{,}900 faint ($r > 19.5$) stars with high proper motion ($\geqsim 200$\,mas\,yr$^{-1}$). These targets were drawn from the sample of \citet{hpm}, which derived proper motions by comparing the positions of sources between the Sloan Digital Sky Survey (SDSS) and Pan-STARRS1.

% ADM Maayane Soumagnac, Peter Nugent, Alex Kim, Segev BenZvi, Tamara Davis, Satya Gontcho a Gontcho, Or Grauer, John Moustakas, Antonella Palmese, Greg Aldering CHECKED
\subsection{{\tt SN\_HOSTS}}

This program targeted $\sim$20{,}000 supernova hosts and nuclear variables from the Nearby Supernova Factory \citep[e.g.][]{snf}, Palomar Transient Factory \citep[e.g.][]{ptf}, SDSS-II Supernova Survey \citep{sdssii} and Zwicky Transient Facility \citep[ZTF; e.g.][]{ztf, bts}. DESI spectroscopy is particularly warranted as the ZTF ``SED machine'' \citep{sedm} can only obtain low-resolution ($R\sim100$) spectra. As about half of the {\tt SN\_HOSTS} data set was already targeted by the DESI BGS, the true {\tt SN\_HOSTS} sample only comprised $\sim$10{,}000 targets. Scientific goals included; probing correlations between the properties of supernovae and their host galaxies; improving cosmological constraints from supernovae; using direct, supernova-based distance measurements to characterize peculiar velocities and the Fundamental Plane of host galaxies; and enhancing populations of changing-look AGN and Tidal Disruption Events.

% ADM Jesse Golden-Marx, Ying Zu, Alexie Leauthaud, Daniel Gruen, Jeff Newman, Jaime Forero-Romero, Xiaohu Yang, Eduardo Rozo, Misha Barth, Justin Myles
\subsection{{\tt GAL\_CLUS\_BCG}, {\tt GAL\_CLUS\_2ND}, {\tt GAL\_CLUS\_SAT}}

This dark-time campaign sought to build a volume-complete sample of galaxy clusters with spectroscopically confirmed members. Targets were compiled from galaxies with a probability of cluster membership of $P_{\rm mem} > 0.90$ in version 6.3 of the SDSS redMaPPer catalog \citep[see, e.g.,][]{redmapper}. The {\tt GAL\_CLUS\_BCG} bit denotes Brightest Cluster Galaxies --- i.e.\ the redMaPPER most-probable central galaxy --- to a redshift of $z < 0.35$. The {\tt GAL\_CLUS\_SCND} bit signifies the {\it second} brightest cluster member candidate. The {\tt GAL\_CLUS\_SAT} bit denotes all other ($P_{\rm mem} > 0.90$) candidate cluster members to $z < 0.30$. {\tt GAL\_CLUS\_BCG} targets were prioritized for DESI observations over {\tt GAL\_CLUS\_2ND}, which were in turn prioritized over {\tt GAL\_CLUS\_SAT} targets. After removing existing DESI targets, the sample comprised a few 100 (each) {\tt GAL\_CLUS\_BCG} and {\tt GAL\_CLUS\_SCND} targets and $\sim$10{,}800 {\tt GAL\_CLUS\_SAT} targets. 

% ADM Greg Aldering, Saul Perlmutter, David Schlegel, David Rubin, Nao Suzuki, Nicolas Regnault
\subsection{{\tt HSC\_HIZ\_SNE}}

This dedicated dark-time program focused on obtaining redshifts for supernova host galaxies identified in a deep cadenced HSC survey in the COSMOS field \citep[see][]{hscsn}. The target sample consisted of 1036 supernova candidates \citep[out of the 1824 candidates detailed in][]{hscsn} that lacked spectroscopy when the {\tt HSC\_HIZ\_SNE} observations were proposed. A little more than 400 of these candidates were expected to be cosmologically important Type Ia supernovae. The main scientific goal of this program was to double the number of known Type~Ia supernovae at redshifts of $z > 1$ and hence improve constraints on the dark energy equation of state.

% ADM Siwei Zou, Linhua Jiang, Zheng Cai, Jiani Ding
\subsection{{\tt ISM\_CGM\_QGP}}

The {\tt ISM\_CGM\_QGP} campaign sought to probe the circumgalactic medium (CGM) by targeting 114 quasars (S/N $>$ 3) with sight-lines that pass within 30\arcsec\ ($\sim$250\, kpc at z = 2.0) of a galaxy in the COSMOS \citep{ismcosmos} or HSC Ultra-Deep \citep{ismhsc} fields. Targets were selected from SDSS DR14 quasars \citep{dr14qso} with $g < 22$. The cool gas galaxy counterparts is selected from the COSMOS2020 catalog \citep{weaver22}. The main goals of the program \citep[see also][]{ism} were to probe the metal budget in the CGM, and to characterize how the CGM is influenced by the properties of proximate galaxies. 

% ADM C. Storfer, D.J. Schlegel, A. Dey, A. Gu, G. Aldering, S. Banka, M. Landriau, D. Lang, A. Meisner, J. Moustakas, A.D. Myers, S. Perlmutter, D. Rubin, E.F. Schlafly, W. Sheu, N. Suzuki, X. Huang
\subsection{{\tt STRONG\_LENS}}

This program sought to obtain redshifts for strong gravitational lenses identified in DESI Legacy Surveys imaging \citep[see][]{stronglens1, stronglens2}. The brightest image of each lensed source and $\sim$20\% of putative lensing galaxies were scheduled for observations, resulting in a total sample of 3588 targets spread throughout the DESI footprint. The main purpose of obtaining spectroscopic redshifts for the {\tt STRONG\_LENS} sample was to improve lensing models for these systems. Scientific goals included probing dark matter halo density profiles and sub-halo abundances, and identifying superior systems to search for multiply imaged supernovae.

% ADM Christophe Yeche (no proposal submitted)
\subsection{{\tt WISE\_VAR\_QSO}}

The {\tt WISE\_VAR\_QSO} bit denotes quasar targets selected via variability estimated using the {\em WISE} ``light curve sweeps'' supplied with DR9 of the Legacy Surveys.\footnote{\url{https://www.legacysurvey.org/dr9/files/\#light-curve-sweeps-9-0-lightcurves-sweep-brickmin-brickmax-lc-fits}} The selection technique was based on cuts in structure-function-space (represented by the parameters $A$ and $\gamma$) in a similar fashion to \cite{ebossqso} (see \S4.2.1) and \citet{sdssvarqso}. The resulting target density was a little more than 100\,deg$^{-2}$ across most of the DESI footprint. As with all quasar-like classes throughout the DESI survey \citep[see \S5 of][]{schlafly23a}, {\tt WISE\_VAR\_QSO} targets were scheduled for 4 total observations (starting with SV3). The main goal of the {\tt WISE\_VAR\_QSO} sample was to expand the pool of quasars recovered by DESI that could be used for studies of the Lyman-$\alpha$ Forest. 

%Ting Li
\subsection{{\tt MWS\_CALIB}, {\tt BACKUP\_CALIB}}

These target classes indicate calibration sources that were adopted for the SV1 stellar survey described in \citet{cooper22a}. {\tt MWS\_CALIB} and {\tt BACKUP\_CALIB} targets were selected from publicly available survey catalogs (e.g.\ SDSS Segue, APOGEE, GALAH, and the {\it Gaia} ESO Survey). {\tt BACKUP\_CALIB} targets were limited to the magnitude range $10 < G < 16$ and {\tt MWS\_CALIB} targets were limited to $16 < G <19$.

%Ting Li
\subsection{{\tt MWS\_MAIN\_CLUSTER\_SV}}

This class flags targets observed as part of the SV1 stellar survey described in \citet{cooper22a}. Targets were selected to be ``likely'' members of Milky Way globular clusters. ``Likely,'' here, corresponds to a membership probability of $P > 0.3$, with $P$ as defined in \citet{Vasiliev2021}. Targets were limited to a magnitude range of $16 <G <20$. 

% Sergey
\subsection{{\tt MWS\_RRLYR}}

This target class formed part of the SV1 stellar survey described in \citet{cooper22a}. The selection targets stars that are likely RR Lyrae variables based on {\it Gaia} DR2. It combines sources that were labeled as RR Lyrae by the Specific Object Study pipeline \citep{clementini2019} and the general variability processing pipeline \citep{holl2018}, limited to $14 < G < 19$. The sample can be reproduced by running the following {\it Gaia} archival query:
\\
\\
{ \tt WITH x as \\
(SELECT vari_classifier_result.source_id \\
FROM gaia_dr2.vari_classifier_result \\
WHERE vari_classifier_result.best_class_name\\
::text ~~ `RR\%'::text \\
UNION SELECT vari_rrlyrae.source_id
        FROM gaia_dr2.vari_rrlyrae) \\
SELECT g.* FROM gaia_dr2.gaia_source as g, x
       where g.source_id = x.source_id
       and phot_g_mean_mag between 14 and 19;
}

% ADM Boris Gansicke, Arjun Dey, Jay Farihi, Joan Najita, MWS Group
\subsection{{\tt WD\_BINARIES\_BRIGHT}, {\tt WD\_BINARIES\_DARK}}
\label{sec:wdbins}

This campaign pursued a representative sample of all types of white dwarf binaries. The broad scientific focus was to characterize the entire dynamical range of bound white dwarfs, from intrinsically bright, high-mass transfer binaries to extremely faint, highly evolved systems. Targets were selected by cross-matching the Galaxy Evolution Explorer ({\em GALEX}) source catalog with {\em Gaia} and retaining sources with an absolute magnitude of $M_{FUV} > 1.5(FUV-G)-0.3$, which generally lie below the main sequence. Here, $M_{FUV}$ is calculated using distances derived from {\em Gaia} parallaxes that are measured to $\geq5\sigma$, and $FUV$ and $G$ represent magnitudes in {\em GALEX} FUV and {\em Gaia} $G$-band. After removing existing DESI targets, this sample comprised $\sim$28{,}300 ($\sim$7{,}400) sources with $16 \leq G \leq 18$ ($G > 18$). The brighter (fainter) of these subsamples is signified using the {\tt WD\_BINARIES\_BRIGHT} ({\tt WD\_BINARIES\_DARK}) bit and scheduled for observations in bright (dark) time.

% ADM Michael J. Wilson, David Schlegel, Kyle Dawson, Marcin Sawicki, Martin White, Arjun Dey, Zheng Cai, Simone Ferraro
\subsection{{\tt DESILBG}}
\label{sec:desilbgearly}

This project sought $\sim$10 hours of dedicated dark-time observations in fields covered by the CFHT Large Area $U$-band Deep Survey \citep[CLAUDS;][]{clauds}. By supplementing CLAUDS with deep $grz$ imaging from HSC the project aimed to target $\sim$5{,}000 Lyman Break Galaxies (LBGs) and LAEs in the redshift range $2 < z < 4$. The main goal was to prepare for future DESI-like experiments by characterizing a population of high-density, high-redshift tracers with which to improve cosmological constraints at times before dark energy began to dominate the Universe. Three different approaches were adopted to target redshifts near $z\sim2$ \citep[the BX technique, see, e.g.][]{BX} and $z\sim3$--4 \citep[$u$- and $g$-dropout techniques, see., e.g.][]{lbgdropout}. The {\tt DESILBG} bit was not introduced until version {\tt 0.51.0} of the {\tt desitarget} code and it was rapidly replaced by the bits described in \S\ref{sec:desilbg}.

% ADM Corentin Ravoux, Christophe Yeche, Nathalie Palanque-Delabrouille, Eric Armengaud, Alex Krolewski, Michael J. Wilson, David Schlegel, Martin White, Kyle Dawson, Arjun Dey, Simone Ferraro, Zheng Cai
\subsection{{\tt LBG\_TOMOG\_XMM}, {\tt LBG\_TOMOG\_COSMOS}, {\tt LBG\_TOMOG\_W3}, {\tt LBG\_TOMOG\_COSMOS\_FINAL}}
\label{sec:lbgtomog}

These targeting bits denote updates to the {\tt LBG\_TOMOG} program described in \S\ref{sec:lbgtomogearly}. Overall, the science program remained the same but the selection was tweaked slightly, or applied in a new field, as outlined at each bit's {\tt docs} link (see the introduction to this appendix). The {\tt LBG\_TOMOG\_XMM} and {\tt LBG\_TOMOG\_W3} bits were introduced in version {\tt 0.51.0} of the {\tt desitarget} code (and targeting files) and were also populated in version {\tt 0.52.0}. The {\tt LBG\_TOMOG\_COSMOS} bit was introduced in {\tt 0.51.0} and quickly deprecated in favor of the {\tt LBG\_TOMOG\_COSMOS\_FINAL} bit for {\tt 0.52.0}.

% ADM Alex Krolewski, Simone Ferraro, Zack Slepian, Eddie Schlafly, Martin White, Rongpu Zhou. CHECKED by Alex Krolewski
\subsection{{\tt UNWISE\_GREEN\_II\_3700}, {\tt UNWISE\_GREEN\_II\_3800}, {\tt UNWISE\_GREEN\_II\_3900}, {\tt UNWISE\_GREEN\_II\_4000}, {\tt UNWISE\_BLUE\_FAINT\_II}, {\tt UNWISE\_BLUE\_BRIGHT\_II}}
\label{sec:unwise}

These bits were introduced in version {\tt 0.51.0} of the {\tt desitarget} code to augment the {\tt UNWISE\_BLUE} and {\tt UNWISE\_GREEN} targets. The science goals outlined in \S\ref{sec:unwiseearly} were unchanged, but adding bits facilitated finer-grained control of how targets were prioritized for DESI observations. The {\tt UNWISE\_GREEN\_II\_4000} targets were assigned the highest priority, followed, in order, by {\tt UNWISE\_GREEN\_II\_3900}, {\tt UNWISE\_GREEN\_II\_3800}, {\tt UNWISE\_GREEN\_II\_3700}, {\tt UNWISE\_BLUE\_FAINT\_II} and, finally, {\tt UNWISE\_BLUE\_BRIGHT\_II} at the lowest priority. The {\tt UNWISE\_BLUE\_BRIGHT\_II} and {\tt UNWISE\_BLUE\_FAINT\_II} targets were split at a Legacy Surveys fiber magnitude of $z_{\rm fiber} = 21.1$. The green targets were split using {\em WISE} colors according to $W1-W2 < 0.8,\,W1 < 17.0$ ({\tt UNWISE\_GREEN\_II\_4000}); $W1-W2 < 0.8,\,W1 < 17.2$ ({\tt UNWISE\_GREEN\_II\_3900}); and $W1-W2\,< 0.8$ ({\tt UNWISE\_GREEN\_II\_3800}); with remaining sources from the original {\tt UNWISE\_GREEN} selection signified by {\tt UNWISE\_GREEN\_II\_3700}. The split in the blue sample allowed for longer exposure times on faint targets, whereas the split in the green sample de-prioritized galaxies at $z > 1.6$, where the redshift completeness is poor due to the \otwo line and 4000 \AA\ break redshifting out of the DESI wavelength range.

% ADM Michael J. Wilson, David Schlegel, Kyle Dawson, Marcin Sawicki, Martin White, Arjun Dey, Zheng Cai, Simone Ferraro
\subsection{{\tt DESILBG\_TMG\_FINAL}, {\tt DESILBG\_G\_FINAL}, {\tt DESILBG\_BXU\_FINAL}}
\label{sec:desilbg}
Starting with version {\tt 0.52.0} of the {\tt desitarget} code, the {\tt DESILBG} sample was split into several subsamples to make it easier to track the provenance of each targeting approach described in \S\ref{sec:desilbgearly}. The BX selection \revisionforreviewer{\citep[using the BX technique, see e.g.][]{BX}} and $u$-dropout targets were signified by the {\tt DESILBG\_BXU\_FINAL} bit; the $g$-dropouts were signified by the {\tt DESILBG\_G\_FINAL} bit; and a new selection that resembled the {\tt LBG\_TOMOG} target class outlined in \S\ref{sec:lbgtomogearly} was built. A detailed code to derive each of these subsamples is available on GitHub.\footnote{See \url{https://github.com/michaelJwilson/DESILBG/blob/1.0.2/gold/README}.}

\subsection{{\tt BRIGHT\_TOO}, {\tt DARK\_TOO}}

These target classes were intended to flag general ``Targets of Opportunity'' (ToOs) during the SV1 phase of DESI. In actuality, ToOs were not tracked until SV3 and these bits were replaced by the bits described in \S\ref{sec:ToO}.

\begin{deluxetable*}{cccccc}[t]
\tablecaption{Updates to secondary targeting bits for SV3}\label{table:sv3sec}
\tablewidth{0pt}
\tablehead{
\colhead{Bit-name} & 
\colhead{Bit-value} &
\colhead{Bit-name} & 
\colhead{Bit-value} &
\colhead{Bit-name} & 
\colhead{Bit-value}
}
\startdata
{\tt LOW\_Z\_TIER1}    & 15 & {\tt PV\_BRIGHT\_MEDIUM} & 44 & {\tt BRIGHT\_TOO\_LOP} & 59  \\
{\tt LOW\_Z\_TIER2}    & 16 & {\tt PV\_BRIGHT\_LOW}    & 45 & {\tt BRIGHT\_TOO\_HIP} & 60  \\
{\tt LOW\_Z\_TIER3}    & 17 & {\tt PV\_DARK\_HIGH}     & 46 & {\tt DARK\_TOO\_LOP}  & 61  \\
{\tt Z5\_QSO}          & 36 & {\tt PV\_DARK\_MEDIUM}   & 47 & {\tt DARK\_TOO\_HIP}  & 62  \\
{\tt PV\_BRIGHT\_HIGH} & 43 & {\tt PV\_DARK\_LOW}      & 48 &   &   \\
\enddata
\begin{center}
\tablenotetext{}{These bits are taken from the {\tt desitarget} code\footnote{See \url{https://github.com/desihub/desitarget/blob/2.5.0/py/desitarget/sv3/data/sv3_targetmask.yaml\#L132-L173}.} and are described in the body of the Appendix. In addition to the changes listed in the table, many bits from SV1 were not retained or set in SV3.}
\end{center}
\end{deluxetable*}

% ADM Elise Darragh-Ford, Risa Wechsler, Jeremy Tinker, John Moustakas, Ethan Nadler, Antonella Palmese, Greg Aldering, Segev BenZvi, Kelly Douglass, ChangHoon Hahn, and David Weinberg
\subsection{{\tt LOW\_Z\_TIER1}, {\tt LOW\_Z\_TIER2}, {\tt LOW\_Z\_TIER3} }
When DESI moved to its SV3 phase, the {\tt LOW\_Z} targets described in \S\ref{sec:lowzsv1} were split into three tiers to allow different target classes to be assigned different observational priorities. The highest priority targets ({\tt LOW\_Z\_TIER1}) contain the most likely low-redshift candidates based on the prediction of a machine learning method (convolutional neural network; see \citealt{wu2022}). The next-highest priority targets ({\tt LOW\_Z\_TIER2}) comprised objects that are in a tighter photometric space where most low-redshift candidates are situated (as introduced in \citealt{mao21}). The {\tt LOW\_Z\_TIER2} sample was designed to {\em not} overlap with BGS targets. Finally, the lowest priority targets ({\tt LOW\_Z\_TIER3}) were based on the remaining targets within the overall photometric cuts (referred to as  ``$z<0.03$-complete cuts'' in  \citealt{lowz}). The {\tt LOW\_Z\_TIER3} sample was allowed to overlap with BGS targets. Again, see \citet{lowz} for a full description of the {\tt LOW\_Z} program.

% ADM Jinyi Yang, Xiaohui Fan, Feige Wang
\subsection{{\tt Z5\_QSO}}
The {\tt Z5\_QSO} program targeted quasars at redshifts of $5.0~\leqsim\,z~\leqsim\,6.5$ based on color cuts applied to Legacy Surveys imaging ($grzW1W2$) supplemented by $i$- and $y$-band from Pan-STARRS1 \citep[see, e.g.,][]{z5qsoimaging1, z5qsoimaging2}. Where available, $J$-band from public NIR surveys was also incorporated to reject stellar contaminants. 
%The main color criteria, which are detailed in \citet{z5qso}, were applied in $r-i$/$i-z$ to target redshifts of $5~\leqsim\,z~\leqsim\,6$, $i-z$/$z-y$ to target redshifts of $6.0~\leqsim\,z~\leqsim\,6.5$, and $z$/$y-W1$/$W1-W2$ to target the entire redshift range of interest. 
The main color criteria, which are detailed in \citet{z5qso}, produce a target sample with a density of $\sim0.5\,{\rm deg}^{-2}$. Science goals included constraining the quasar luminosity function at high redshift, building a sample to study the intergalactic medium near the epoch of reionization, and probing supermassive black hole growth in the early Universe. As with all DESI quasar-like targets \citep[see \S5 of][]{schlafly23a}, the {\tt Z5\_QSO} sample was scheduled for 4 total observations.

\subsection{{\tt PV\_BRIGHT\_HIGH}, {\tt PV\_BRIGHT\_MEDIUM}, {\tt PV\_BRIGHT\_LOW}, {\tt PV\_DARK\_HIGH}, {\tt PV\_DARK\_MEDIUM}, {\tt PV\_DARK\_LOW}}

Starting with SV3, the {\tt PV\_BRIGHT} and {\tt PV\_DARK} samples described in \S\ref{sec:pvsv1} were split into three sub-classes (each) to facilitate them being scheduled for observations at different priorities \citep[for details, see Table 1 of][]{pvtarget}. Broadly, the highest-priority targets ({\tt PV\_BRIGHT\_HIGH}, {\tt PV\_DARK\_HIGH}) included ``FP'' (Fundamental Plane) targets and the subset of ``TF'' (Tully-Fisher) targets that were positioned along the axes of SGA galaxies. Then, the medium-priority targets ({\tt PV\_BRIGHT\_MEDIUM}, {\tt PV\_DARK\_MEDIUM}) included all other ``TF'' targets. Finally, any additional PV targets were signified by the lowest-priority targeting bits ({\tt PV\_BRIGHT\_LOW}, {\tt PV\_DARK\_LOW}). {\tt PV\_BRIGHT\_HIGH} and {\tt PV\_DARK\_HIGH} were scheduled for extra observations during SV3 (a total of five) to improve spectral signal-to-noise.

\subsection{{\tt BRIGHT\_TOO\_LOP}, {\tt BRIGHT\_TOO\_HIP}, {\tt DARK\_TOO\_LOP}, {\tt DARK\_TOO\_HIP}}
\label{sec:ToO}

These bits were used to handle ``Targets of Opportunity'' (ToOs) in SV3 --- i.e.\ targets such as transients that need to be observed at short notice. Transient ToOs were selected from the DECam Survey of Intermediate Redshift Transients (DESIRT; \citealt{2022TNSAN.107....1P}), and will be described in \citet{palmese23}.
In the bit-names, {\tt DARK} signifies a ToO to be observed in dark time and {\tt BRIGHT} denotes a ToO that can be scheduled for {\em either} dark- {\em or} bright-time observations. {\tt LOP} indicates a ``low priority'' ToO, which would be prioritized below all primary targets and most secondaries.\footnote{The ``filler'' targets such as the {\tt LOW\_Z} and {\tt PSF\_OUT} targets are scheduled at a priority lower than low-priority ToOs.} {\tt HIP} signifies a ``high priority'' ToO, which would be prioritized above {\em all} other DESI targets, including primaries. The general mechanisms by which ToOs are handled are discussed more in \S3.2.2 of \citet{myers23a} and \S5.4 of \citet{schlafly23a}.

% \end{appendix}

\bibliographystyle{aasjournal}
\bibliography{references}

\end{document}